\documentclass[smallextended]{svjour3}     

\usepackage[left=2cm,top=2cm,right=2cm,bottom=2cm]{geometry}
\usepackage{mathptmx}
\usepackage{helvet}
\usepackage{courier}
\usepackage{makeidx}
\usepackage{multicol}
\usepackage{amsfonts}
\usepackage{graphicx}
\usepackage{amsmath}
\usepackage[authoryear]{natbib}
\usepackage{setspace}
\usepackage{fancybox}
\usepackage[bottom]{footmisc}
\usepackage{color}
\usepackage{url}
\usepackage{rotating}

\begin{document}

\title{Null Models of Economic Networks: The Case of the World Trade Web
}

\author{Giorgio Fagiolo \and Tiziano Squartini \and Diego Garlaschelli}

\institute{Giorgio Fagiolo (Corresponding Author) \at Sant'Anna School of Advanced Studies, Laboratory of Economics and Management, Piazza Martiri della Libert\`{a} 33, I-56127 Pisa, Italy. Tel: +39-050-883356 Fax: +39-050-883343 \email{giorgio.fagiolo@sssup.it} \and Tiziano Squartini \at CSC and Department of Physics, University of Siena, Via Roma 56, 53100 Siena (Italy) and Instituut-Lorentz for Theoretical Physics, Leiden Institute of Physics, University of Leiden, Niels Bohrweg 2, 2333 CA Leiden (The Netherlands) \email{squartini@lorentz.leidenuniv.nl}\and Diego Garlaschelli \at Instituut-Lorentz for Theoretical Physics, Leiden Institute of Physics, University of Leiden, Niels Bohrweg 2, 2333 CA Leiden (The Netherlands). \email{garlaschelli@lorentz.leidenuniv.nl}}

\titlerunning{}

\maketitle
\begin{abstract}
\noindent In all empirical-network studies, the observed properties of economic networks are informative only if compared with a well-defined null model that can quantitatively  predict the behavior of such properties in constrained graphs. However,  predictions of the available null-model methods can be derived analytically only under assumptions (e.g., sparseness of the network) that are unrealistic for most economic networks like the World Trade Web (WTW). In this paper we study  the evolution of the WTW using a recently-proposed family of null network models. The method allows to analytically obtain the expected value of any network statistic across the ensemble of networks that preserve on average some local properties, and are otherwise fully random. We compare expected and observed properties of the WTW in the period 1950-2000, when either the expected number of trade partners or total country trade is kept fixed and equal to observed quantities. We show that, in the binary WTW, node-degree sequences are sufficient to explain higher-order network properties such as disassortativity and clustering-degree correlation, especially in the last part of the sample. Conversely, in the weighted WTW, the observed sequence of total country imports and exports are not sufficient to predict higher-order patterns of the WTW. We discuss some important implications of these findings for international-trade models.

\vskip 0.6cm


\vskip 0.6cm


\end{abstract}


\onehalfspacing

\section{Introduction\label{Section:Introduction}}

In the last years an increasing number of contributions have been addressing the study of economic systems and their dynamics in terms of networks \citep{ScienceNets2009}. The description of an economic system as a network requires to characterize economic units (e.g., countries, industries, firms, consumers, individuals, etc.) as nodes and their market and non-market relationships as links between them. Successive snapshots of these interacting structures can give us a clue about how networked systems evolve in time. Heterogeneity of agent and link types can be easily considered. Nodes can be tagged with different characteristics or properties (e.g., economic size) and links may be defined to be directed or undirected, binary or weighted, etc., according to the focus of the analysis.        

The study of economic networks has recently proceeded along three main complementary directions. First, a large body of empirical contributions have investigated the topological properties of many real-world economic and social networks \citep{Caldarelli_2007_book}, ranging from macroeconomic networks where nodes are countries and linkages represent trade or financial transactions, all the way to firm and consumer networks where links represent knowledge or information exchange. This empirical-research program has generated a very rich statistical evidence, pointing to many differences and similarities in the way economic networks are shaped. As a consequence, a very fertile ground for theoretical work has emerged.

Second, a stream of theoretical research has explored efficiency properties of equilibrium networks arising in cooperative and non-cooperative game-theoretic setups, where players have the possibility to choose both their strategy in the game and whom to play the game with \citep{Goyal_2007_book,Vega-Redondo_2007_book,Jackson_2008_book}. Despite such models have been very successful in highlighting the role of network structure in explaining aggregate outcomes, they fell short from providing a framework where observed network regularities can be reproduced and explained. 

Third, a large number of contributions coming from a complex-network perspective have been developing simple stochastic models of graph evolution where nodes hold very stylized and myopic probabilistic rules determining their future connectivity patterns in the network \citep{Newman_2010_book}. The two foremost examples of such an approach are \cite{WattsStrogatz1998} small-world model and \citet{AlbertBarabasi2002} preferential-attachment model. Despite this family of stochastic models are able to reproduce observed economic-network patterns, the extent to which such stylized representations can be employed to understand causal relations between incentive-based choices made in strategic contexts and the overall efficiency of the long-run equilibrium networks is still under scrutiny.      

All that hints to the dramatic need for theoretical models that are able to reproduce and economically explain the observed patterns of topological properties in real-world networks. Despite we know a great deal about how economic networks are shaped in reality and what that means for dynamic processes going on over networked structures \citep[e.g., diffusion of shocks and contagion effects, cf. for example][]{Allen_Gale_2001,Battiston_etal_2009_NBERw15611}, we still lack a clear understanding of why real-world network architectures looks like they do, and how all that has to do with individual incentives and social welfare.

This paper contributes to the aforementioned debate by exploring an alternative approach to the trade-off between explanation and reproduction of topological properties, grounded in the generation of \textit{null (random) network models}. The idea is not new. Instead of building economically- or stochastically-based micro-foundations for explaining observed patterns, one tries to ask the question whether observed statistical-network properties may be simply reproduced by simple processes of network generation that only match some (empirically-observed) constraints, but are otherwise fully random. If they do, then the researcher may conclude that such regularities are not that interesting from an economic point of view, as no alternative, more structural, model would pass any test discriminating against the random counterpart. Conversely, if observed regularities cannot be reproduced by the null random model, we are led to argue that some more structural economic process may be responsible for what we observe. Null random-network models may therefore serve as a sort of sieve that can help us to discriminate between interesting and useless observed-network properties. Exactly as in statistics and econometrics one performs significance tests, null network models are very helpful to understand the distributional properties of a given network statistics, under very mild null hypotheses for the underlying network-generation process.\footnote{In economics the use of purely-random models is not new. Examples range from industrial agglomeration \citep{Ellison_Glaeser_1997,Rysman_Greenstein_2005} to international trade \citep{Armenter_Miklos_2010}.} 

Null (random) network models have been extensively used in the recent past \citep[see][for a review]{Squartini_Garlaschelli_2011}. Since the seminal work of \citet{Erdos_Renyi_1960} on random graphs, many alternative null network models have been proposed.\footnote{See for example \citet{Katz_Powell_1957,Holland_Leinhardt_1976,Snijders_1991,Rao_etal_1996,Kannan_etal_1999,Roberts_2000,Newman_2001,Shen-Orr_etal_2002,Maslov_etal_2004_rewiring_physa,Ansmann_Lehnertz_2011,Bargigli_Gallegati_2011}.} A useful way of classifying them is according to the constraints they pose in the way the otherwise-random mechanism of network construction works. A large number of contributions, for example, have been focusing on generating random networks able to control (exactly or on average) for the degree sequence in binary graphs, or for the strength sequence in weighted ones.\footnote{In what follows, given a statistic $X$ computed on the $N$ nodes of graph, we call ``sequence'' of $X$ the collection $\{x_i\}_{i=1}^{N}$. The degree of a node is defined as the number of links of a node. The strength of a node is the sum of all weights of the links of a given node. See Tables \ref{Tab:BinaryStats} and \ref{Tab:WeightedStats} for formal definitions.} This is reasonable, as degree and strength sequences are one of the most basic statistics characterizing graphs. It is therefore very important to study the properties of network statistics (other than degrees and strengths) in ensembles of otherwise fully-random graphs preserving those basic topological quantities (and thus looking somewhat similar to the observed one).

However, most of the existing network null-model methods suffer from
important limitations. A large class of algorithms generates randomized variants of a network computationally, through iterated ``moves'' that locally rewire the
original connections in such a way that the desired constraints remain
unchanged \citep{Shen-Orr_etal_2002,Maslov_etal_2002_rewiring_science,Maslov_etal_2004_rewiring_physa}. These approaches are extremely demanding in terms of
computation time. In order to obtain expectations from the null model,
one has indeed to constructively build many alternative random graphs
belonging to the desired family, then measure any target topological
property on each of such randomized graphs, and finally perform a
final sample average of this property.
At the opposite extreme, analytical approaches have been proposed in
order to obtain mathematical expressions characterizing the expected
properties, thus avoiding time-consuming randomizations \citep{Newman_2001,Chung_Lu_2002,Serrano_Boguna_2005,Bargigli_Gallegati_2011}. The problem with the latter approaches is that they
are only valid under specific hypotheses about the structure of the
original network. For instance, methods based on probability
generating functions are generally only valid for sparse and locally
tree-like (thus with vanishing clustering) networks \citep{Newman_2001}. Similarly, models predicting factorized connection
probabilities in binary graphs \citep{Chung_Lu_2002} or factorized
expected weights in weighted networks \citep{Serrano_Boguna_2005,Bargigli_Gallegati_2011} make (either explicitly or implicitly)
the assumption of sparse networks, as has been shown recently
\citep{Squartini_Garlaschelli_2011}.
Additionally, each method or algorithm is generally designed to
generate random networks satisfying a specific set of constraints
(e.g., degree sequence) and cannot be easily extended to cover
different sets of constraints (e.g., strength sequence, possibly in
directed-graph contexts).
For instance, a problem that inherently pervades random models of
weighted networks is the simplifying assumption of real-valued edge
weights \citep{Bhatta2007a,Ansmann_Lehnertz_2011,Fronczak_Fronczak_2011}. When made in models that specify the
strength sequence, this assumption leads to randomized ensembles of
networks where edges with zero weight have zero probability, so that
the typical networks are fully connected \citep{Ansmann_Lehnertz_2011,Fronczak_Fronczak_2011}. This actually makes the original
network an unlikely outcome of the model, rather than one with the
same probability as all other instances with the same sufficient
statistics (e.g. with the same strength sequence).

In this paper we employ a recently-proposed method that overcomes
all the above restrictions simultaneously \citep{Squartini_Garlaschelli_2011}. The method is analytical and therefore does not require
simulations to generate the family of all randomized variants of the
target network. This important property makes the method very fast and
strongly facilitates exhaustive analyses which require the analysis of
many networks, e.g. in order to track the temporal evolution of a
particular system or to study all the individual components of a
multi-network with many layers, or both \citep{Squartini_etal_2011a_pre,Squartini_etal_2011b_pre}.
At the same time, the method does not make assumptions about the
structure of the original network, and therefore works also for dense
and clustered networks.
Furthermore, the method can deal with binary graphs and weighted
networks in a unified fashion (in both cases, edges can be either
directed or undirected). In the weighted case, it exploits the natural
notion of a fundamental unit of weight to treat edge weights as
discrete and integer-valued, preventing randomized networks from
becoming fully connected. This property ensures that, even when
dealing with randomized weighted networks, the expected bare topology
is nontrivial and allows comparisons with that of the original
network.
In general, the method allows to set any given target topological
property of interest and to obtain the expected values and standard
deviations of the corresponding quantity over the family of all
randomized variants of the network that preserve some arbitrary local
structural properties.

We apply the method to the World Trade Web (WTW) network, also known as the International Trade Network (ITN). The WTW is a weighted-directed network, where nodes are countries and directed links represent the value trade (export) flows between countries in that year. We also study the binary projection of this network, where a directed link between country $i$ and country $j$ is in place if and only if $i$ exports to $j$. Therefore, the binary WTW maps trade relationships, whereas the weighted WTW accounts for heterogeneity of bilateral trade flows associated to trade partnerships.      

The study of the WTW has received a lot of attention in the last years.\footnote{See for example \citet{LiC03,SeBo03,Garla2004,Garla2005,Garla2007,serrc07,Bhatta2007a,Bhatta2007b,Fagiolo2008physa,Fagiolo2009pre,Fagiolo2008acs,Fagiolo2009jee,BariFagiGarla2010,Fagiolo2010jeic,Fagiolo_etal_2011physa,DeBene_Tajoli_2011}.} Despite we know a great deal about statistical regularities of the WTW, we still lack a clear understanding of whether such regularities can be really meaningful, or, conversely, whether they are just the effect of randomness, i.e. whether a simple null-network model could easily explain that evidence. 

This issue was already tackled in \citet{Squartini_etal_2011a_pre,Squartini_etal_2011b_pre}, who show that, for the 1992-2002 period, much of the binary WTW architecture (both at the aggregate and product-specific level) can be reproduced by a null model controlling for in- and out-degree, whereas weighted-network regularities cannot be fully explained by node-strength sequences. More specifically, observed patterns of network disassortativity and clustering can be fully predicted by degree sequences, whereas they become non-trivially deducible from null-network models controlling for node strengths.

These results have important consequences for international-trade issues. Indeed, controlling for in- or out-degree and strength means fixing local-country properties (e.g., involving direct bilateral relations only) that give us information about the number of trade partnerships and country total imports and exports. These are statistics that are traditionally employed by international-trade economists to fully characterize country-trade profiles. Conversely, higher-order network properties like assortativity or clustering are non-local, as they refer to indirect trade relations involving trade partners of a country's partners, and so on. The fact that higher-order properties cannot be explained by random-network models controlling for local-properties only implies that a network approach to the study of the WTW is able to discover fresh statistical regularities. In turn, this suggests that we require more structural models to explain why such higher-order property do emerge.

In this paper, we extend the analysis in \citet{Squartini_etal_2011a_pre,Squartini_etal_2011b_pre} and we analyze a longer time frame (1950-2000). This allows us to better understand if subsequent globalization waves have changed the structure of the WTW and whether local properties like node degrees and strengths have been playing the same role in explaining higher-order properties. We compare observed and expected directed-network statistics in both binary and weighted aggregate WTW for the period under analysis. Our results show that, in the binary WTW, knowing the sequence of node degrees, i.e. number of import and export partners of a country, is largely sufficient to explain higher-order network properties related to disassortativity and clustering-degree correlation, especially in the last part of the sample (i.e., after 1965). We also find that in the first part of the sample (before 1965) local binary properties badly predict the structure of the network, which however does not present any clear evident structural correlation pattern. We interpret this result in terms of pre-globalization features of the web of international-trade relations, mostly ruled by geographical constraints and political barriers. Our weighted network analysis conveys instead an opposite message: observed local properties (i.e. country total imports and exports) hardly explain any observed higher-order weighted property of the WTW. This implies that in the binary case node-degree sequences (local properties) become maximally informative and higher-order properties of the network turn out to be statistically irrelevant as compared to the null model. Conversely, in the weighted case, the observed sequence of total country imports and exports are never able to explain higher-order patterns of the WTW, making the latter fresh statistical properties in search of a deeper explanation.     

The rest of the paper is organized as follows. Section \ref{Sec:WTW} briefly reviews the recent literature on the WTW. Section \ref{Sec:Method} introduces the null model. Data and methodology are described in Section \ref{Sec:Data}. Sections \ref{Sec:Results} and \ref{Sec:Discussion} present and discuss the main results. Finally, Section \ref{Sec:Conclusions} concludes. 

\section{The World Trade Web: A Complex-Network Approach}\label{Sec:WTW}
The idea that international trade flows among countries can be
conceptualized by means of a network has been originally put forth
in sociology and political sciences to test some flavor of ``world system'' or ``dependency'' theory. According to the latter, one can distinguish between core and peripheral countries: the former would appropriate most of the surplus value added produced by the latter, which are thus prevented
from developing. Network analysis is then used to validate this
polarized structure of exchanges.\footnote{Cf., among others, \citet{SnKi79}, \citet{NeSm85}, \citet{SackC01}, \citet{Brei81}, \citet{SmWi92}, \citet{KiSh02}.}

More recently, the study of international trade as a relational network has been revived in the field of econophysics, where a number of
contributions have explored the (notionally) complex nature of the
WTW. The common goal of these studies is to
empirically analyze the mechanics of the international trade network
and its topological properties, by abstracting from any social and
economic causal relationships that might underlie them (i.e., a sort
of quest for theory-free stylized facts). 

From a methodological perspective, a great deal of contributions
carry out their analysis using a binary approach. In other words, a
link is either present or not in the network according to whether the trade flow that it carries is larger than a given lower
threshold.\footnote{There is no agreement whatsoever on the way this
threshold should be chosen \citep[see for example][]{KiSh02,SeBo03,Garla2004,Garla2005}. In what follows, in line with much of the existing literature, we straightforwardly
define a link whenever a non-zero trade flow occurs.} For instance,
\citet{SeBo03} and \citet{Garla2004} study the WTW using binary
undirected and directed graphs. They show that the WTW is
characterized by a disassortative pattern: countries with many trade
partners (i.e., high node degree) are on average connected with countries
with few partners (i.e., low average nearest-neighbor degree). Furthermore, partners of well connected countries are less interconnected than those of poorly connected ones, implying some hierarchical arrangements. In other
words, a negative correlation emerges between clustering and degree sequences. Remarkably, \citet{Garla2005} show that this evidence
is quite stable over time. This casts some doubts on whether
economic integration (globalization) has really increased in the
last 20 years. Furthermore, node-degree distributions appear to be very
skewed. This implies the coexistence of few countries with many
partners and many countries with only a few partners.

These issues are taken up in more detail in a few subsequent studies adopting a weighted-network approach to the study of the WTW. The motivation is that a binary approach may not be able to fully extract the wealth of information about the intensity of the trade relationship carried by each edge and
therefore might dramatically underestimate the role of heterogeneity
in trade linkages. This seems indeed to be the case:
\citet{Fagiolo2008physa,Fagiolo2009pre,Fagiolo2009jee} show that the statistical properties of the WTW viewed as a weighted undirected network crucially differ from those exhibited by its binary counterpart. For example, the strength distribution is highly left-skewed, indicating that a few intense trade connections co-exist with a majority of low-intensity
ones. This confirms the results obtained by \citet{Bhatta2007b} and
\citet{Bhatta2007a}, who find that the size of the  group of
countries controlling half of the world's trade has decreased in the
last decade. Furthermore, weighted-network analyses show that the WTW architecture has been extremely stable in the 1981-2000 period and highlights some interesting regularities \citep{Fagiolo2009pre}. For example, WTW countries holding many trade
partners (and/or very intense trade relationships) are also the
richest and most (globally) central; they typically trade with many
partners, but very intensively with only a few of them (which turn
out to be themselves very connected); and form few but
intensive-trade clusters (triangular trade patterns). 

Such observed WTW topological properties turn out to be important in \textit{explaining} macroeconomics dynamics. For example,
\citet{Kali_Reyes_2007_growth} and \citet{Kali_Reyes_2010_contagion} have shown that country positions in the trade network (e.g., in terms of their node degrees) has indeed substantial implications for economic growth and a good potential for explaining episodes of financial contagion. Furthermore, network position appears to be a substitute for physical capital but a complement for human capital.

In a nutshell, the existing literature adopting a complex-network approach to the study of international trade emphasizes the emergence of a few relevant regularities in the way the WTW is shaped, and posits that such peculiarities can be useful to explain what happens over time in the international global \textit{macroeconomic} network. However, we do not currently have network-formation models that are able to explain why the WTW is shaped the way it is.\footnote{The work-horse model in international trade is the so-called gravity equation. \citet{Fagiolo2010jeic} shows that a gravity model can explain a great deal of WTW architecture, but that a still relevant amount of information is left in the weighted network built using the residuals of gravity-equation estimation.} Therefore, the question whether observed WTW topological properties may be the result of randomness, constrained by some mild local features, or of more structural network-generation processes, remains unanswered. 

In what follows, we shall take up this question by estimating the expected value of the most important network statistics of the WTW under the null hypothesis that the network belongs to the ensemble of random structures satisfying on average some local constraints. We shall focus on two related local constraints: node in/out degree and node in/out strength sequences. In the specific case of the WTW, focusing on these local constraints is also important in order to assess whether the network formalism is really conveying additional, nontrivial information with respect to traditional international-economics analyses, which instead explain the empirical properties of trade in terms of country-specific macroeconomic variables alone. Indeed, the standard economic approach to the empirics of international trade \citep{Feenstra2004} has traditionally focused its analyses on the statistical properties of country-specific indicators like total trade, number of trade partners, etc., that can be easily mapped to what, in the jargon of network analysis, one denotes as local properties or first-order node characteristics. Ultimately, understanding whether network analyses go a step beyond with respect to standard trade theory amounts to assess the effects of indirect interactions in the world trade system. In fact, a wealth of results about the analysis of international trade have already been derived in the macroeconomics literature without making explicit use of a network description, and focusing on the above country-specific quantities alone. Network features like assortativity and clustering patterns do instead depend on indirect trade relationships, i.e. second or higher-order links between any two country not necessarily connected by a direct-trade relationship. 

\section{The Randomization Method}\label{Sec:Method}
Given a network with $N$ nodes, there are various ways to generate a family of randomized variants of it.\footnote{See for example \citet{Katz_Powell_1957,Holland_Leinhardt_1976,Snijders_1991,Rao_etal_1996,Kannan_etal_1999,Roberts_2000,Newman_2001,Shen-Orr_etal_2002,Maslov_etal_2004_rewiring_physa,Ansmann_Lehnertz_2011,Bargigli_Gallegati_2011}.} The most popular one is the \emph{local rewiring algorithm} proposed by Maslov and Sneppen \citep{Maslov_etal_2002_rewiring_science,Maslov_etal_2004_rewiring_physa}. In this method, one starts with the real network and generates a series of randomized graphs by iterating a fundamental rewiring step that preserves the desired properties. In the binary undirected case, where one wants to preserve the degree of every vertex, the steps are as follows: choose two edges, say $(i,j)$ and $(k,l)$; rewire these connections by swapping the end-point vertices and producing two new candidate edges, say $(i,l)$ and $(k,j)$; if these two new edges are not already present, accept them and delete the initial ones. After many iterations, this procedure generates a randomized variant of the original network, and by repeating this exercise a sufficiently large number of times, many randomized variants are obtained. By construction, all these variants have exactly the same degree sequence as the real-world network, but otherwise random. In the directed and/or weighted case, the rewiring steps defined above still work, but of course they are able to preserve in and out degrees of each vertex only (vertex strengths may change).\footnote{See \citet{serrc07,Opsahl_etal_2008} for extensions of this method that control for average vertex strengths in undirected and directed networks. Note that by controlling for either degree or strength sequences, one automatically fixes the ``volume'' of the network, in terms of either the total number of links or the sum of link weights. Conversely, controlling for the volume only, means assuming equi-probability among links and weights. This typically destroys the topology of the underlying network and thus turns out to be a bad null model for most observed networks.} Maslov and Sneppen's method allows one to check whether the enforced properties are partially responsible for the topological organization of the network. For instance, one can measure the degree correlations, or the clustering coefficient, across the randomized graphs and compare them with the empirical values measured on the real network.\footnote{This method has been applied to several networks, including the Internet and protein networks\citep{Maslov_etal_2002_rewiring_science,Maslov_etal_2004_rewiring_physa}. Different webs have been found to be affected in very different ways by local constraints, making the problem interesting and not solvable \emph{a priori}.}

The main drawback of the local rewiring algorithm is its computational requirements. Since the method is entirely numerical, and analytical expressions for its results are not available, one needs to explicitly generate several randomized graphs, measure the properties of interest on each of them (and store their values), and finally perform an average. This average is an approximation for the actual expectation value over the entire set of allowed graphs. In order to have a good approximation, one needs to generate a large number $M$ of network variants. Thus, the time required to analyze the impact of local constraints on any structural property is $M$ times the time required to measure that property on the original network, plus the time required to perform many rewiring steps producing each of the $M$ randomized networks. The number of rewiring steps required to obtain a single randomized network is $O(L)$, where $L$ is the number of links, and $O(L)=O(N)$ for sparse networks while $O(L)=O(N^2)$ for dense networks.\footnote{It must be noted that the WTW is a very dense network. Density in the aggregate directed network indeed oscillates in the range $[0.32,0.56]$. As a result, in the case of the WTW, applying a local rewiring algorithm would be rather expensive.} Thus, if the time required to measure a given topological property on the original network is $O(N^\tau)$, the time required to measure the randomized value of the same property is $O(M\cdot L)+O(M\cdot N^\tau)$, which is $O(M\cdot N^\tau)$ as soon as $\tau \ge 2$.

A recently-proposed alternative method, which is relatively faster due to its analytical character, is based on the maximum-likelihood estimation of maximum-entropy models of graphs \citep{Squartini_Garlaschelli_2011}. Unlike other analytical methods \citep{Newman_2001,Chung_Lu_2002,Serrano_Boguna_2005,Bargigli_Gallegati_2011},
this method does not require assumptions (such as sparseness and/or
low clustering) about the structure of the original empirical network. In this method, one first specifies the desired set of local constraints $\{C_a\}$. Second, one writes down the analytical expression for the probability $P(\mathbf{G})$ that, subject to the constraints $\{C_a\}$, maximizes the entropy
\begin{equation}
S\equiv -\sum_\mathbf{G} P(\mathbf{G})\ln P(\mathbf{G})
\label{eq_entropy}
\end{equation}
where $\mathbf{G}$ denotes a particular graph in the ensemble, and $P(\mathbf{G})$ is the probability of occurrence of that graph. This probability defines the ensemble featuring the desired properties, and being maximally random otherwise. Depending on the particular description adopted, the graphs $\mathbf{G}$ can be either binary or weighted, and either directed or undirected. Accordingly, the sum in Eq.~(\ref{eq_entropy}), and in similar expressions shown later on, runs over all graphs of the type specified.
The formal solution to the entropy maximization problem can be written in terms of the so-called Hamiltonian $H(\mathbf{G})$, representing the energy (or cost) associated to a given graph $\mathbf{G}$. The Hamiltonian is defined as a linear combination of the specified constraints $\{C_a\}$:
\begin{equation}
H(\mathbf{G})\equiv\sum_a \theta_a C_a(\mathbf{G})
\label{eq_H}
\end{equation}
where $\{\theta_a\}$ are free parameters, acting as Lagrange multipliers controlling the expected values $\{\langle C_a\rangle\}$ of the constraints across the ensemble. The notation $C_a(\mathbf{G})$ denotes the particular value of the quantity $C_a$ when the latter is measured on the graph $\mathbf{G}$.
In terms of $H(\mathbf{G})$, the maximum-entropy graph probability  $P(\mathbf{G})$ can be shown to be
\begin{equation}
P(\mathbf{G})=\frac{e^{-H(\mathbf{G})}}{Z}
\end{equation}
where the normalizing quantity $Z$ is the \emph{partition function}, defined as
\begin{equation}
Z\equiv\sum_\mathbf{G} e^{-H(\mathbf{G})}
\label{eq_partition}
\end{equation}
Third, one maximizes the likelihood $P(\mathbf{G^*})$ to obtain the particular graph $\mathbf{G^*}$, which is the real-world network that one wants to randomize. This steps fixes the values of the Lagrange multipliers that finally allow to obtain the numerical values of the expected topological properties averaged over the randomized ensemble of graphs.
The particular values of the parameters $\{\theta_a\}$ that enforce the local constraints, as observed on the particular real network $\mathbf{G^*}$, are found by maximizing the log-likelihood
\begin{equation}
\lambda\equiv \ln P(\mathbf{G^*})=-H(\mathbf{G^*})-\ln Z
\label{eq_likelihood}
\end{equation}
to obtain the real network $\mathbf{G^*}$. It can be shown \citep{mylikelihood} that this is equivalent to the requirement that the ensemble average $\langle C_a\rangle$ of each constraint $C_a$ equals the empirical value measured on the real network:
\begin{equation}
\langle C_a\rangle=C_a(\mathbf{G^*})\quad \forall a
\end{equation}
Note that, unless explicitly specified, in what follows we simplify the notation and simply write $C_a$ instead of $C_a(\mathbf{G^*})$ for the empirically observed values of the constraints. 

Once the parameter values are found, they are inserted into the formal expressions yielding the expected value
\begin{equation}
\langle X\rangle\equiv \sum_\mathbf{G} X(\mathbf{G}) P(\mathbf{G})
\label{eq_X}
\end{equation}
of any (higher-order) property of interest $X$. The quantity $\langle X\rangle$ represents the average value of the property $X$ across the ensemble of random graphs with the same average (across the ensemble itself) constraints as the real network. For simplicity, we shall sometimes denote $\langle X\rangle$ as a \emph{randomized property}, and its value as the \emph{randomized value} of $X$.\footnote{See Tables \ref{Tab:BinaryStats} and \ref{Tab:WeightedStats} for a detailed account of the expressions for the randomized properties appearing in the following analysis. Cf. \citet{Squartini_Garlaschelli_2011} for a discussion on how standard deviations of topological properties under the random null model are obtained.}

Technically, while the local rewiring algorithm generates a \emph{microcanonical} ensemble of graphs, containing only those graphs for which the value of each constraint $C_a$ is exactly equal to the observed value $C_a(\mathbf{G}^*)$, the maximum-likelihood method generates an expanded \emph{grandcanonical} ensemble where all possible graphs with $N$ vertices are present, but where the ensemble average of each constraint $C_a$ is equal to the observed value $C_a(\mathbf{G}^*)$. One can show that the two methods tend to converge for large networks \citep[for a detailed comparison between the two methods, see][]{Squartini_Garlaschelli_2011}. However, the maximum-likelihood one is remarkably faster. More importantly, enforcement of local constraints only implies that $P(\mathbf{G})$ factorizes as a simple product over pairs of vertices. This has the nice consequence that the expression for $\langle X\rangle$ is generally only as complicated as that for $X$. Furthermore, this implies that, e.g. in the binary case, the only random variables whose expected values over the \emph{grandcanonical} ensemble need to be calculated are the $a_{ij}$s, i.e. the entries of the binary adjacency matrix, with expected values equal to $p_{ij}(\{\theta_a\})$. In other words, after the preliminary maximum-likelihood estimation of the parameters $\{\theta_a\}$, in this method the time required to obtain the exact expectation value of an $O(N^\tau)$ property across the entire randomized graph ensemble is the same as that required to measure the same property on the original real network, i.e. still $O(N^\tau)$. Therefore, as compared to the local rewiring algorithm, which requires a time $O(M\cdot N^\tau)$, the maximum-likelihood method is $O(M)$ times faster, for arbitrarily large $M$. Using this method allows us to perform a detailed analysis of the WTW, covering all possible representations across several years, which would otherwise require an impressive amount of time. 

\section{Data and Methodology}\label{Sec:Data}
We employ international-trade flow data taken from Kristian \citet{GledData2002} database\footnote{Data are freely available at \url{http://privatewww.essex.ac.uk/~ksg/data.html}} to build a time-sequence of weighted directed networks for the period 1950-2000. In each year, we keep in the sample a country only if its total imports or total exports (or both) are non zero. Therefore, in each year $t$, the size of the network $N^t$ may change.

To build adjacency and weight matrices, we follow the flow of goods. This means that rows represent exporting countries, whereas columns stand for importing countries. The $N^t\times N^t$ time-$t$ weight matrix is therefore defined as $E^t=\{e_{ij}^t\}$, where $e_{ij}^t$ represents current-value exports in USD (millions) from $i$ to $j$ in year $t$ (rounded to the nearest integer). To build the binary WTW, we define a ``trade relationship'' by setting the generic entry $a_{ij}^t$ of the adjacency matrix $A^t$ to 1 if and only if $e_{ij}^t>0$ (and zero otherwise). Thus, the sequence of $N^t\times N^t$ adjacency and weight matrices $\{A^t,E^t\}$, $t=1950,...,2000$ fully describes the dynamics of the WTW. 

\begin{figure}[t]
	\begin{center}
	\begin{minipage}[t]{7.5cm}
		\includegraphics[width=1\textwidth]{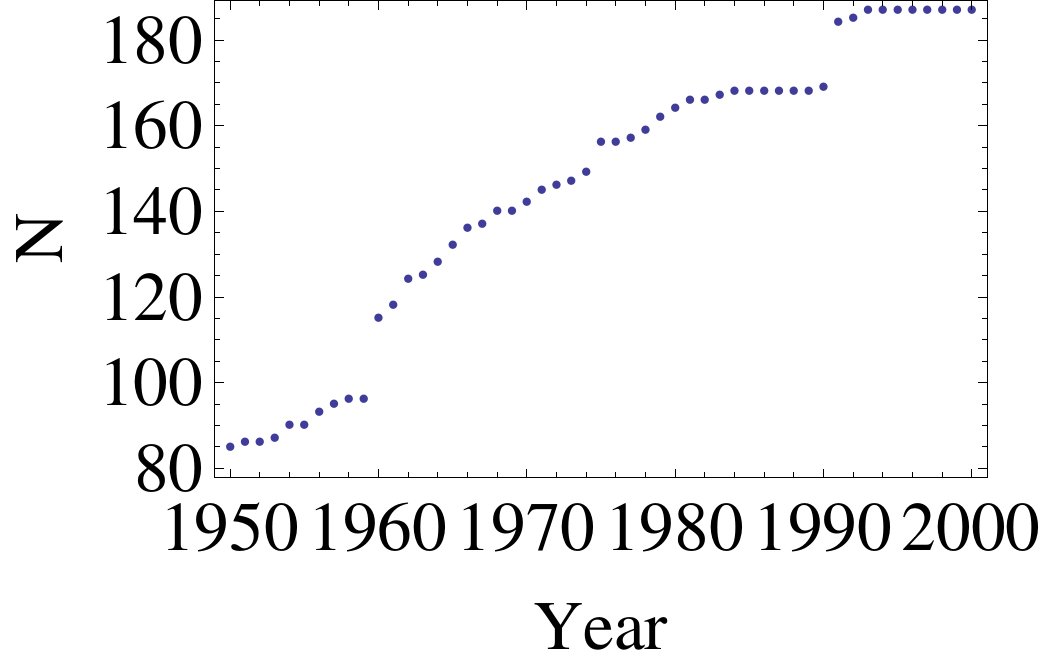}
	\end{minipage}
	\begin{minipage}[t]{7.5cm}
		\includegraphics[width=1\textwidth]{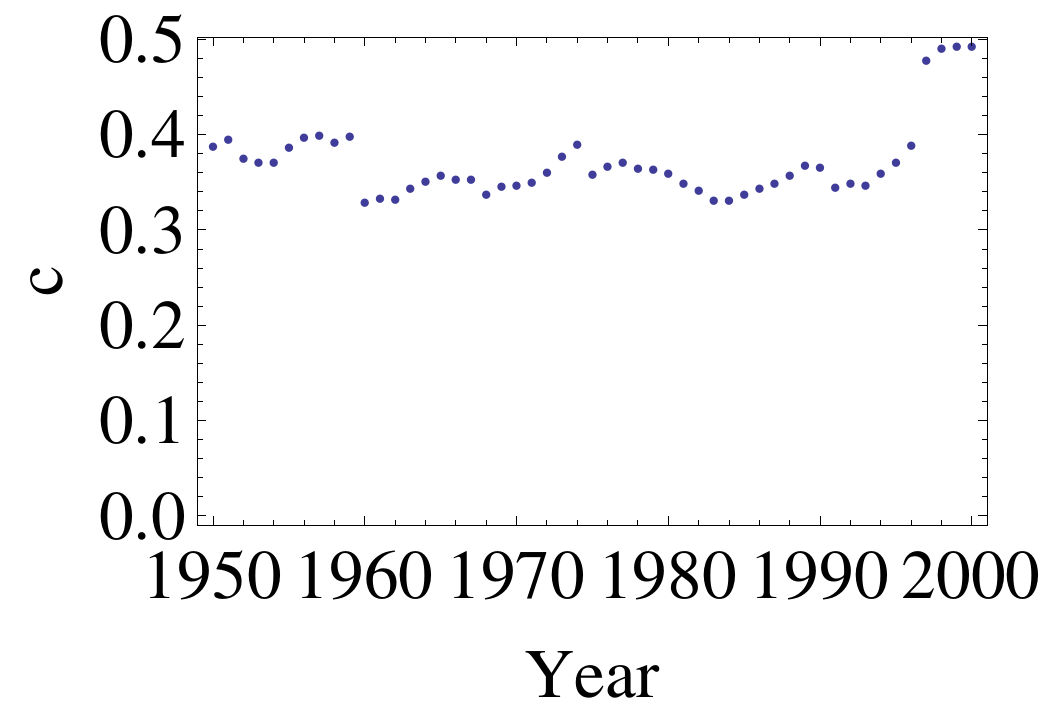}
	\end{minipage}
	\end{center}
	\caption{WTW size (N) and density (c) over time.} \label{Fig:size_dens}
\end{figure}

Figure \ref{Fig:size_dens} shows the evolution of network size and density (defined as the proportion of filled directed links) for the database under study. The number of countries in the network constantly increases over time. In the fifties, only about 80 countries where present in the network. Notice that a zero in a given trade flow may be due to both unreported entries or to missing trade.\footnote{This is a well-known issue in international trade statistics. Most of the literature on gravity has indeed tried to estimate models where one accounts for the large number of zero entries, see for example \citet{Santos2006}. Nevertheless, only a few papers have addressed the problem of discriminating between sheer zeros and missing valus, cf. \citet{Baranga_2009} for some progress in this direction.} As a result, growth and jumps in network size over time may arise because of sheer entry/exit in the international-trade market or because new data become available. Often, entry/exit is the consequence of some geo-political change, causing an increase in the number of countries reported in the dataset (e.g., independence of some African colonies around the Sixties, the fall of Soviet Union around 1990, etc.). By contrast, network density ($c^t$) stays relatively constant until the second part of the nineties. This means that country entering is not balanced by a strong increase in new trade links, i.e. the number of links in the network $c^t N^t(N^t-1)$ has approximately grown as $N^2$ until a very recent jump and upward trend close to year 2000. Note also that, since in general bilateral imports and exports may differ and trade relations may not be reciprocated, both binary and weighted versions of the WTW configure themselves as directed networks. One can therefore compute the relative frequency of reciprocated links (i.e. the frequency of times $a_{ij}^t=1$ and $a_{ji}^t=1$). This statistics is very high in the WTW (around 0.8), and is almost constant throughout the entire time period \citep[see also][]{myreciprocity,Fagiolo_2006_ecobull}, hinting to a strong symmetry in binary trade relationships. 

We study the architecture of the WTW over time employing a set of standard topological properties (i.e., network statistics), see \citet{Fagiolo2009pre} for a discussion. As Tables \ref{Tab:BinaryStats} and \ref{Tab:WeightedStats} show, we focus on three families of properties. First, total node-degree and total node-strength, measure, for binary and weighted networks respectively, the number of node partners and total trade intensity. In a directed network, one can also distinguish between node in-degree/in-strength (i.e., number of markets a country imports from, and total imports) and node out-degree/out-strength (i.e., number of markets a country exports to, and total exports). Second, total average nearest-neighbor degree (ANND) and strength (ANNS) measure, respectively, the average number of trade partners and total trade value of trade partners of a given node. This gives us an idea of how much a country is connected with other very well-connected countries. ANND and ANNS statistics can be disaggregated so as to account for both import/export partnerships of a country, and import/export partnerships of its partners. More precisely, one can compute four different measures of average nearest-neighbor degree/strength, obtained by coupling the two ways in which a node X can be a partner of a given target country Y (importer or exporter) and the two ways in which the partners of X may be related to it (as exporters or importers). Finally, we consider clustering coefficients (CCs), see \citet{Fagiolo2007pre} for a discussion. In the binary case, a node overall CC returns the probability that any two trade partners of that node are themselves partners. In the weighted case, these probabilities are computed taking into account link weights to proxy how strong are the edges of the triangles that are formed in the neighborhood of a node. Again, in the directed case one can disaggregate total node CC according to the four different shapes that directed triangular motifs can possess.\footnote{These are labelled \textit{cycle} (if $i$ exports to $j$, who exports to $h$, who exports to $i$), \textit{in} (if both $j$ and $h$, who are trade partners, exports to $i$), \textit{out} (if both $j$ and $h$, who are trade partners, imports from $i$) and \textit{mid} (if $i$ imports from $h$ and exports to $j$, and $j$ and $h$ are trade partners).}

We are interested not only in node average of such statistics over time, but also in the way node statistics correlate, and how such correlation patterns evolve across the years. 

To avoid meaningless comparisons over time of nominal variables, we compute all weighted topological quantities after having renormalized trade flows (observed and expected under our null model) by yearly total trade $T^t=\sum_{ij}{e_{ij}^t}$. We label renormalized link weights by $w_{ij}^t=e_{ij}^t/T^t$ and the corresponding weight matrix sequence by $W^t$.

After having computed network statistics on the observed data using $\{A^t,W^t\}$, we fit our null model to both binary and weighted directed WTW representations. More precisely, in the binary case, we compute expected values of all statistics (and their correlation) subject two sets of local constraints: (i) expected in-degrees equal to observed in-degree sequence $k_{i}^{in}$; (ii) expected out-degrees equal to observed out-degree sequence $k_{i}^{out}$. More precisely, we firstly compute the entries of the adjacency matrix $\{a_{ij}^{t}\}=\{\Theta[w_{ij}^{t}]\}$. Then we find the maximum of the likelihood function solving:

\begin{equation}
\left\{ \begin{array}{ll}
k_{i}^{in} &= \sum_{j\neq i}\frac{x_{i}^{in}x_{j}^{out}}{1+x_{i}^{in}x_{j}^{out}} \\
k_{i}^{out} &= \sum_{j\neq i}\frac{x_{i}^{out}x_{j}^{in}}{1+x_{i}^{out}x_{j}^{in}}
\end{array} \right. 
\end{equation}
to get the hidden variables $\{x^{out}_{i}\},\:\{x^{in}_{i}\}$. These are substituted back in the expression $p_{ij}=\frac{x_i y_j}{1+x_i y_j}$, which enters in the definition of random variables $a_{ij}$s. Finally, we compute the relevant topological properties. We use a linear approximation method for all the binary quantities that are functions of linear powers of the $a_{ij}$s. This allows us to get expected values of fractions as fractions of expected values, i.e. expected vale of the numerator divided by expected value of the denominator.

A similar procedure is applied in the weighted case, where we compute expected values of all weighted statistics (and their correlation) subject two sets of local constraints: (i) expected in-strengths equal to observed in-strengths sequence $s_{i}^{in}$; (ii) expected out-strengths equal to observed out-strengths sequence $s_{i}^{out}$. More precisely, one solves:

\begin{equation}
\left\{ \begin{array}{ll}
s_{i}^{in} &= \sum_{j\neq i}\frac{y_{i}^{in}y_{j}^{out}}{1-y_{i}^{in}y_{j}^{out}} \\
s_{i}^{out} &= \sum_{j\neq i}\frac{y_{i}^{out}y_{j}^{in}}{1-y_{i}^{out}y_{j}^{in}}
\end{array} \right. 
\end{equation}
to find the hidden variables $\{y^{out}_{i}\},\:\{y^{in}_{i}\}$.

In addition to expected average values of any given network statistics, we compute their standard deviations. In general, given a node-statistic X computed on a $N$-sized network and its observed sequence $\underline{x}_i=\{x_1,\dots,x_N\}$, one can compute expected sequence-values $\langle \underline{x}_i \rangle=\{\langle x_1\rangle,\dots,\langle x_N\rangle\}$. As a consequence, expected population average will simply read:

\begin{equation}
m_{\langle \underline{x} \rangle} =\frac{\sum_i{\langle x_i \rangle}}{N},
\end{equation}
whereas standard deviation reads:
\begin{equation}
s_{\langle \underline{x} \rangle}=\sqrt{\frac{\sum_i{[\langle x_i \rangle-m_{\langle \underline{x} \rangle}]^2}}{N-1}}.
\end{equation}
This easily allows one to compute 95\% confidence intervals for both $m(\langle \underline{x}_i \rangle)$ and $s(\langle \underline{x}_i \rangle)$, using respectively $t-$Student and $\chi^2$ distributions with $N-1$ degrees of freedom.
    
Note that, in the binary case, the set of constraints employed here allows us to compare observed average topological properties (and their correlation) over time with their expected values in trade networks that, on average, replicate the observed sequence of trade partnerships, both in the import and in the export market (and are otherwise fully random). In the weighted WTW, by fixing strength constraints, one can control for the sequence of total imports and exports (properly normalized), and consequently for all observed trade unbalances. 

As a result, the reference null model employed below is able notionally to generate an ensemble of fully-random alternatives of the observed WTW that are nevertheless in line with some baseline observed properties of the ``local'' structure of international trade. Indeed, by fixing degrees and strengths one is constraining only the ``volume'' of a node neighborhood, either in terms of trade partnerships or trade values, but allows for random reshuffling of ``local'' quantities that remain consistent throughout the network. Most of these random alternatives will probably be economically unfeasible. Nevertheless, they may serve as a benchmark to understand whether the patterns of ``higher-order'' network statistics like ANND/ANNS or clustering coefficients can be reproduced by the null model, or they persistently deviate from it. 

Furthermore, by constraining the null model to ``local'' quantities such as the number of trade partnerships or country trade value one can also address the question whether a complex-network approach to international trade is really able to convey additional, non-trivial information as compared to traditional international-trade empirical analyses. Indeed, traditional empirical international-trade studies have mostly focused on the statistical properties of country-specific indicators like total country trade and number of trade partners, which correspond to node degree and strength in the network jargon \citep{Feenstra2004}. Focusing on these two sets of statistics only will not add anything new to what we already know about the web of trade between countries.\footnote{Note that these quantities are trivially reproduced by our null models where, by definition, $\langle k_i^{out}\rangle=k_i^{out}(A)$, $\langle k_i^{in}\rangle=k_i^{in}(A)$, $\langle s_i^{out}\rangle=s_i^{out}(W)$ and $\langle s_i^{in}\rangle=s_i^{in}(W)$.} What network theory does is instead focusing also on indirect interactions in the world trade system, involving higher-order statistics like ANND and clustering, which take into account trade interactions occurring between trade partners of a country's trade partners, and so on. It is therefore crucial to understand whether, by controlling for local properties only, one can replicate statistical properties involving higher-order statistics. If this is not the case, the we can conclude that the latter are conveying some fresh and statistically relevant information on the structure of world trade.

A final remark before turning to our results is in order. As discussed in the Introduction, our null-model analysis is not involved in explaining the underlying causal mechanisms shaping the network. Therefore, throughout this paper, we shall use the term ``explaining'' in a very weak term. For example, finding that a local network statistics X ``explains'' a higher-order network statistics Y in our null model will signal the presence of a strong correlation between the two statistics, so that X can be sufficient to fully reproduce Y in the network. Of course, we do not aim at using our null model to identify subtle causal links between X and Y, which in the real-world may be caused e.g. by some omitted variables that cause in a proper way the high observed correlation between X and Y.

\section{Results}\label{Sec:Results}
In this Section, we ask two main related questions. First, we are interested in assessing to what extent the null model works in replicating the most important topological features characterizing the WTW over time. We mostly focus on node-average ANND and clustering coefficients (see Tables \ref{Tab:BinaryStats}-\ref{Tab:WeightedStats}). We are also interested in (Pearson) correlation coefficients between ANND and ND (ANNS and NS in the weighted case), and between binary (resp. weighted) CCs and ND (resp. NS). Recall that a positive (resp. negative) and high ANND-ND or ANNS-NS correlation hints to an assortative (disassortative) network structure. Likewise, a high and positive correlation between (binary or weighted) CCs and NS or ND indicates that more and better connected countries are also more clustered, i.e. that their neighbors are also well connected between them.        

\subsection{The Binary Directed WTW}\label{SubSec:BDN_res}
\begin{figure}[b]
	\begin{center}
	\begin{minipage}[t]{6.5cm}
		\includegraphics[width=1\textwidth]{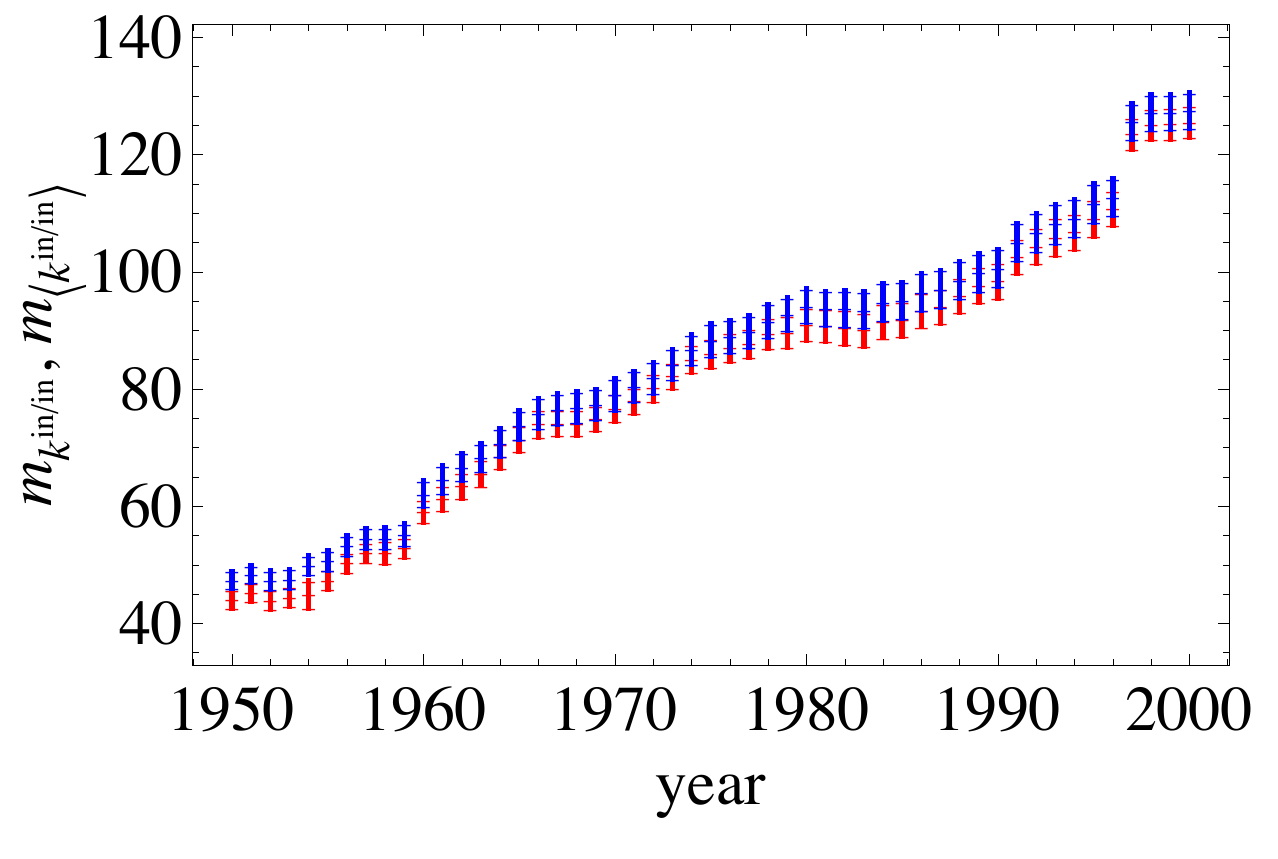}
	\end{minipage}
	\begin{minipage}[t]{6.5cm}
		\includegraphics[width=1\textwidth]{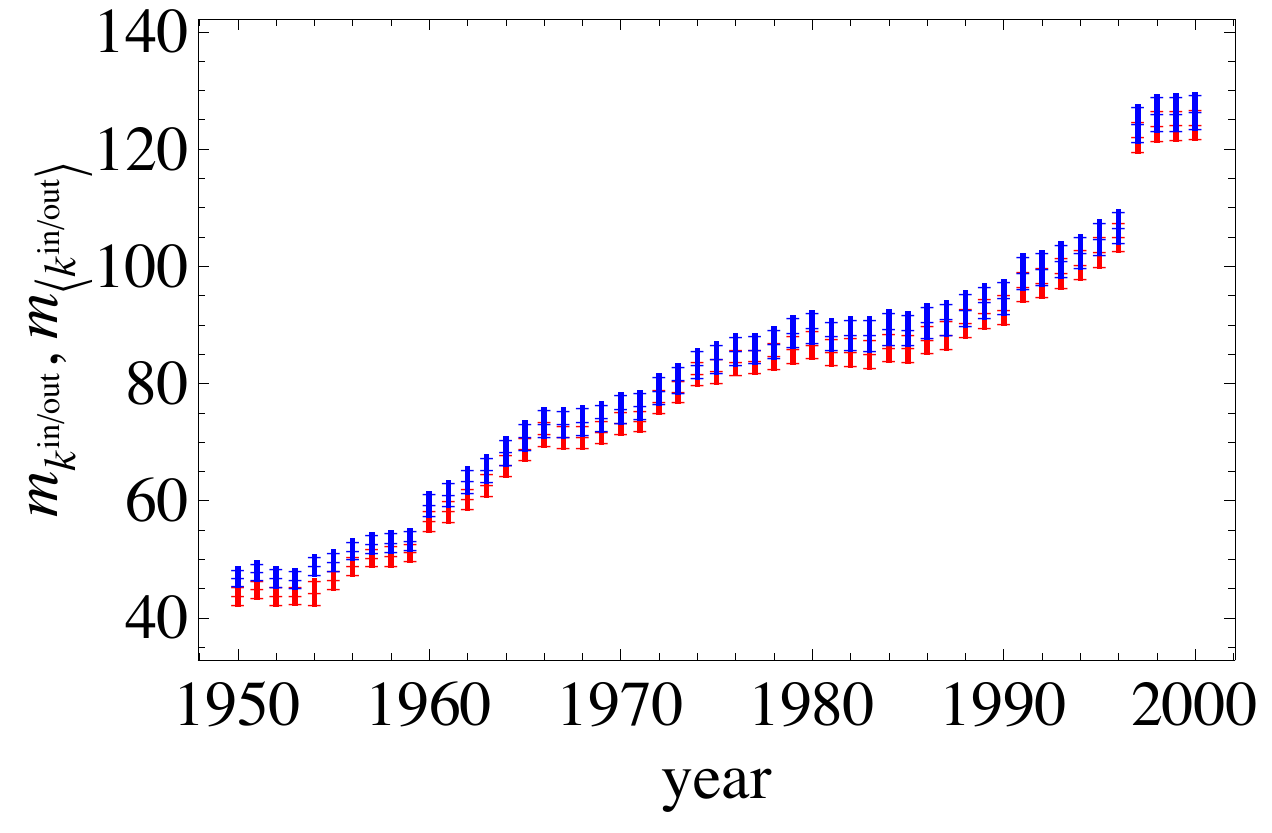}
	\end{minipage}
	\begin{minipage}[t]{6.5cm}
		\includegraphics[width=1\textwidth]{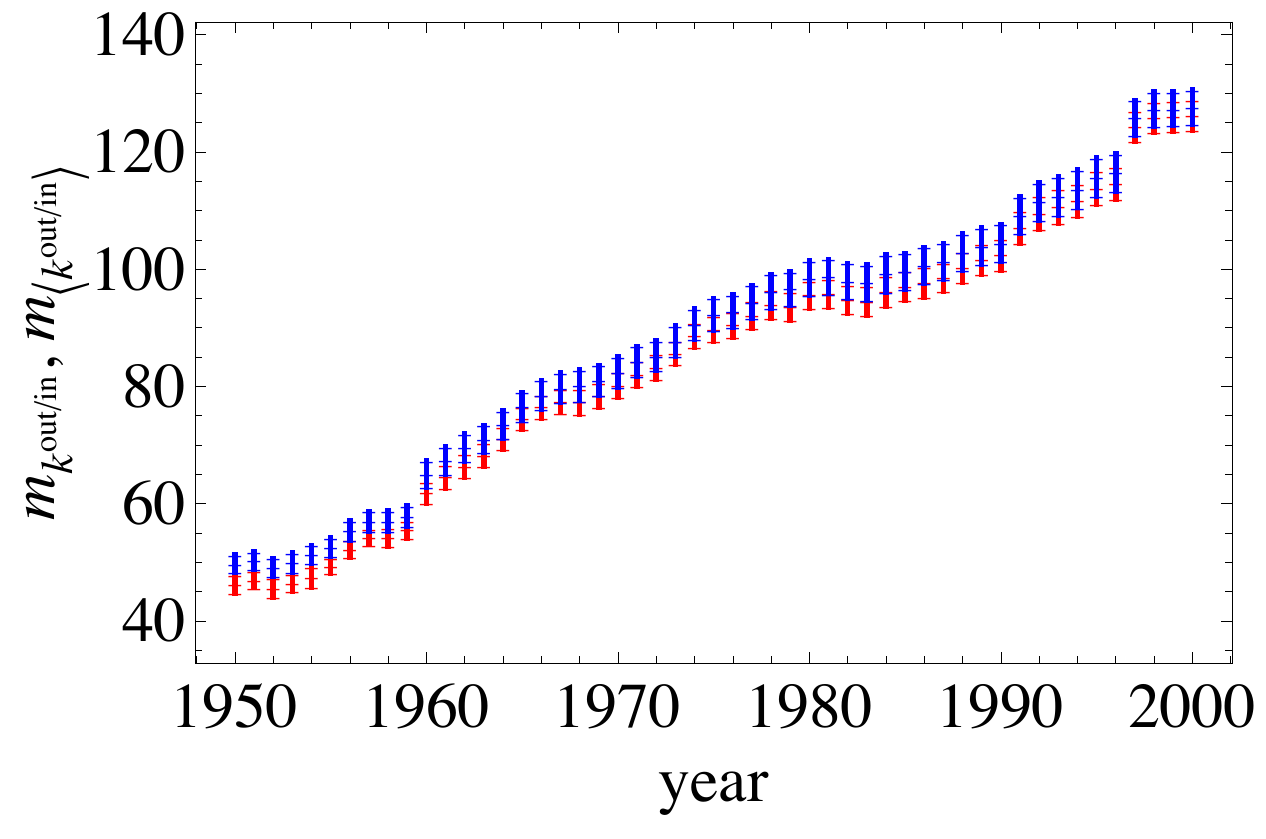}
	\end{minipage}
	\begin{minipage}[t]{6.5cm}
		\includegraphics[width=1\textwidth]{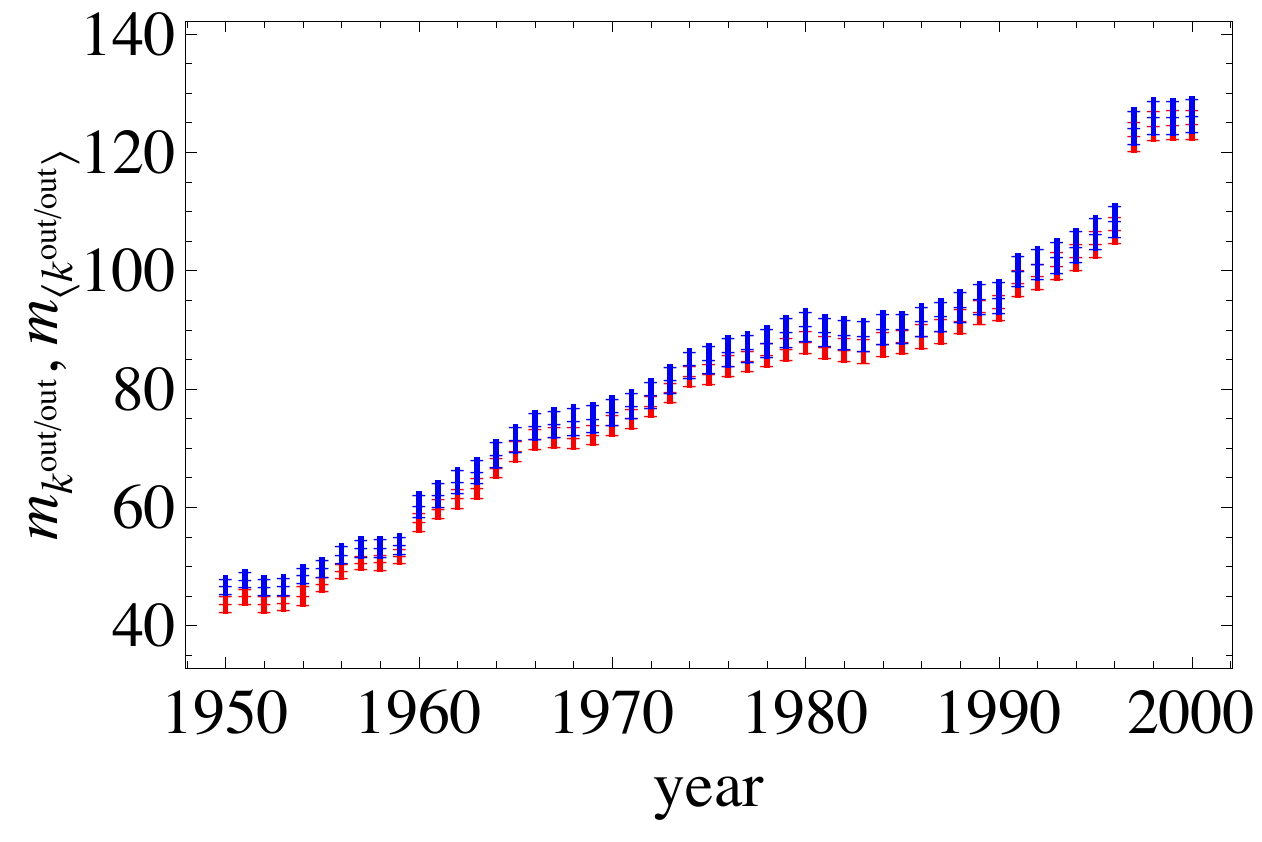}
	\end{minipage}
	\end{center}
	\caption{The binary-directed WTW: average nearest neighbor degrees and 95\% confidence bands. Red: observed quantities. Blue: null-model fit. Top-left: IN-IN ANND. Top-right: IN-OUT ANND. Bottom-left: OUT-IN ANND. Bottom-right: OUT-OUT ANND. } \label{Fig:ave_annd}
\end{figure}

We begin by investigating average ANND patterns over time. As Figure \ref{Fig:ave_annd} shows, average ANND displays increasing, almost linear, trends over time. This is mostly due to the increase in network size. There are two clear structural breaks emerging, one around 1960 ---which coincides with a huge drop in reported countries--- and another one around 1996, which instead occurs despite network size remains constant and therefore may be solely due to an increase in average neighbor connectivity. Note also that, qualitatively, ANND evolution over time is similar in the four plots, hinting to a strong symmetry in the binary directed network. 

\begin{figure}[t]
	\begin{center}
	\begin{minipage}[t]{6.5cm}
		\includegraphics[width=1\textwidth]{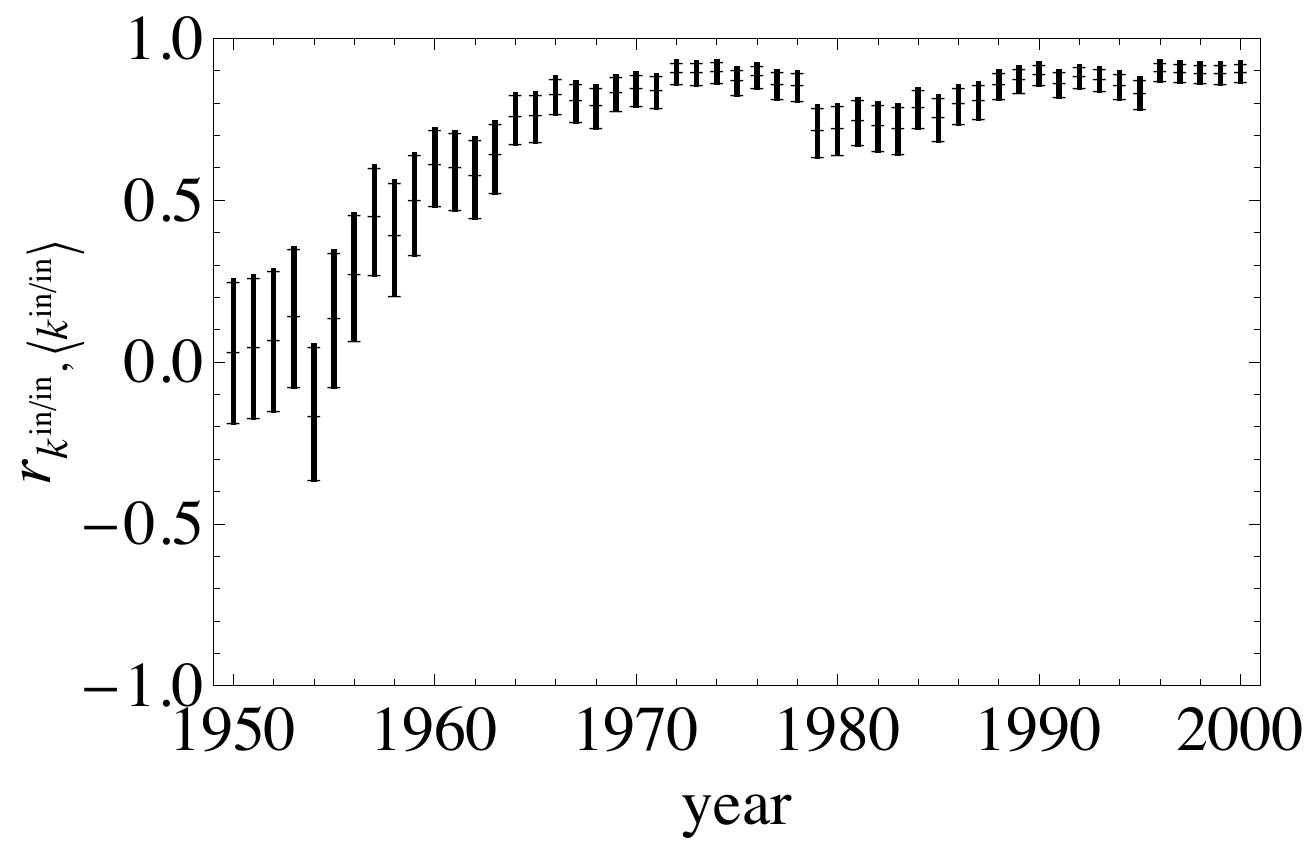}
	\end{minipage}
	\begin{minipage}[t]{6.5cm}
		\includegraphics[width=1\textwidth]{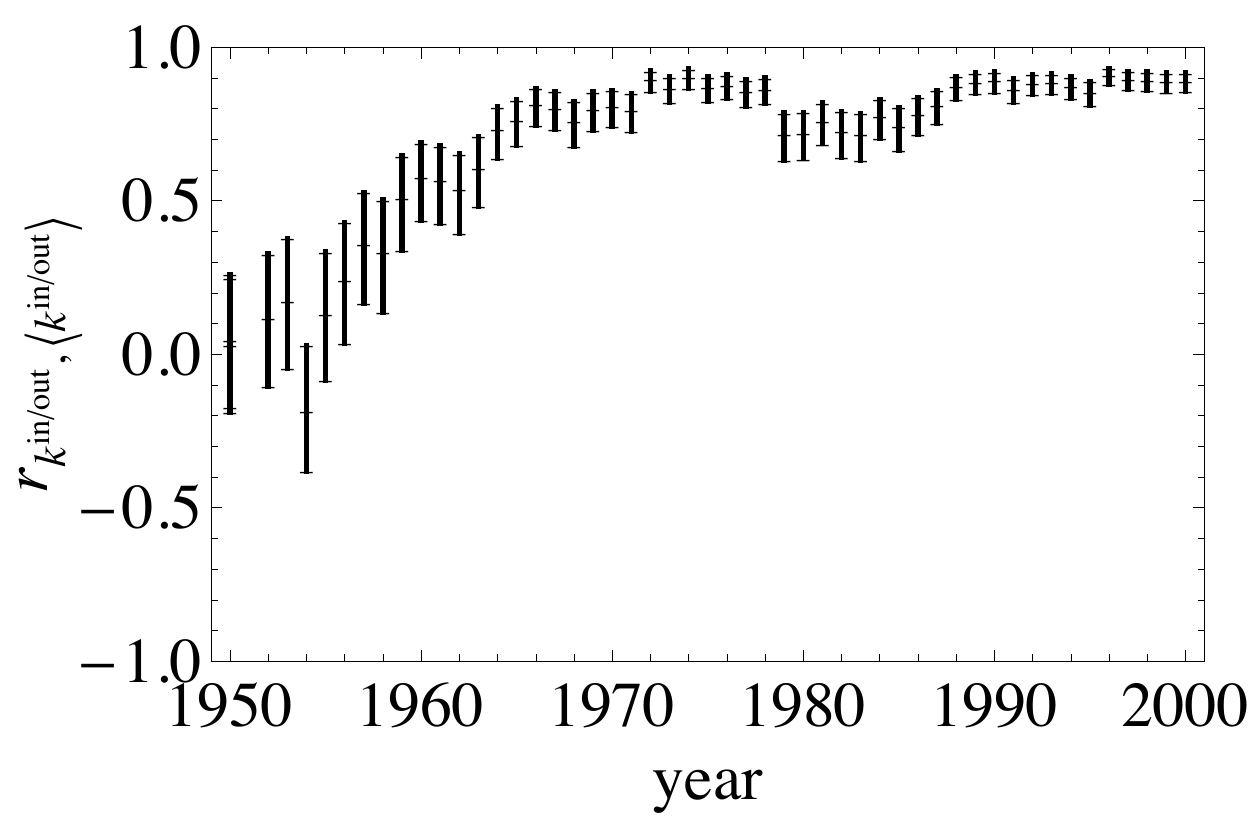}
	\end{minipage}
	\begin{minipage}[t]{6.5cm}
		\includegraphics[width=1\textwidth]{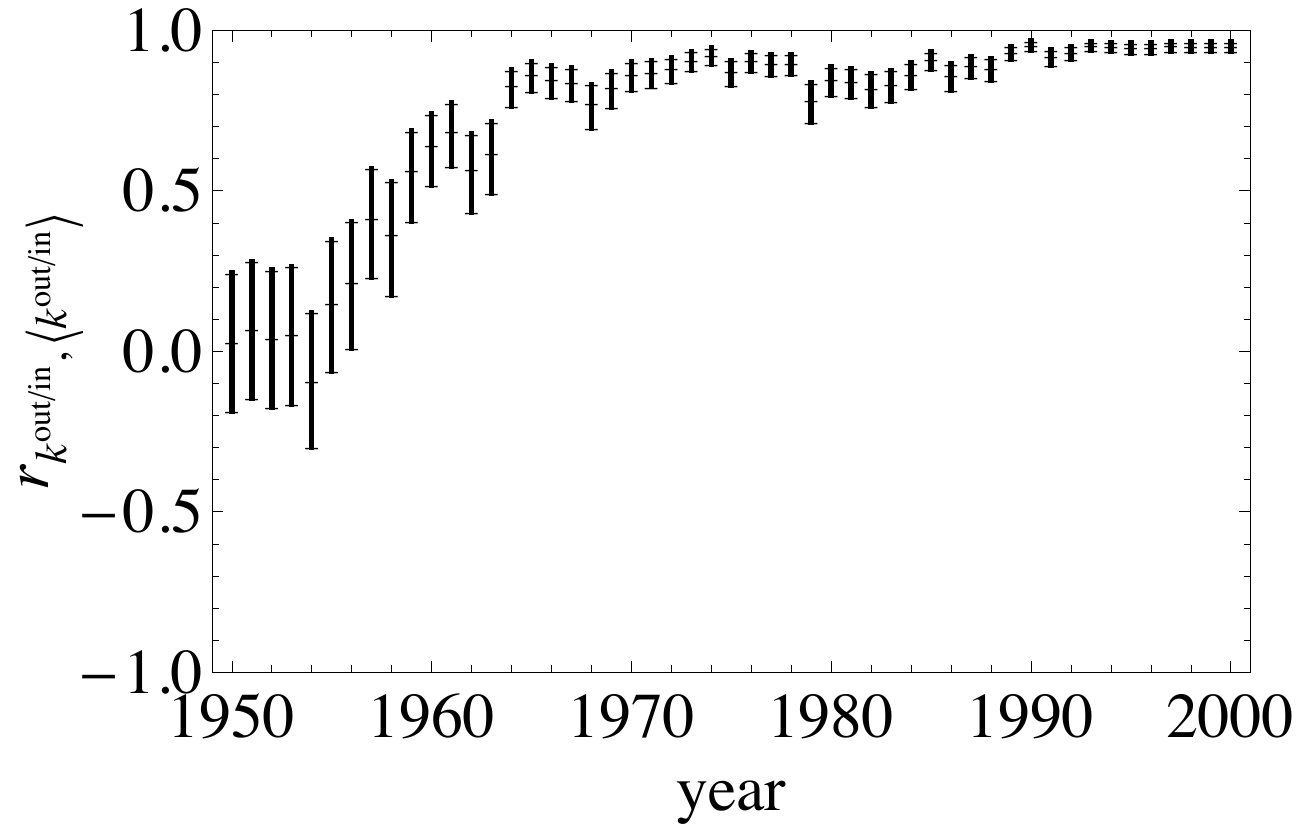}
	\end{minipage}
	\begin{minipage}[t]{6.5cm}
		\includegraphics[width=1\textwidth]{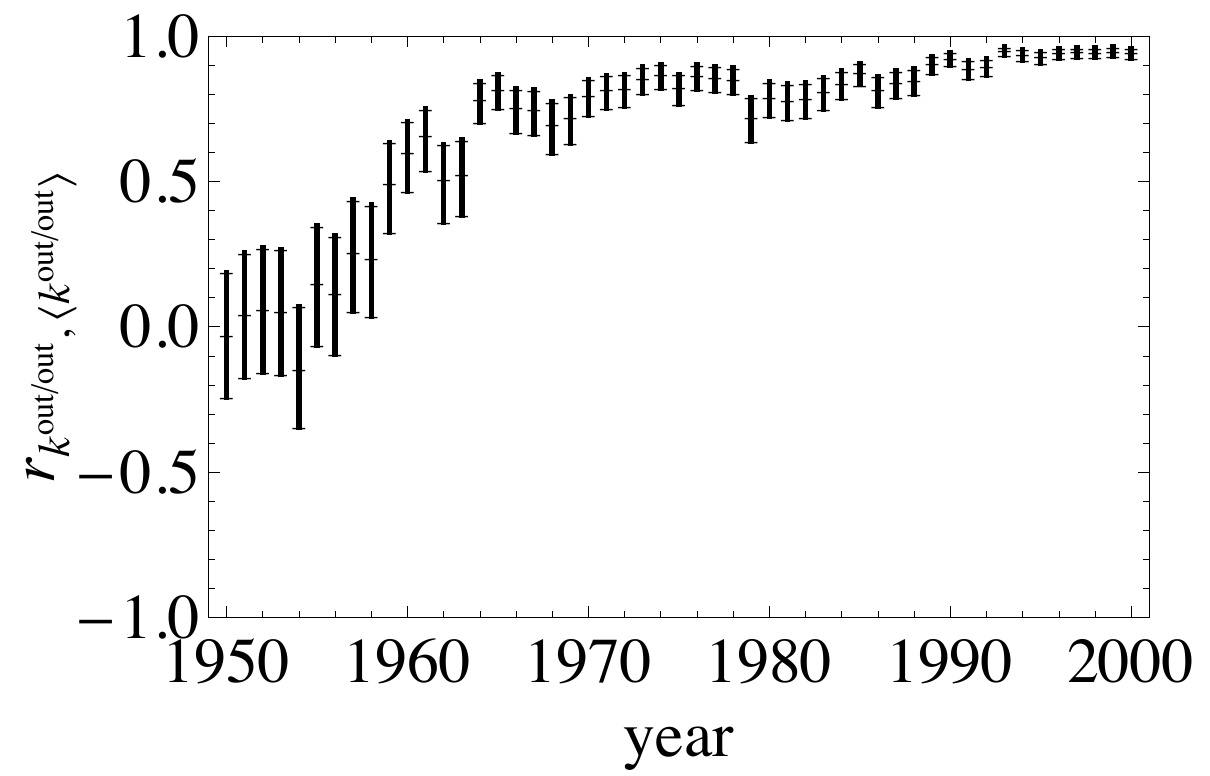}
	\end{minipage}
	\end{center}
	\caption{The binary-directed WTW: Pearson correlation coefficient between observed and null-model node ANND. Top-left: IN-IN ANND. Top-right: IN-OUT ANND. Bottom-left: OUT-IN ANND. Bottom-right: OUT-OUT ANND. } \label{Fig:corr_annd}
\end{figure}

More importantly, all plots show a good accordance between observed and null-model estimates for average ANND, in all four possible directed versions, especially as we approach year 2000. This means that average ANND patterns can be fully explained by observed in- and out-degree sequences, which are our constraints in the maximum-likelihood binary problem. To further explore this issue, we report correlation coefficients between observed and null-model node ANND statistics over the years. A positive and significant value for this correlation means that the null-model replicates observed ANNDs not only on average, but on a node-by-node basis. As Figure \ref{Fig:corr_annd} suggests, until 1965 within-year accordance between observed and expected ANND levels was not so satisfying: observed and expected ANND were almost uncorrelated and confidence bands were very large. From 1965 on, the null model is perfectly able to match observed country ANND values. 

\begin{figure}[b]
	\begin{center}
		\includegraphics[width=10cm]{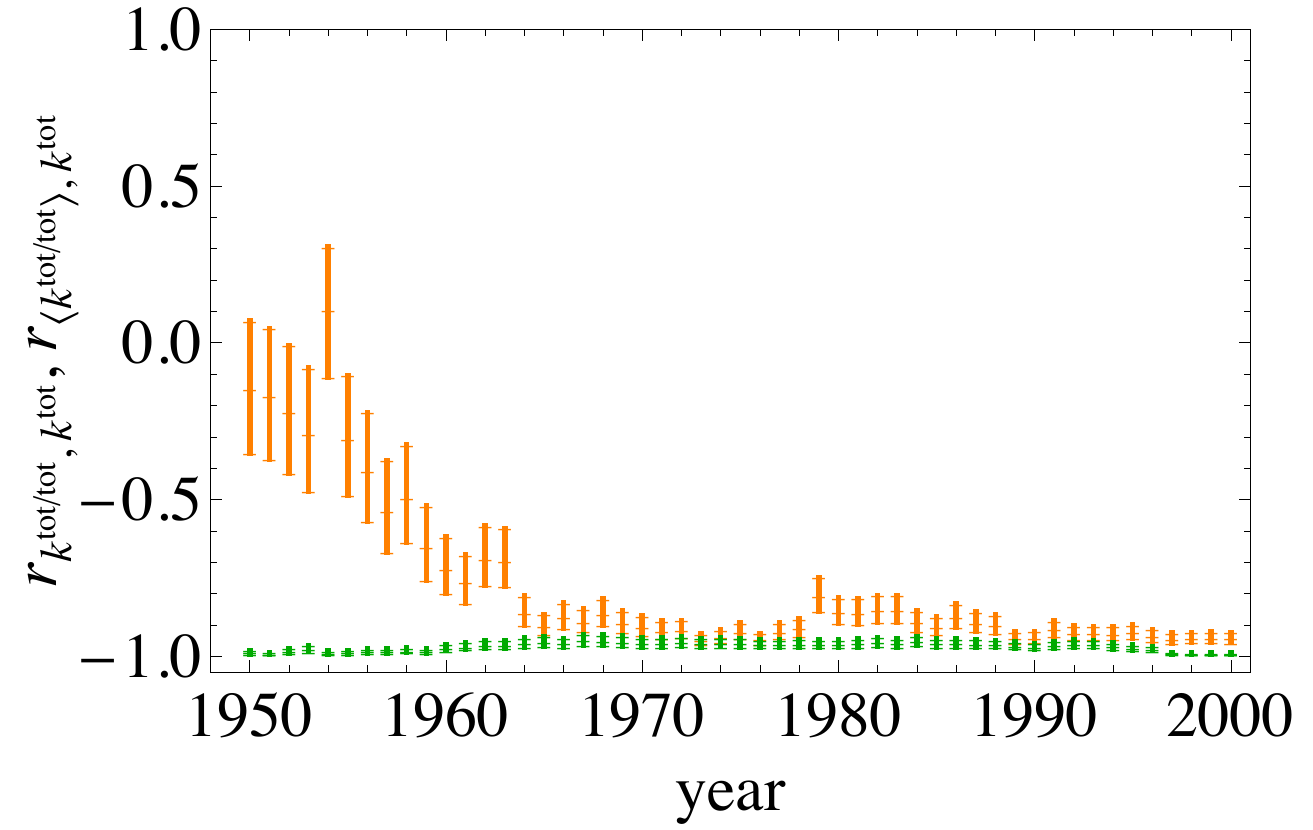}
	\end{center}
	\caption{Disassortativity in the binary-directed WTW. Orange: Observed correlation between total ANND and total ND. Green: Correlation between expected total ANND and observed total ND.} \label{Fig:binary_disass}
\end{figure}

Such a pattern is even more evident looking at network disassortativity. Figure \ref{Fig:binary_disass} plots the correlation coefficient between total ANND and total ND vs. time for both the observed and the expected binary WTW. In the expected case, the correlation is computed by considering observed NDs, which represents our constraints. As expected \citep{Garla2005,Fagiolo2009pre}, observed disassortativity is very marked in the binary WTW, but only from the second part of the sixties on. The null model is quite able to account for that strong disassortativity in that period. However, in the first 15 years of our sample, the binary WTW is not disassortative and the expected correlation strongly overestimates the observed one.

This evidence indicates that degree sequences are not enough to explain disassortativity in the whole sample. However, when the null model fails in replicating the observed WTW structure, the latter was not characterized by a strong disassortative or assortative pattern, as observed ANND/ND correlations were statistically not different from zero. This also suggests that after 1965 the marked observed disassortativity was not conveying additional meaningful information, as it can be easily reproduced by a null random model where in- and out-degrees where the only explaining factors. A possible economic interpretation can be rooted into the observation that, early in the sample period, geographical barriers and trade costs played a greater role. Before subsequent waves of globalization occurred, the WTW was organized in more disconnected-communities structures, where geography was mainly driving trade partnerships. As a larger number of countries started to enter global trade markets, and more links were added in the WTW, geographical constraints became less important, and strong disassortative patterns emerged where poorly-connected countries linked to very-connected ones. However, this process led to a network statistically indistinguishable from a similar one where links were placed at random and only the in- and out-degree sequence were preserved. 

\begin{figure}[t]
	\begin{center}
	\begin{minipage}[t]{6.5cm}
		\includegraphics[width=1\textwidth]{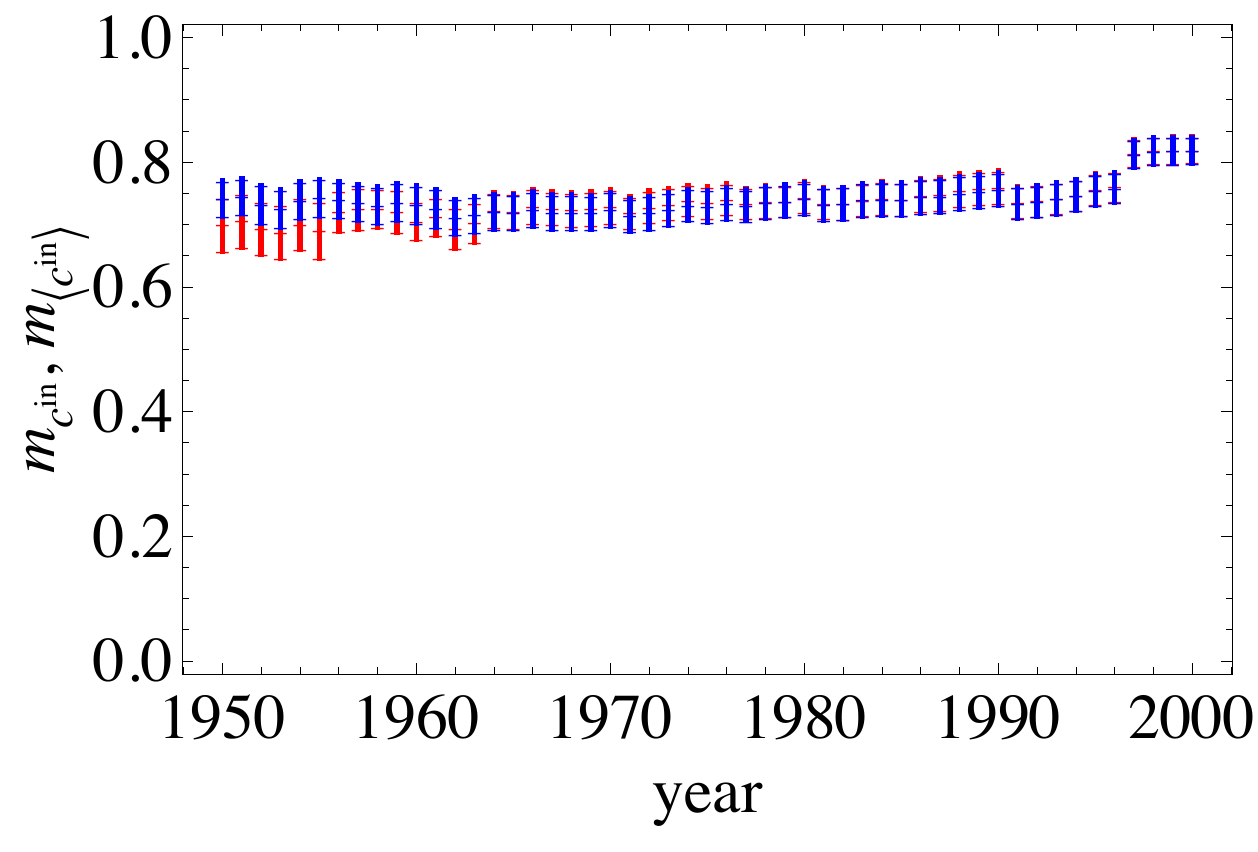}
	\end{minipage}
	\begin{minipage}[t]{6.5cm}
		\includegraphics[width=1\textwidth]{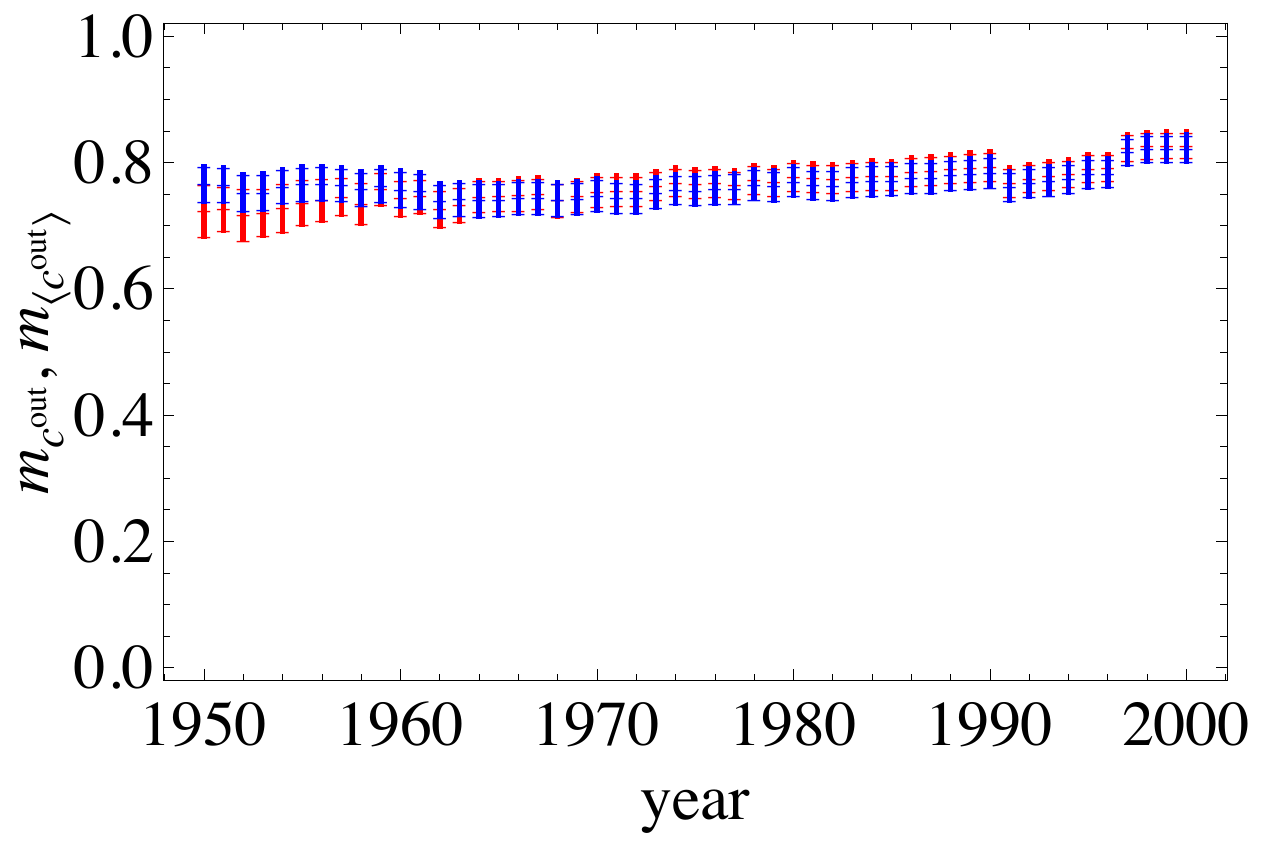}
	\end{minipage}
	\begin{minipage}[t]{6.5cm}
		\includegraphics[width=1\textwidth]{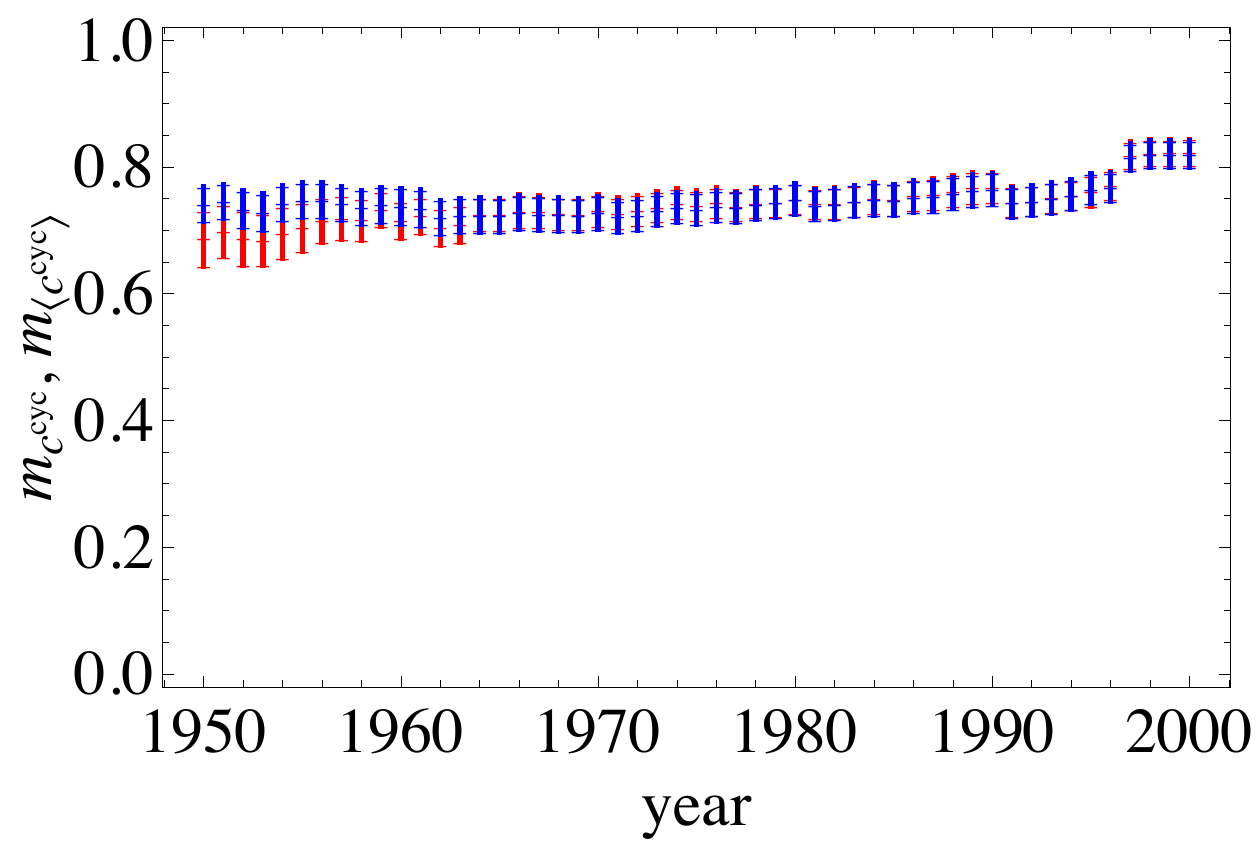}
	\end{minipage}
	\begin{minipage}[t]{6.5cm}
		\includegraphics[width=1\textwidth]{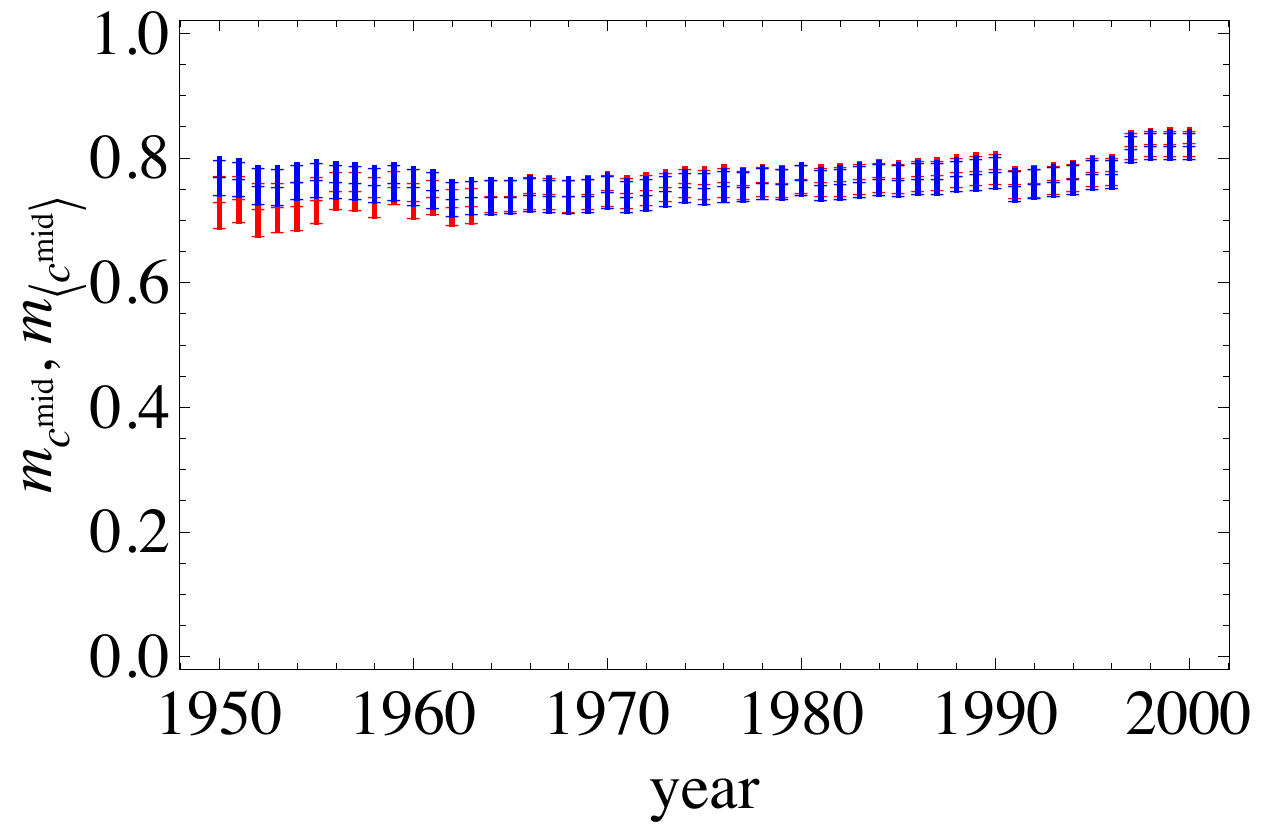}
	\end{minipage}
	\end{center}
	\caption{The binary-directed WTW: average clustering coefficients and 95\% confidence bands. Red: observed quantities. Blue: null-model fit. Top-left: BCC In. Top-right: BCC Out. Bottom-left: BCC Cycle. Bottom-right: BCC Middleman.} \label{Fig:ave_bcc}
\end{figure}

\begin{figure}[b]
	\begin{center}
	\begin{minipage}[t]{6.5cm}
		\includegraphics[width=1\textwidth]{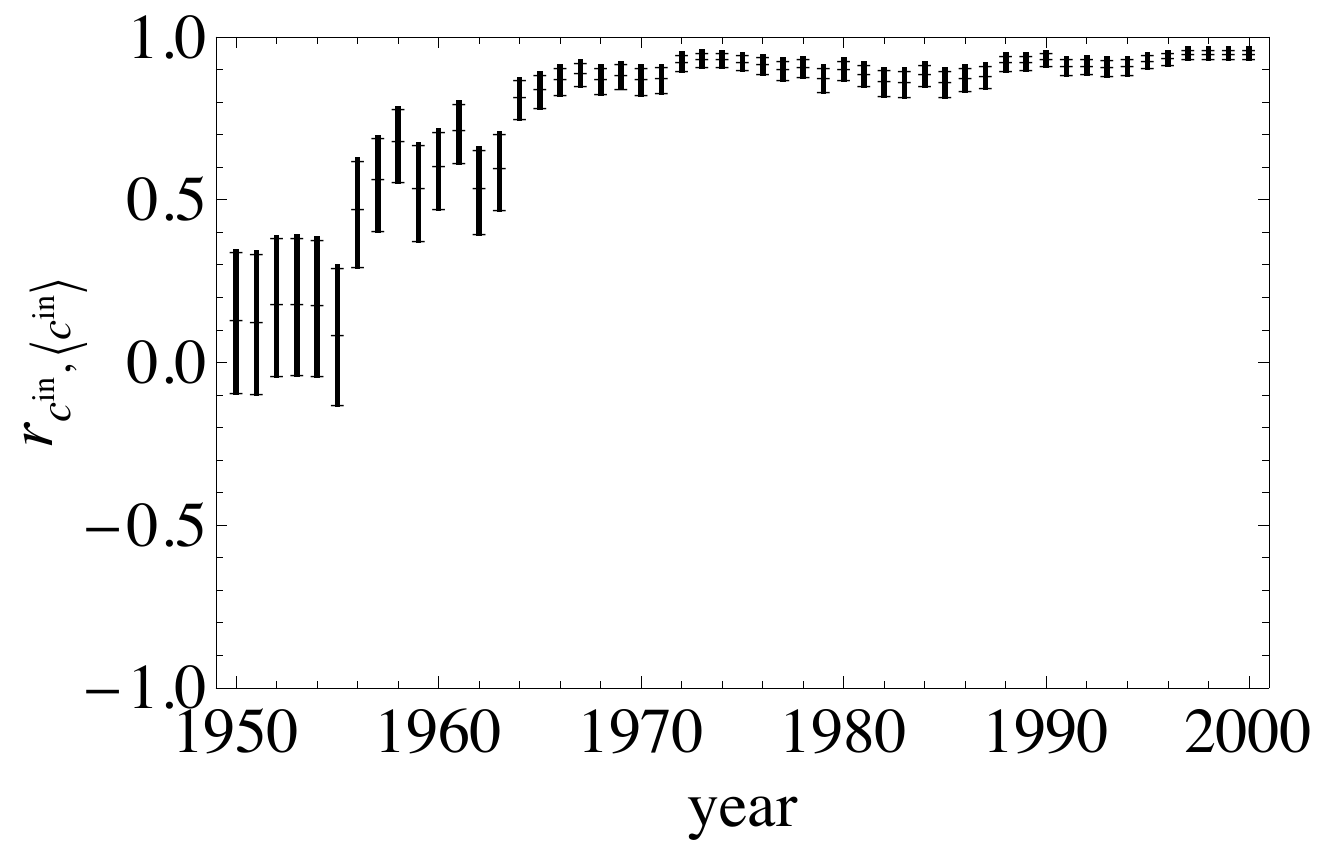}
	\end{minipage}
	\begin{minipage}[t]{6.5cm}
		\includegraphics[width=1\textwidth]{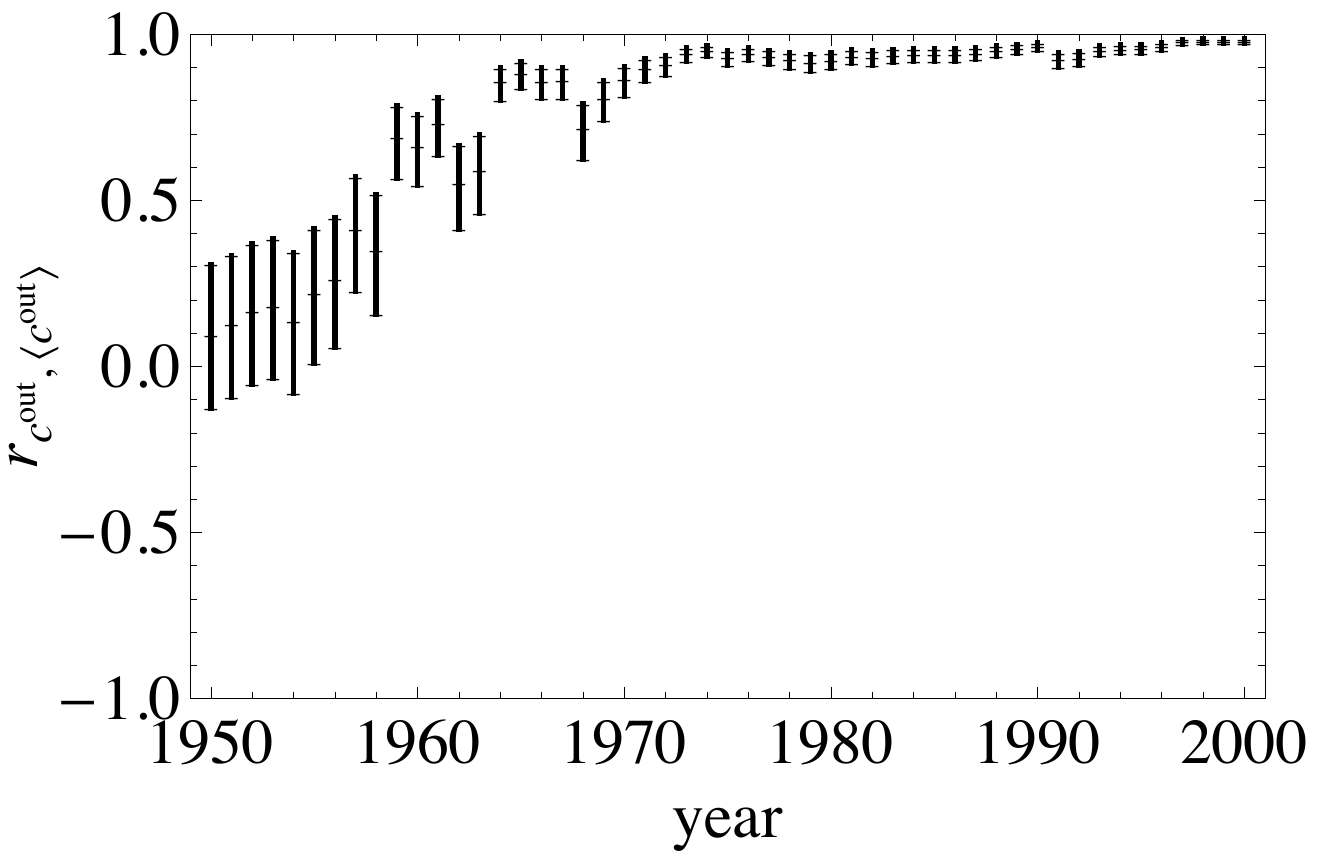}
	\end{minipage}
	\begin{minipage}[t]{6.5cm}
		\includegraphics[width=1\textwidth]{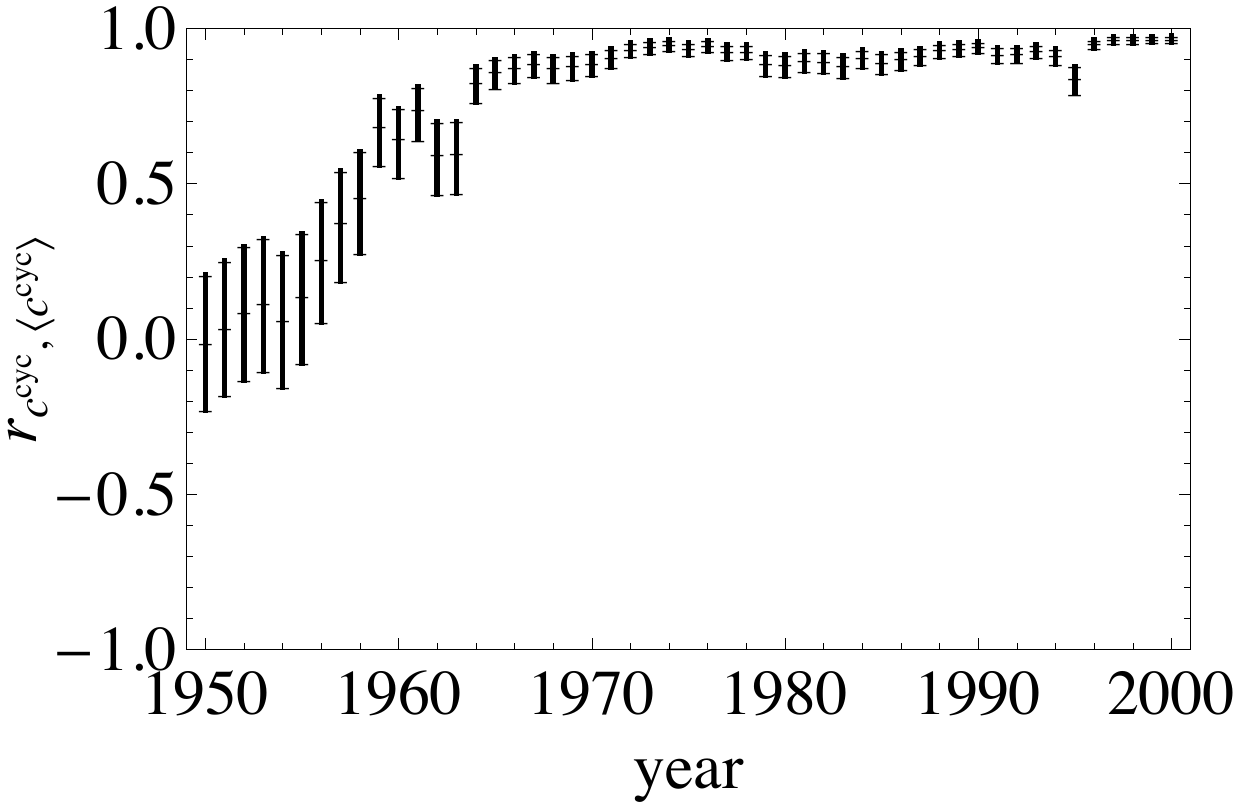}
	\end{minipage}
	\begin{minipage}[t]{6.5cm}
		\includegraphics[width=1\textwidth]{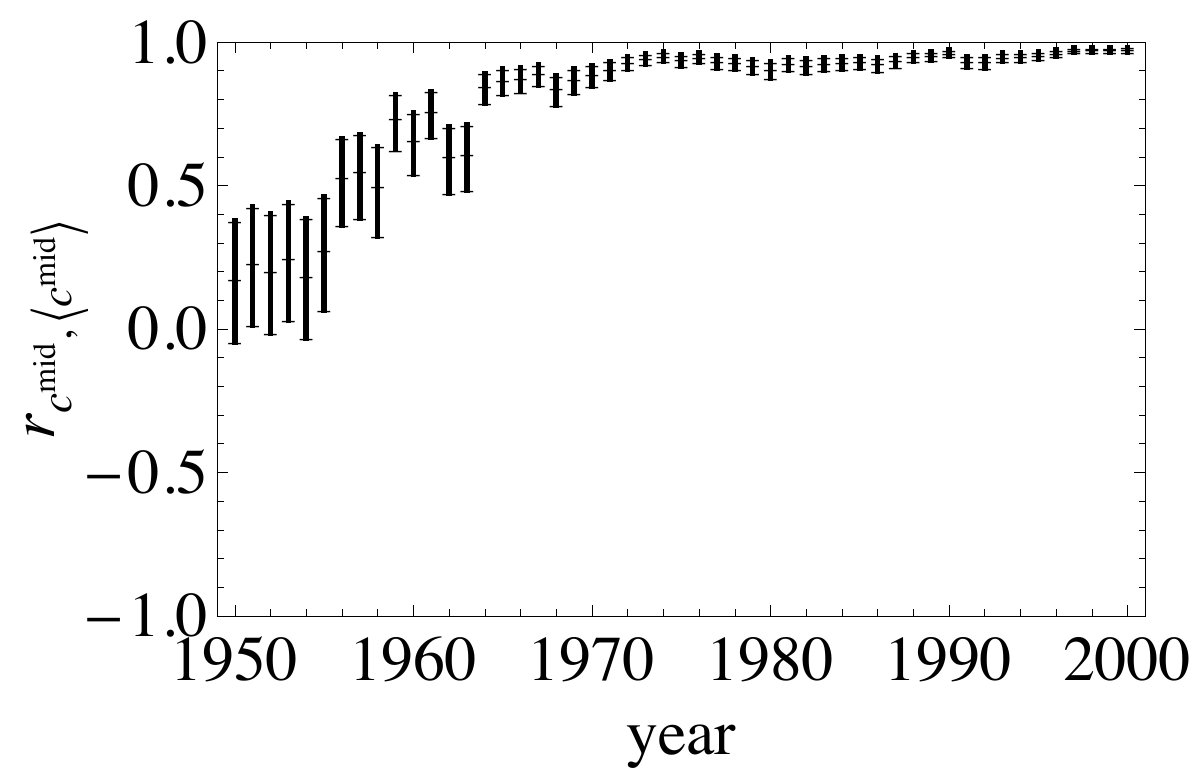}
	\end{minipage}
	\end{center}
	\caption{The binary-directed WTW: Pearson correlation coefficient between observed and null-model node clustering coefficients and 95\% confidence bands. Top-left: BCC In. Top-right: BCC Out. Bottom-left: BCC Cycle. Bottom-right: BCC Middleman.} \label{Fig:corr_bcc}
\end{figure}

\begin{figure}[h]
	\begin{center}
		\includegraphics[width=10cm]{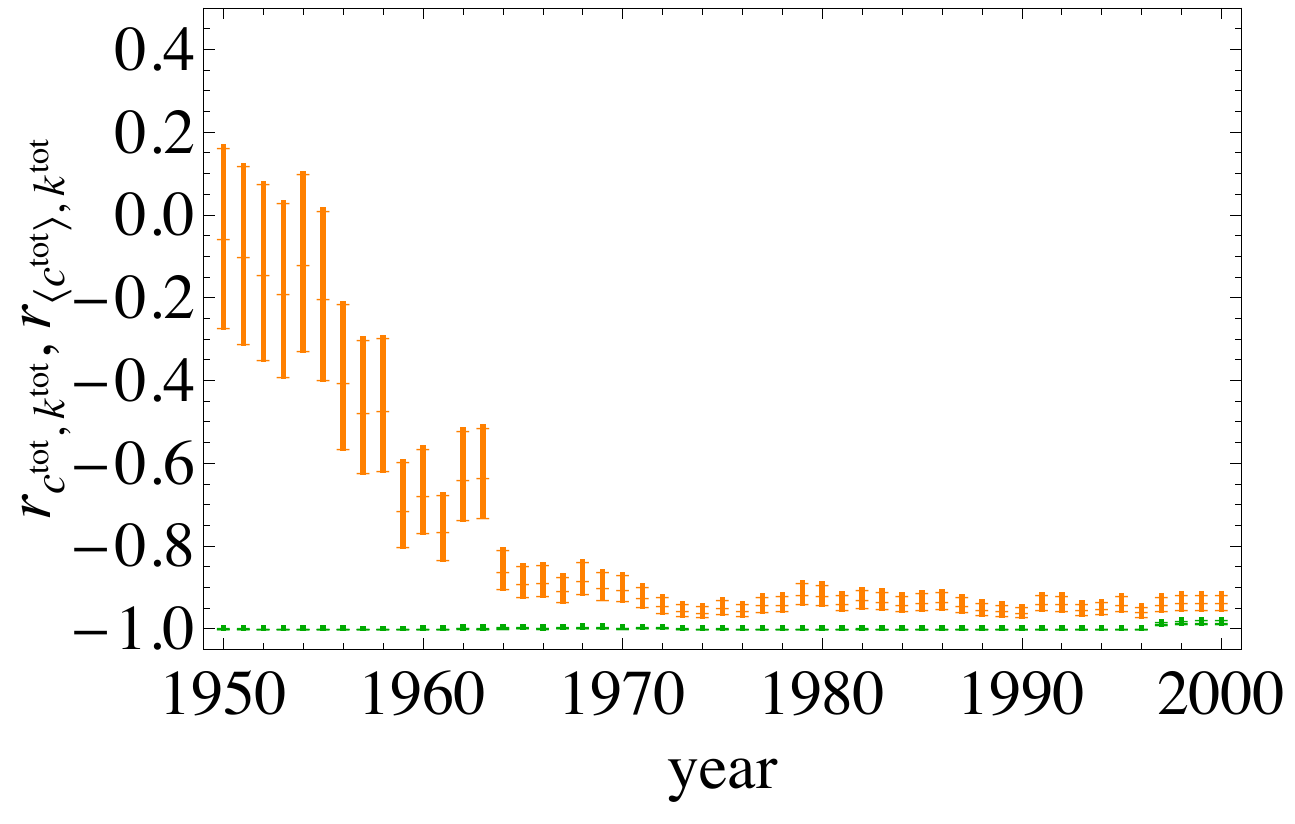}
	\end{center}
	\caption{Correlation between total binary clustering coefficient and node degree in the binary-directed WTW. Orange: Observed correlation between total BCC and total ND. Green: Correlation between expected total BCC and observed total ND.} \label{Fig:binary_clustdeg}
\end{figure}

A similar pattern also characterizes binary clustering coefficients. In this case, average BCC displays a flat trend over time (Figure \ref{Fig:ave_bcc}) around very high levels. This is because of the high density in the binary WTW, which makes every pair of partners of a node to be very likely partners themselves. The null model is perfectly able to match this average pattern: given in- and out-degree sequences, also density is preserved, and therefore average clustering coefficients. However, this does not automatically imply that each single node preserves its clustering level. In fact, as Figure \ref{Fig:corr_bcc} shows, an almost perfect agreement between observed and expected BCC sequences is reached only since the end of the sixties on. Again, in the '50s and early '60s, the null model was only able to match BCC on average but is was not very good at predicting the BCC level of each single country. More importantly, observed and expected correlation between BCC and ND still show a mismatch in the first part of the sample (Figure \ref{Fig:binary_clustdeg}). Indeed, well-connected countries tend to act as centers of a star network in the binary WTW only after 1965, with pairs of partners very unlikely to be trade partners themselves. The null model predicts this behavior also in the very first part of the sample (1950-1965), where, instead, the observed WTW was centered around geographically-close countries where no clear BCC-ND correlation pattern was emerging. As happens for disassortativity, however, the strong and negative BCC-ND correlation gradually emerging after 1965 turn out to be a statistically irrelevant phenomenon, impossible to distinguish from what a purely-random degree-constrained network model could predict.

\begin{figure}[h]
	\begin{center}
	\begin{minipage}[t]{6.5cm}
		\includegraphics[width=1\textwidth]{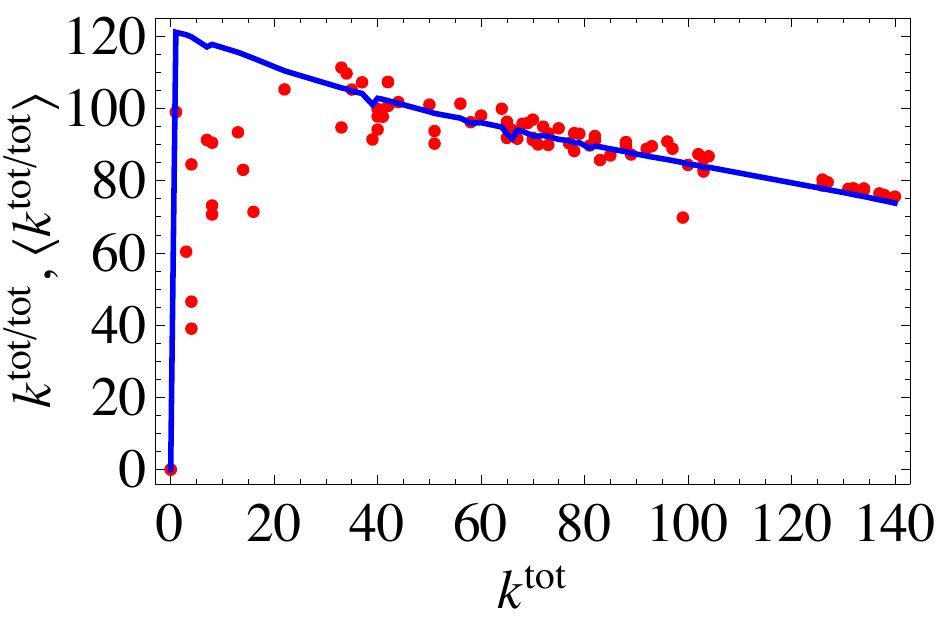}
	\end{minipage}
	\begin{minipage}[t]{6.5cm}
		\includegraphics[width=1\textwidth]{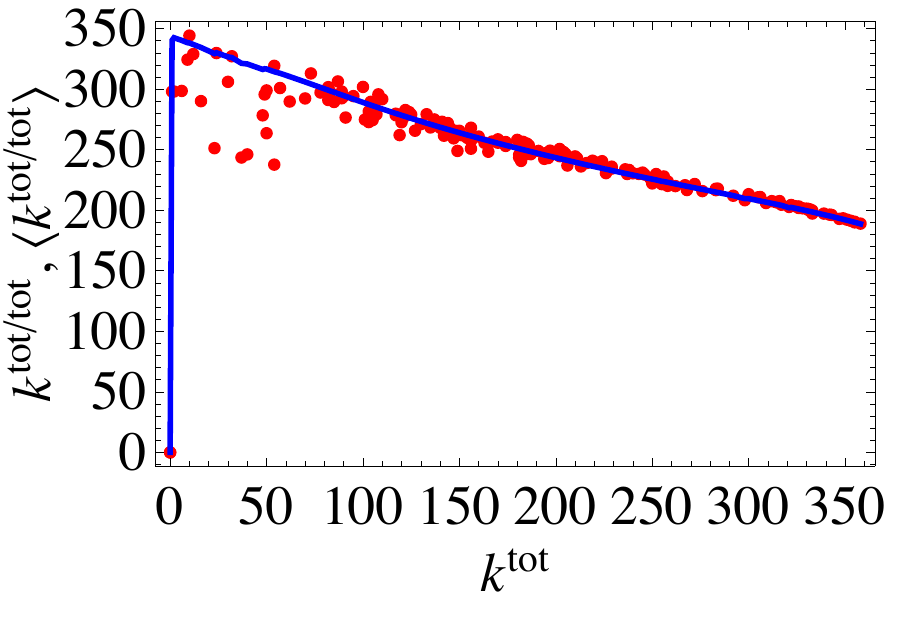}
	\end{minipage}
	\end{center}
	\caption{Disassortativity in the binary WTW. Scatter plots of total ANND vs. observed total node degree in 1950 (left) and 2000 (right). Red: observed quantities. Blue: null-model fit.} \label{Fig:scatters_tot_disassortativity_binary}
\end{figure}

\begin{figure}[h]
	\begin{center}
	\begin{minipage}[t]{6.5cm}
		\includegraphics[width=1\textwidth]{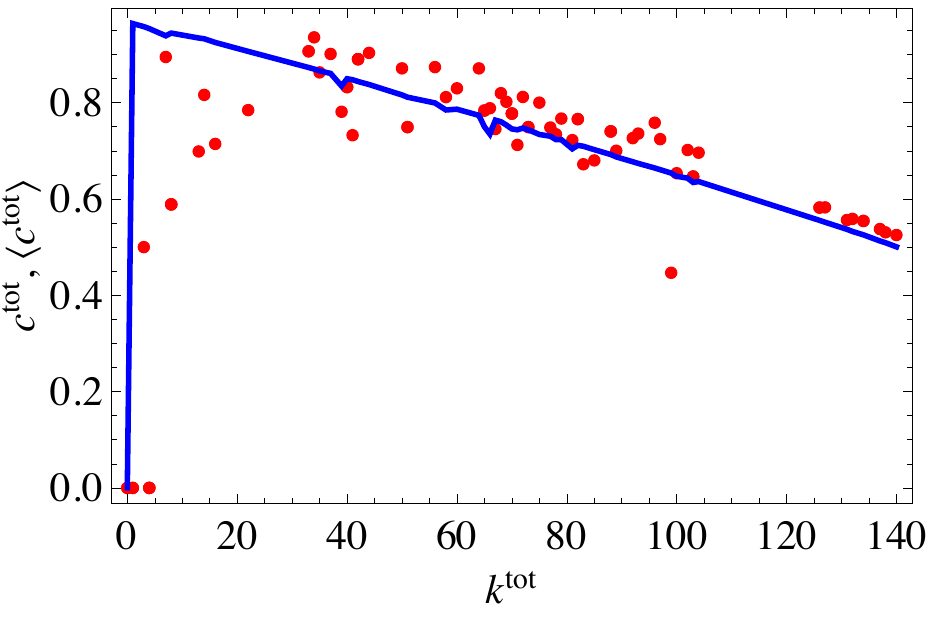}
	\end{minipage}
	\begin{minipage}[t]{6.5cm}
		\includegraphics[width=1\textwidth]{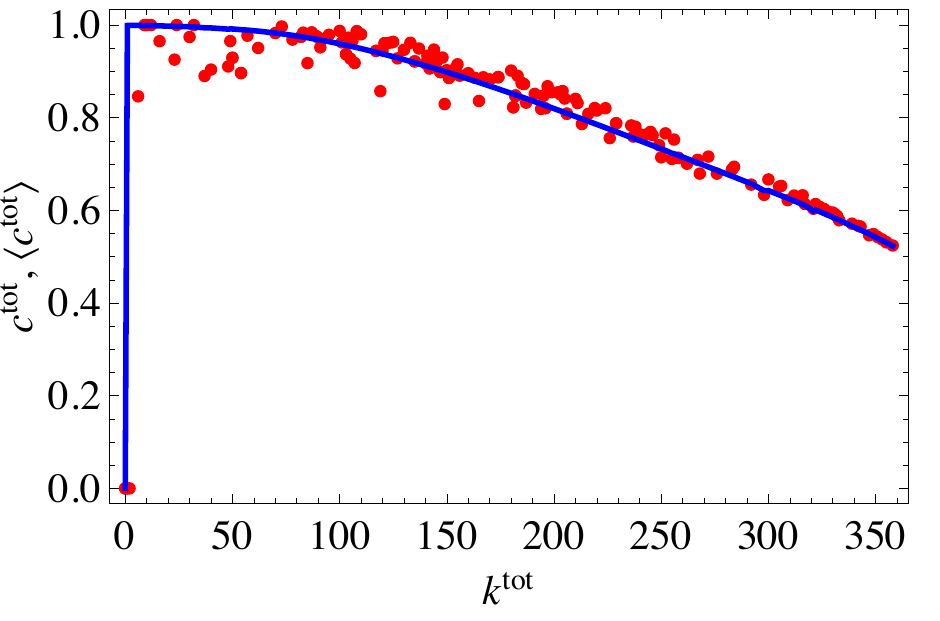}
	\end{minipage}
	\end{center}
	\caption{Clustering coefficient vs. observed total node degree in the binary WTW. Scatter plots of total BCC vs. ND in 1950 (left) and 2000 (right). Red: observed quantities. Blue: null-model fit.} \label{Fig:scatters_tot_bcc_k}
\end{figure}

To further explore the mismatch observed in the first part of the sample, Figures \ref{Fig:scatters_tot_disassortativity_binary} and \ref{Fig:scatters_tot_bcc_k} show scatter plots of observed (red) and expected (blue) total ANND and total BCC in 1950 vs. 2000. It is easy to see that the null model perfectly matches both ANND and BCC at all ND levels. Conversely, a statistically-detectable difference between observed and null-model quantities emerges when trying to predict the behavior of poorly-connected countries, where the null model persistently overestimates both ANND and BCC. For positive node degrees, the null model is not able to pick up the strong non-linearities emerging between ND and higher-order statistics.    

To sum up, the analysis of the WTW as a binary network indicates that the null model is well-equipped to reproduce most of the topological properties of the WTW after year 1965. Therefore, evidence on disassortativity or clustering-degree correlation, despite strongly emerging from the data, may be simply the result of random effects in networks where in- and out-degree sequences are preserved on average. In the first part of the sample, conversely, such a strong evidence about disassortativity and clustering-degree correlation is not empirically detected and the null model is not able to replicate the absence of strong correlation (especially for poorly connected countries). This suggests to look for alternative explanations for the observed topological structure, rooted either in richer null models or in more structural models involving independent variables that are not network-related, such as ---in this case--- geographical distance or economic size. We shall come back to this point in our concluding remarks.          

\subsection{The Weighted Directed WTW}\label{SubSec:WDN_res}
We turn now to a weighted-network analysis of the WTW. It is well-known that weighted and binary properties of the WTW do not always coincide \citep{Fagiolo2008physa}. For example, the WTW viewed as a weighted network
is only \textit{weakly} disassortative. Furthermore, better connected
countries tend to be \textit{more} clustered. It is therefore interesting to see if a null model controlling for in- and out-strength sequences can also explain the weighted-network architecture of the WTW, and in which sub-samples of the time window under analysis.

\begin{figure}[t]
	\begin{center}
	\begin{minipage}[t]{6.5cm}
		\includegraphics[width=1\textwidth]{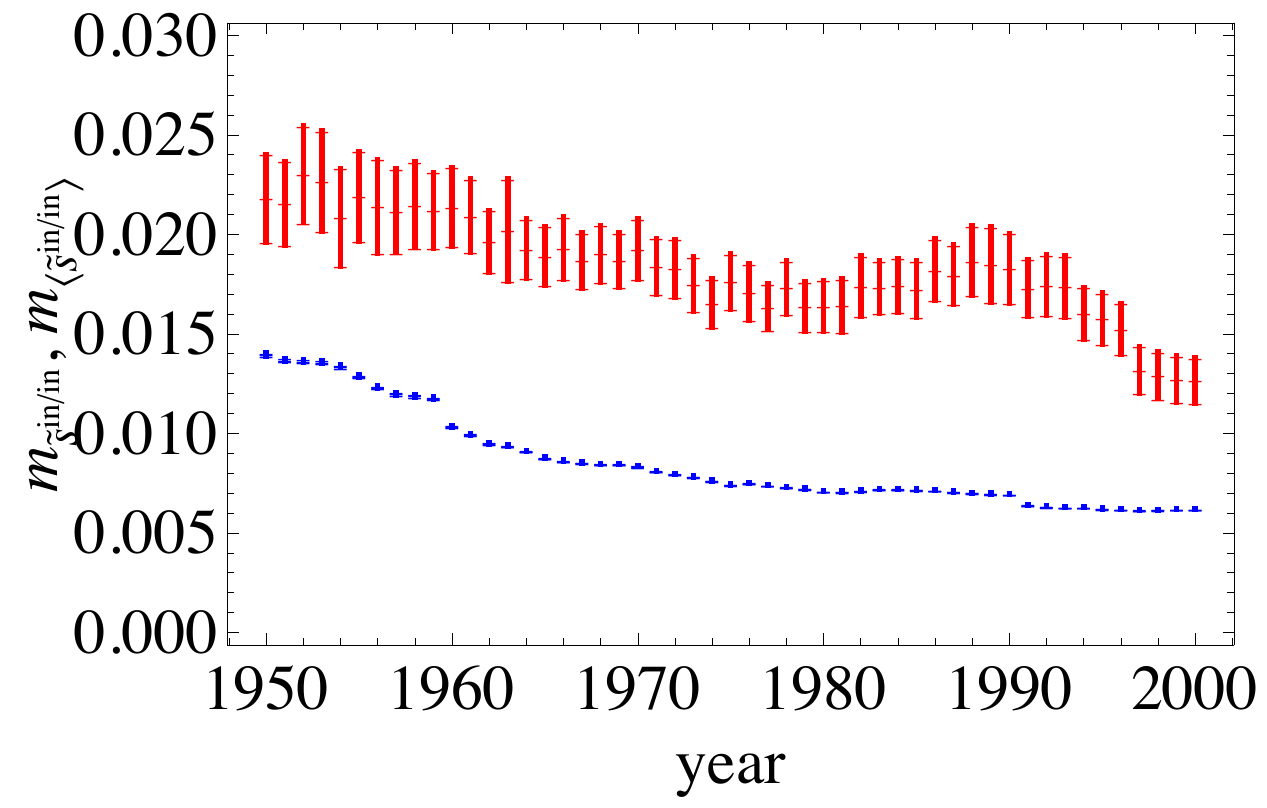}
	\end{minipage}
	\begin{minipage}[t]{6.5cm}
		\includegraphics[width=1\textwidth]{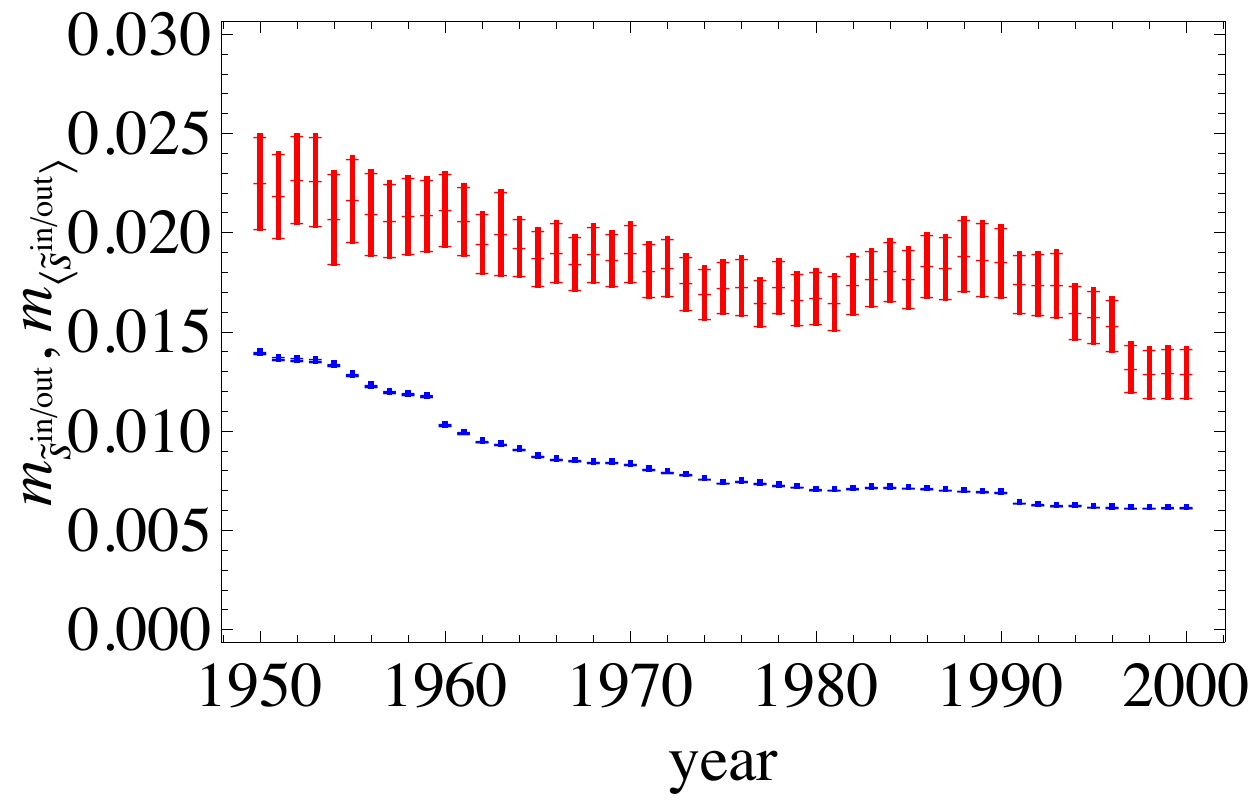}
	\end{minipage}
	\begin{minipage}[t]{6.5cm}
		\includegraphics[width=1\textwidth]{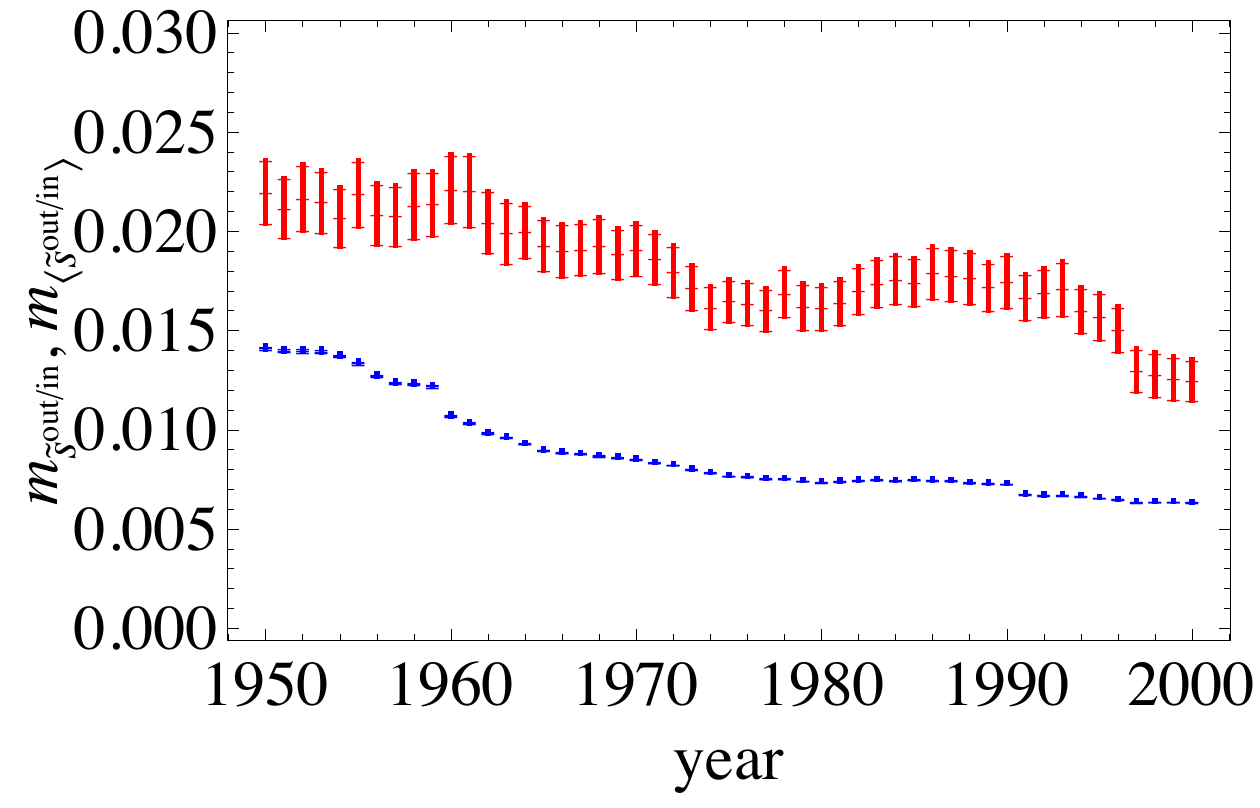}
	\end{minipage}
	\begin{minipage}[t]{6.5cm}
		\includegraphics[width=1\textwidth]{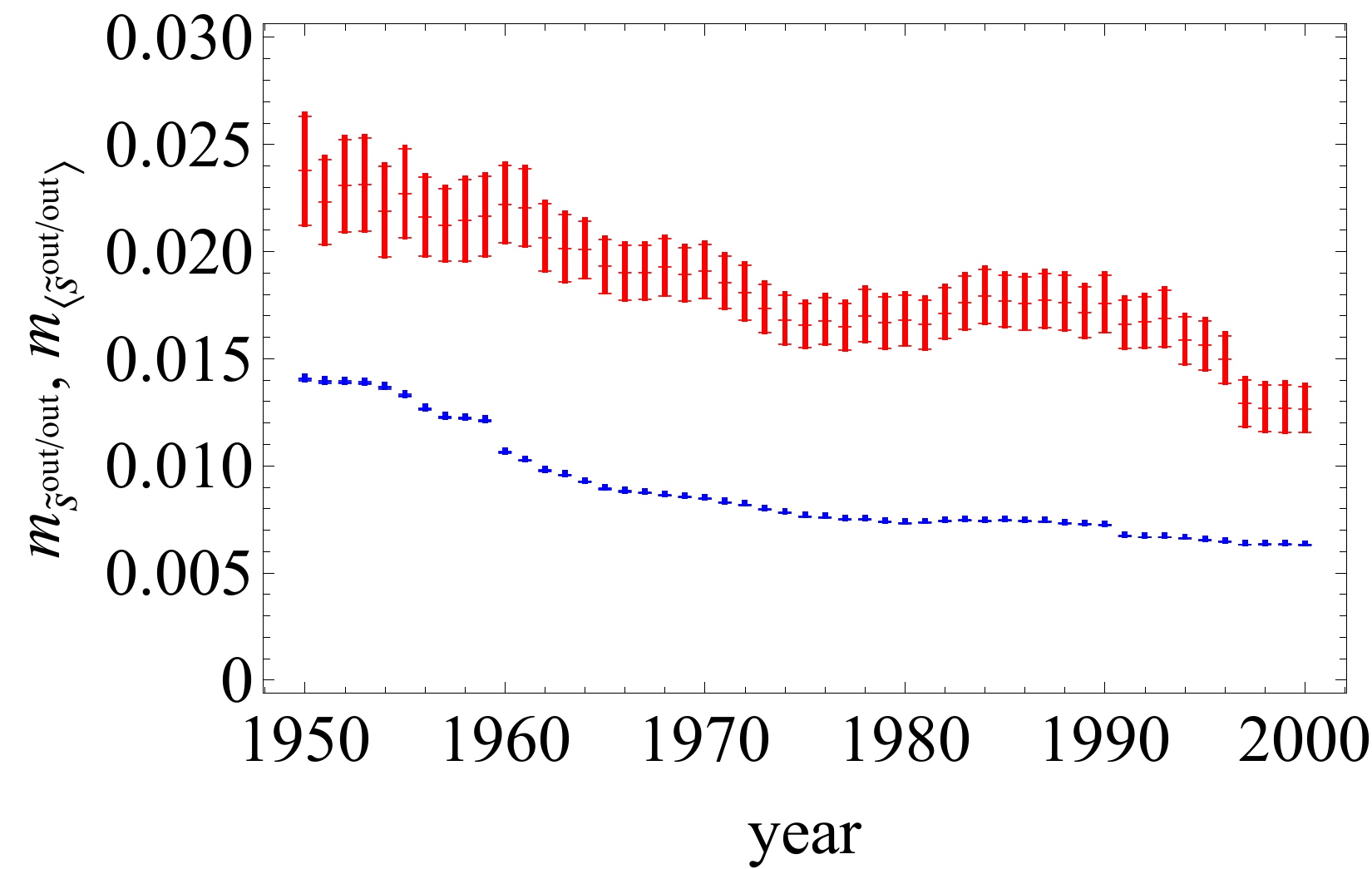}
	\end{minipage}
	\end{center}
	\caption{The weighted-directed WTW: average nearest neighbor strengths and 95\% confidence bands. Red: observed quantities. Blue: null-model fit. Top-left: IN-IN ANNS. Top-right: IN-OUT ANNS. Bottom-left: OUT-IN ANNS. Bottom-right: OUT-OUT ANNS. } \label{Fig:ave_anns}
\end{figure}

\begin{figure}[htbp]
	\begin{center}
	\begin{minipage}[t]{6.5cm}
		\includegraphics[width=1\textwidth]{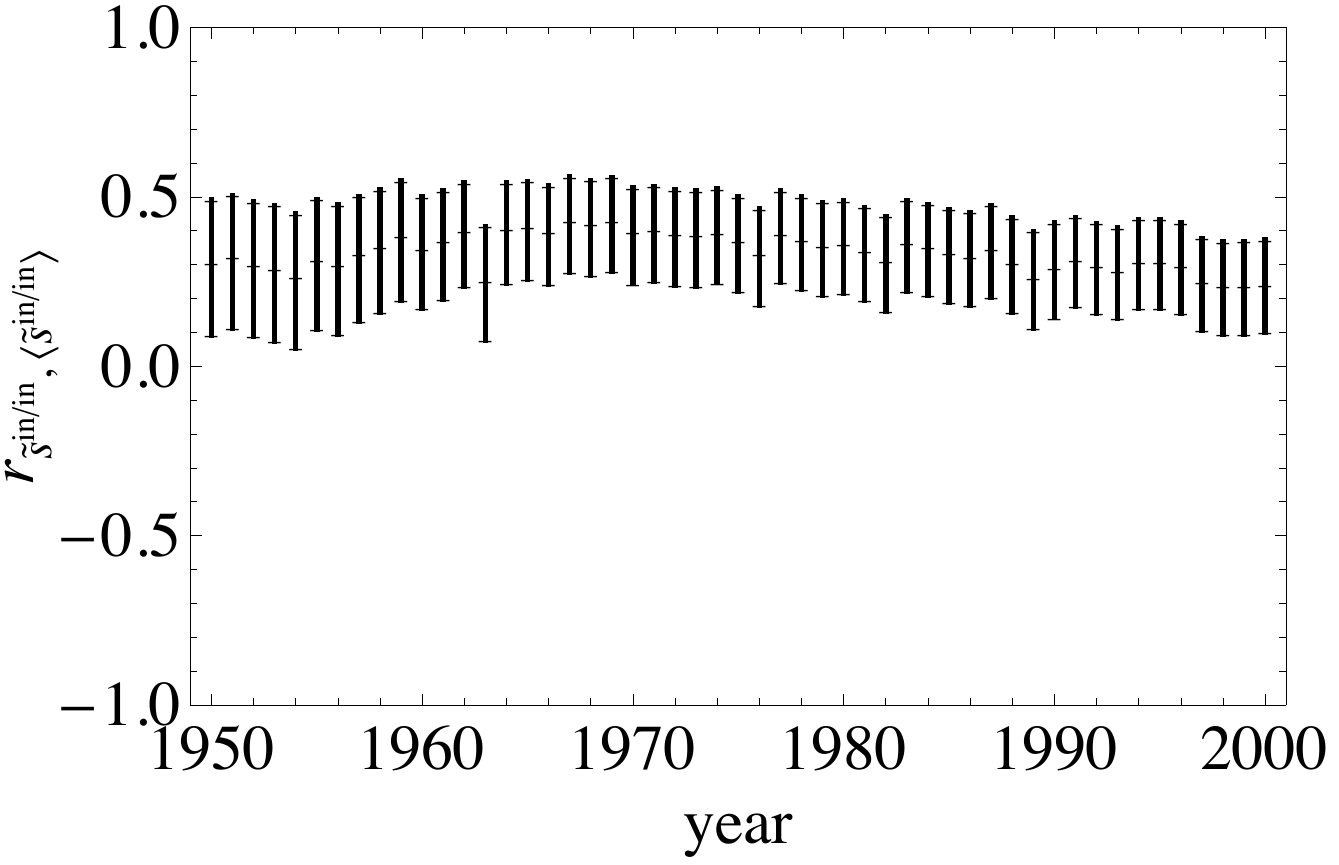}
	\end{minipage}
	\begin{minipage}[t]{6.5cm}
		\includegraphics[width=1\textwidth]{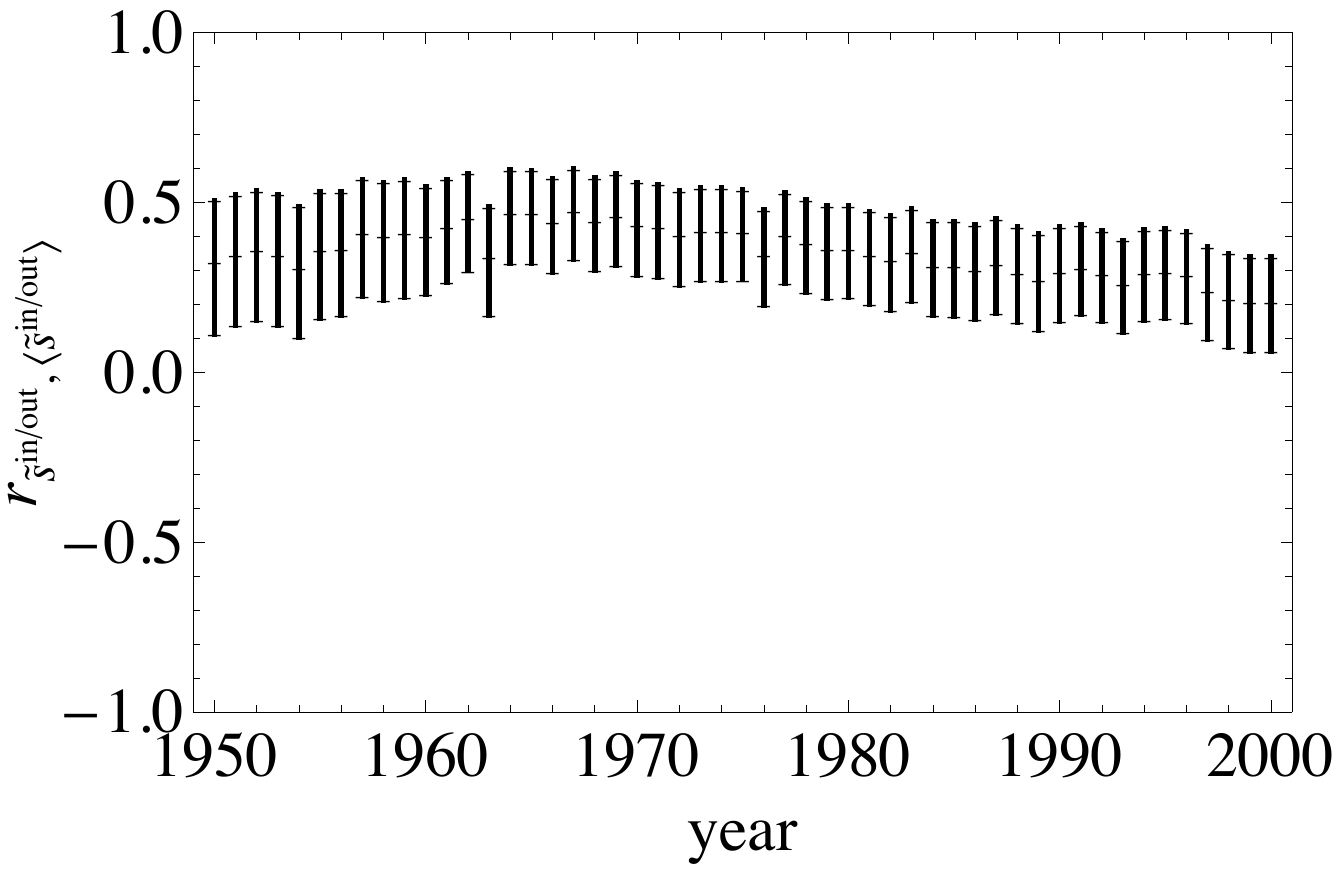}
	\end{minipage}
	\begin{minipage}[t]{6.5cm}
		\includegraphics[width=1\textwidth]{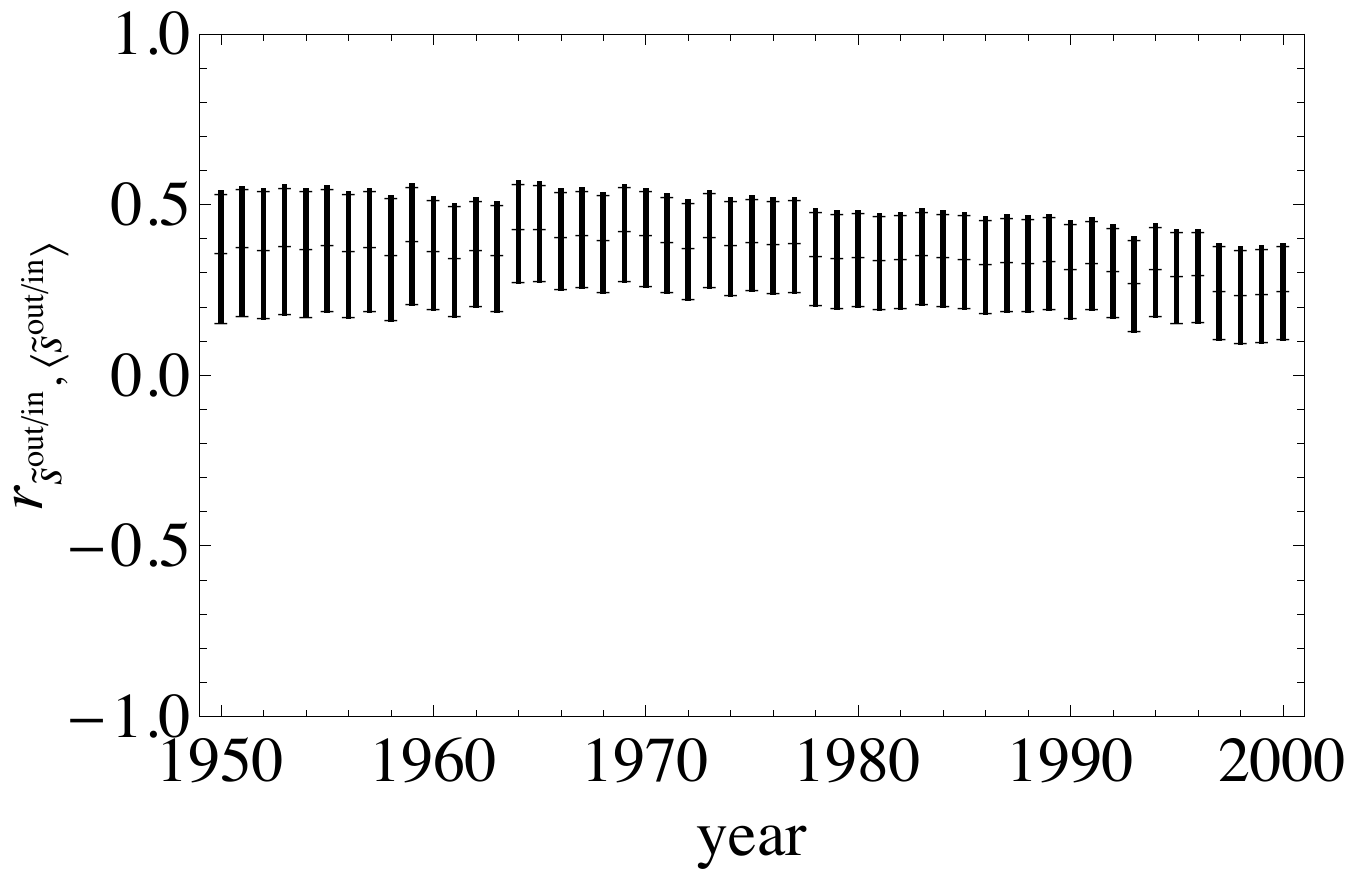}
	\end{minipage}
	\begin{minipage}[t]{6.5cm}
		\includegraphics[width=1\textwidth]{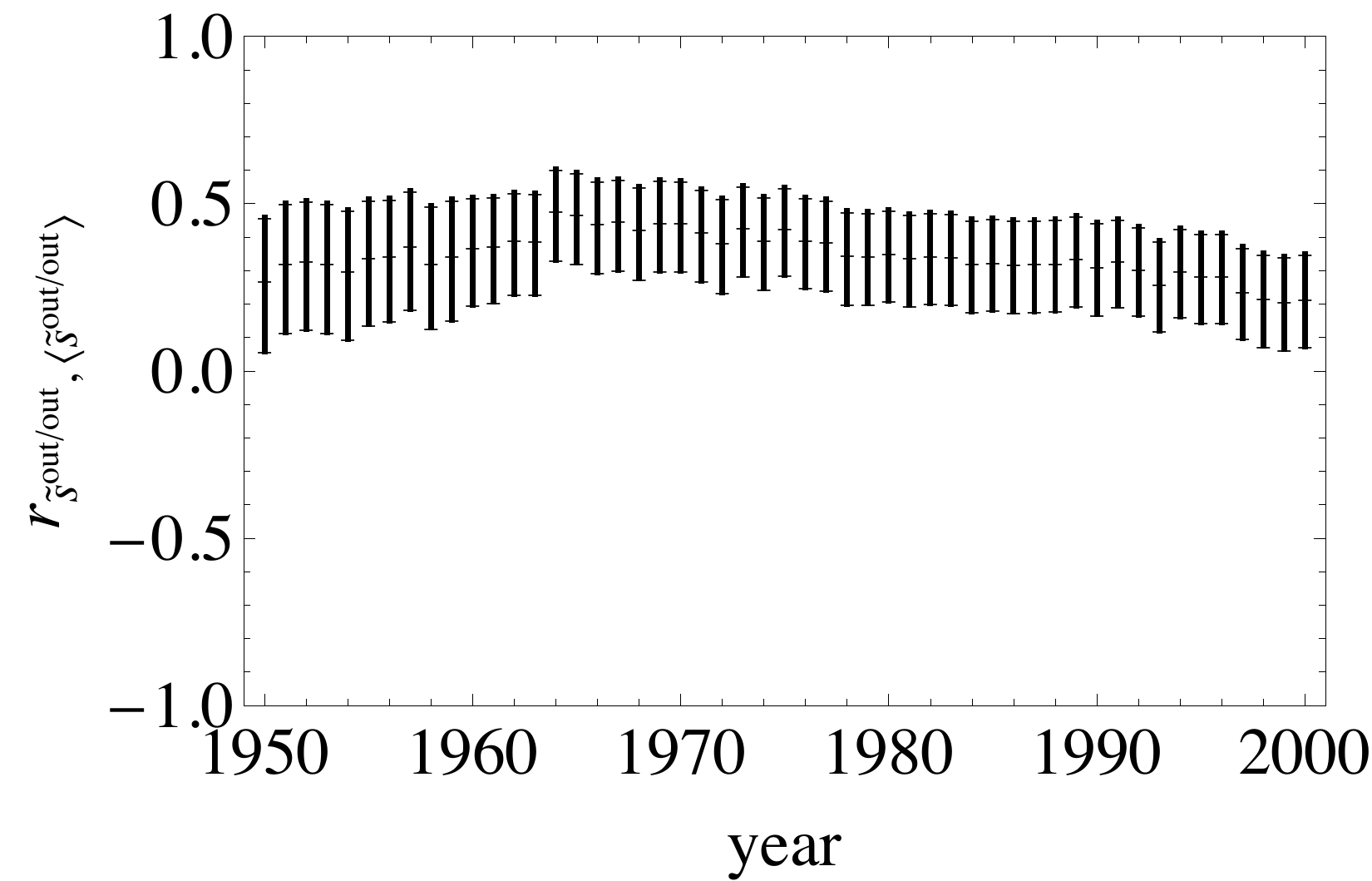}
	\end{minipage}
	\end{center}
	\caption{The weighted-directed WTW: Pearson correlation coefficient between observed and null-model node ANNS. Top-left: IN-IN ANNS. Top-right: IN-OUT ANNS. Bottom-left: OUT-IN ANNS. Bottom-right: OUT-OUT ANNS. } \label{Fig:corr_anns}
\end{figure}

\begin{figure}[htbp]
	\begin{center}
		\includegraphics[width=10cm]{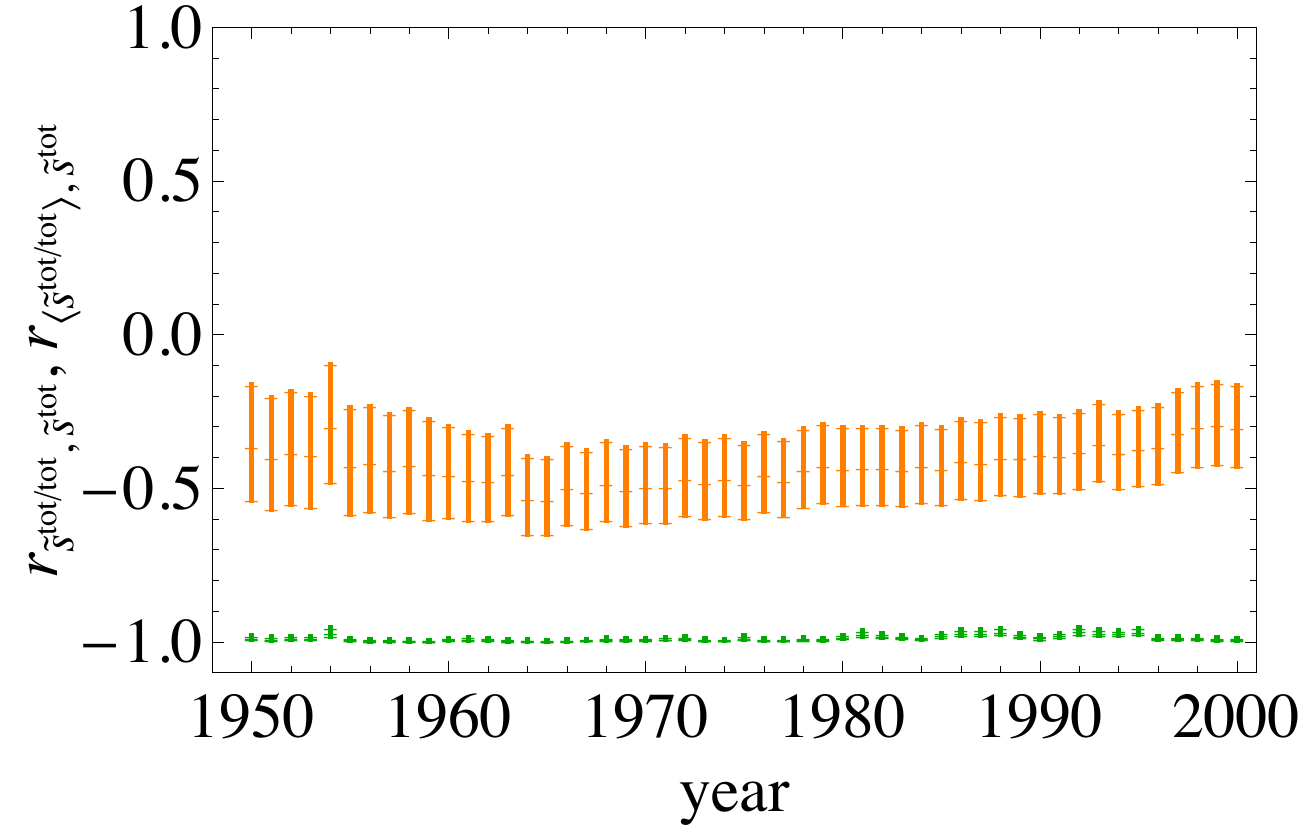}
	\end{center}
	\caption{Disassortativity in the weighted-directed WTW. Orange: Observed correlation between total ANNS and total NS. Green: Correlation between expected total ANNS and observed total NS.} \label{Fig:weighted_disass}
\end{figure}

\begin{figure}[htbp]
	\begin{center}
	\begin{minipage}[t]{6.5cm}
		\includegraphics[width=1\textwidth]{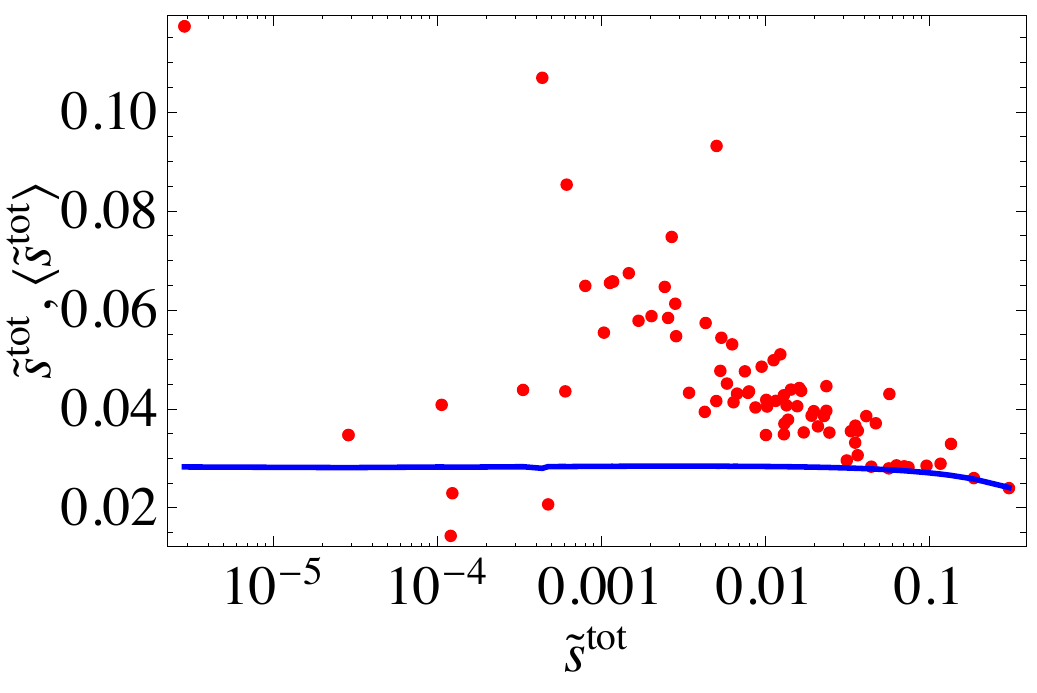}
	\end{minipage}
	\begin{minipage}[t]{6.5cm}
		\includegraphics[width=1\textwidth]{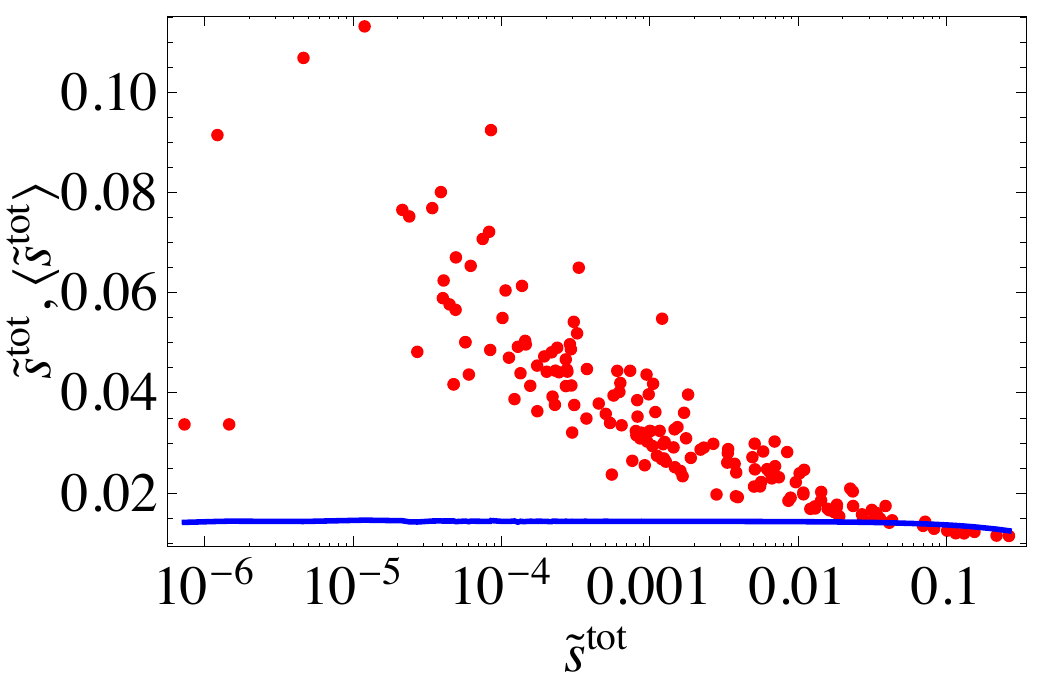}
	\end{minipage}
	\end{center}
	\caption{Disassortativity in the weighted WTW. Scatter plots of total ANNS vs. observed total node strength in 1950 (left) and 2000 (right). Red: observed quantities. Blue: null-model fit.} \label{Fig:scatters_tot_disassortativity_weighted}
\end{figure}

To begin with, note that over the years average ANNS has been slightly decreasing, hinting to a process where better connected countries (i.e., those with higher NS) have been gradually connecting with weakly-connected countries. The null model can replicate this trend but fails completely to predict the level of average ANNS, see Figure \ref{Fig:ave_anns}. Indeed, irrespective of the ANNS disaggregation we consider, the null model persistently predicts a lower population-average ANNS. The bad agreement between observed and expected ANNS can be also appreciated by looking at the correlation coefficients between observed and expected node ANNS in each year (Figure \ref{Fig:corr_anns}), which fluctuate between 0 and 0.5 and exhibit very large error bars. This indicates that the null model controlling for in- and out-strength sequences possesses a very poor ability in matching ANNS figures over time, irrespective of the year considered. As a consequence, also disassortativity patterns cannot be well predicted by the null model. Figure \ref{Fig:weighted_disass} plots how the correlation coefficients between total (observed vs. expected) ANNS and observed NS (i.e. a measure of assortativity in weighted networks) change through time. It is easy to see that, contrary to what happens in the binary WTW, the null model always predict an extreme disassortativity also for the weighted-network characterization of the WTW, which instead displays a weakly disassortative pattern in the entire sample period. The bad agreement between observed data and null-model predictions occurs in the whole sample period, cf. the scatter plots in Figure \ref{Fig:scatters_tot_disassortativity_weighted} for the cases of 1950 and 2000.\footnote{Note that the null model misses not only the scale of disassortativity in the network, but also the scale of ANNS levels associated to every observed NS. Indeed, the blue line in Figure \ref{Fig:scatters_tot_disassortativity_weighted}, which describes expected ANNS, appears flat only because it attains values in a very narrow ANNS range, and not because the ANNS-NS correlation is close to zero.} This confirms and extends results previously obtained for the period 1991-2000 by \citet{Squartini_etal_2011a_pre,Squartini_etal_2011b_pre}.

\begin{figure}[htbp]
	\begin{center}
	\begin{minipage}[t]{6.5cm}
		\includegraphics[width=1\textwidth]{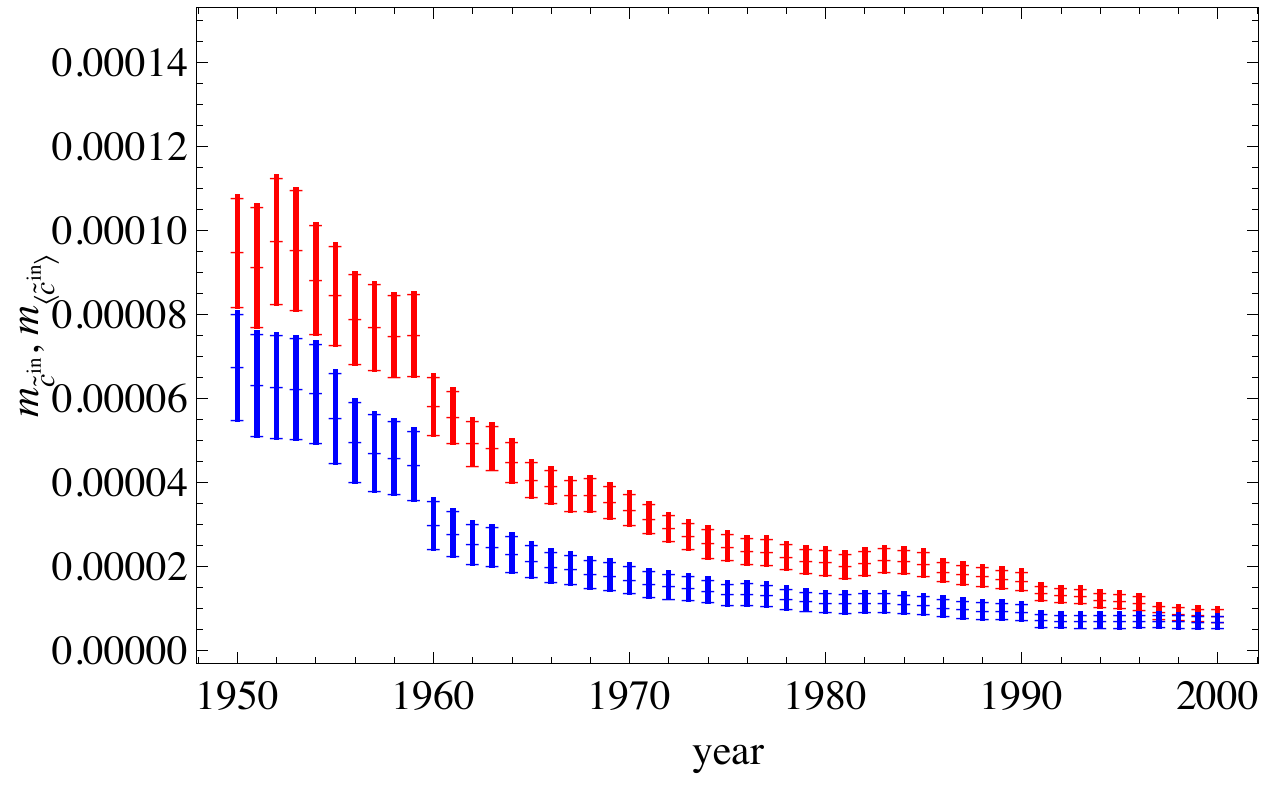}
	\end{minipage}
	\begin{minipage}[t]{6.5cm}
		\includegraphics[width=1\textwidth]{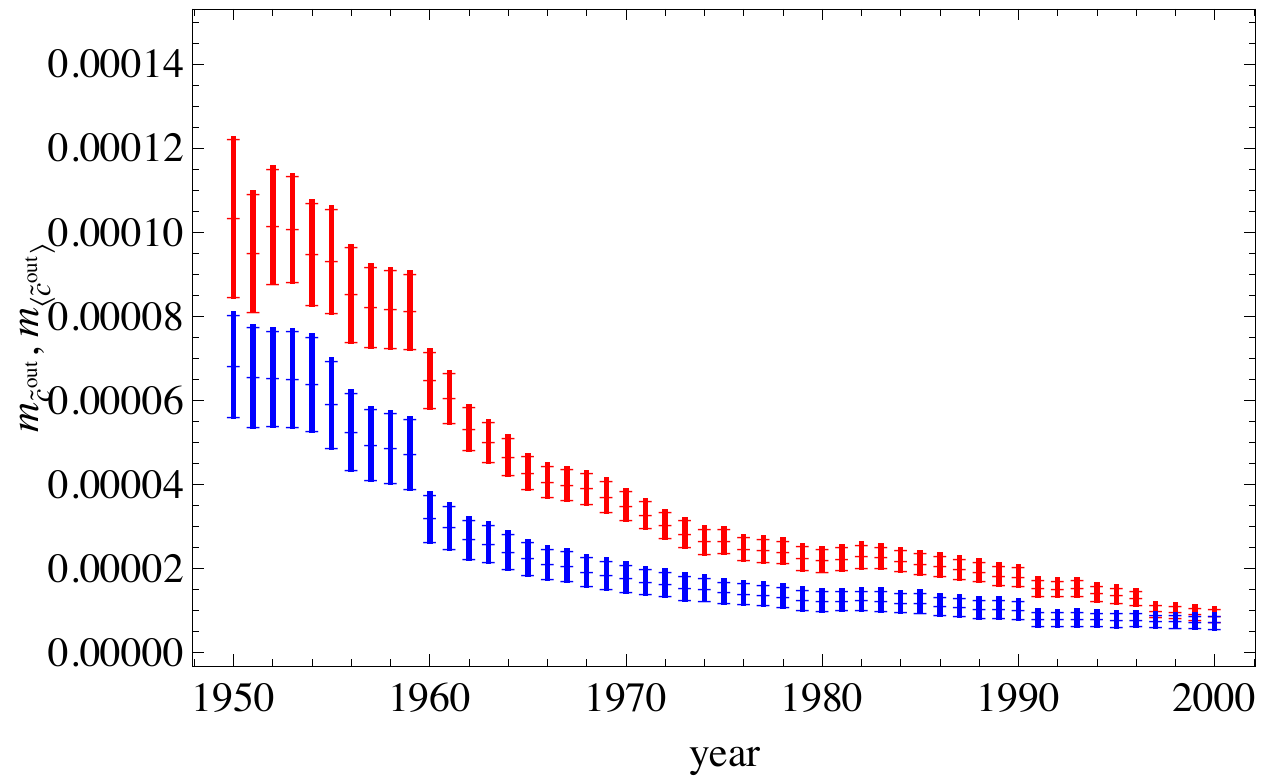}
	\end{minipage}
	\begin{minipage}[t]{6.5cm}
		\includegraphics[width=1\textwidth]{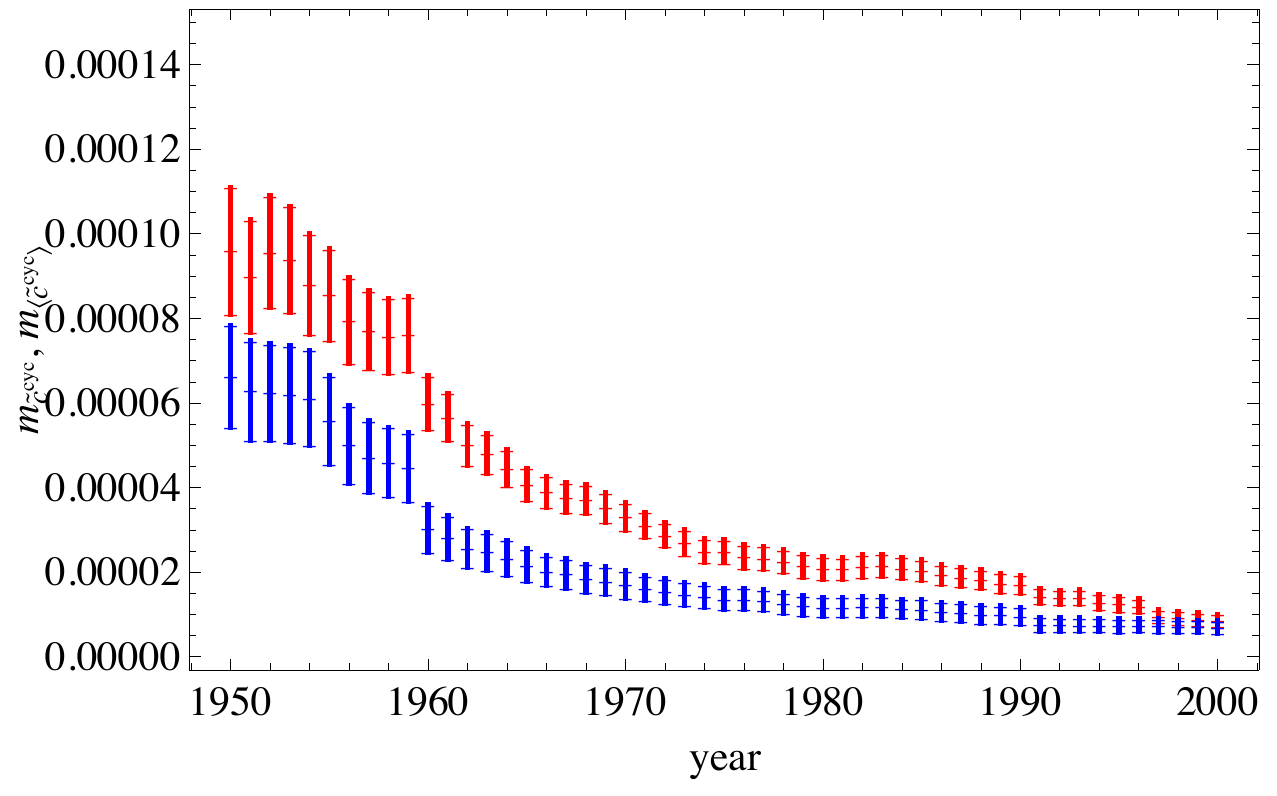}
	\end{minipage}
	\begin{minipage}[t]{6.5cm}
		\includegraphics[width=1\textwidth]{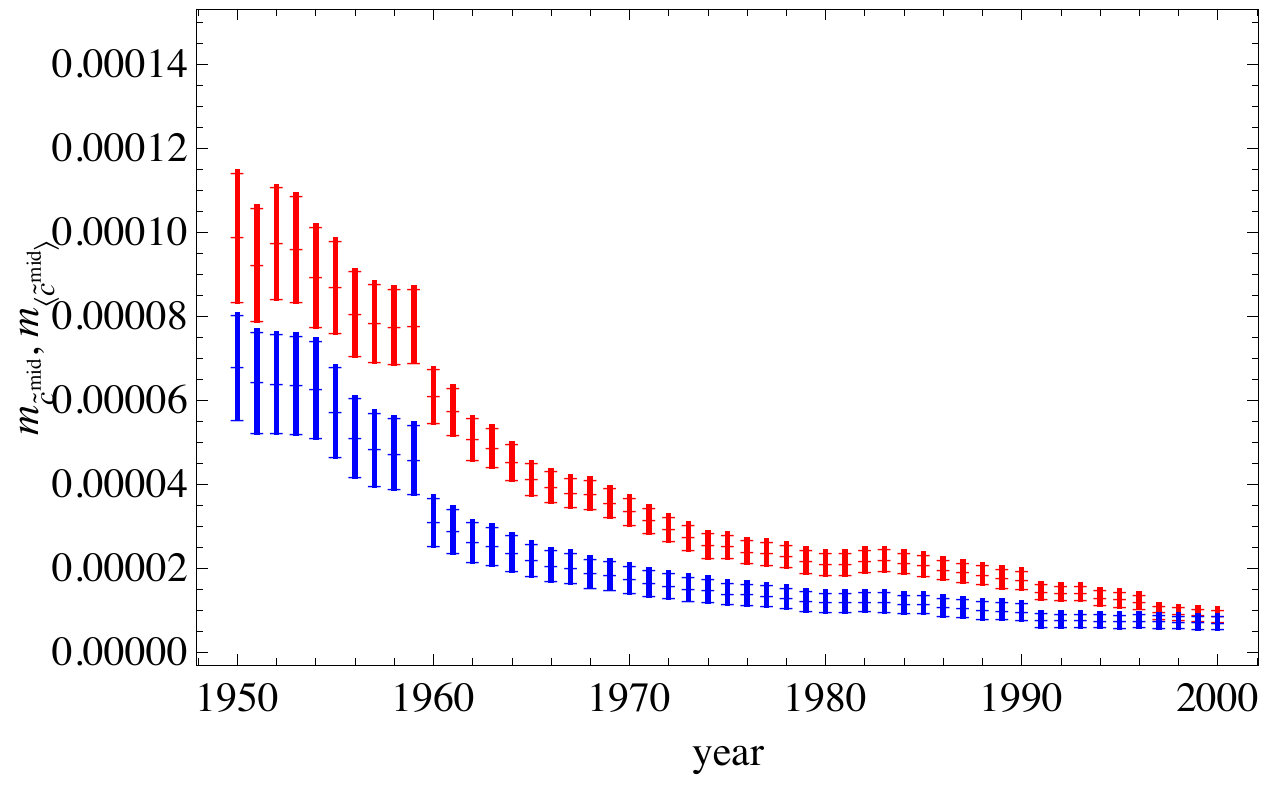}
	\end{minipage}
	\end{center}
	\caption{The weighted-directed WTW: average weighted clustering coefficients and 95\% confidence bands. Red: observed quantities. Blue: null-model fit. Top-left: WCC In. Top-right: WCC Out. Bottom-left: WCC Cycle. Bottom-right: WCC Middleman.} \label{Fig:ave_wcc}
\end{figure}

\begin{figure}[htbp]
	\begin{center}
	\begin{minipage}[t]{6.5cm}
		\includegraphics[width=1\textwidth]{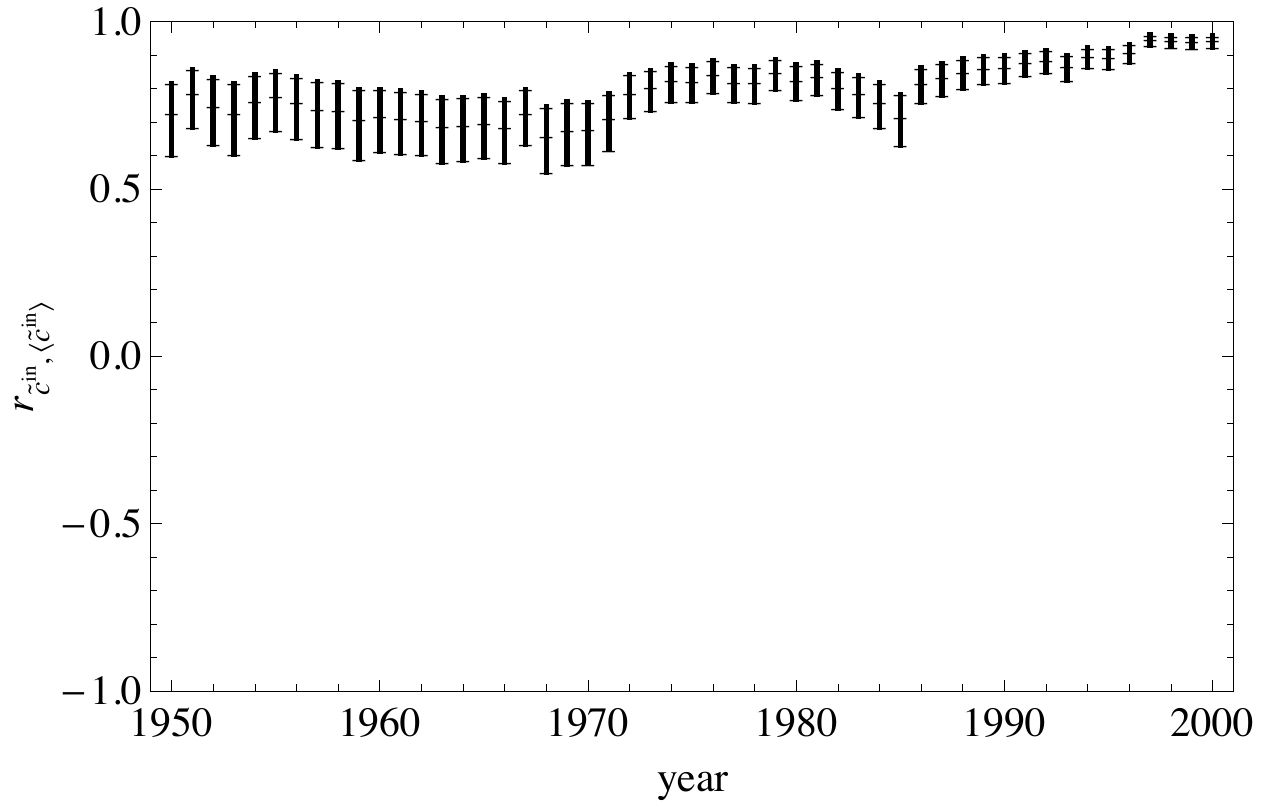}
	\end{minipage}
	\begin{minipage}[t]{6.5cm}
		\includegraphics[width=1\textwidth]{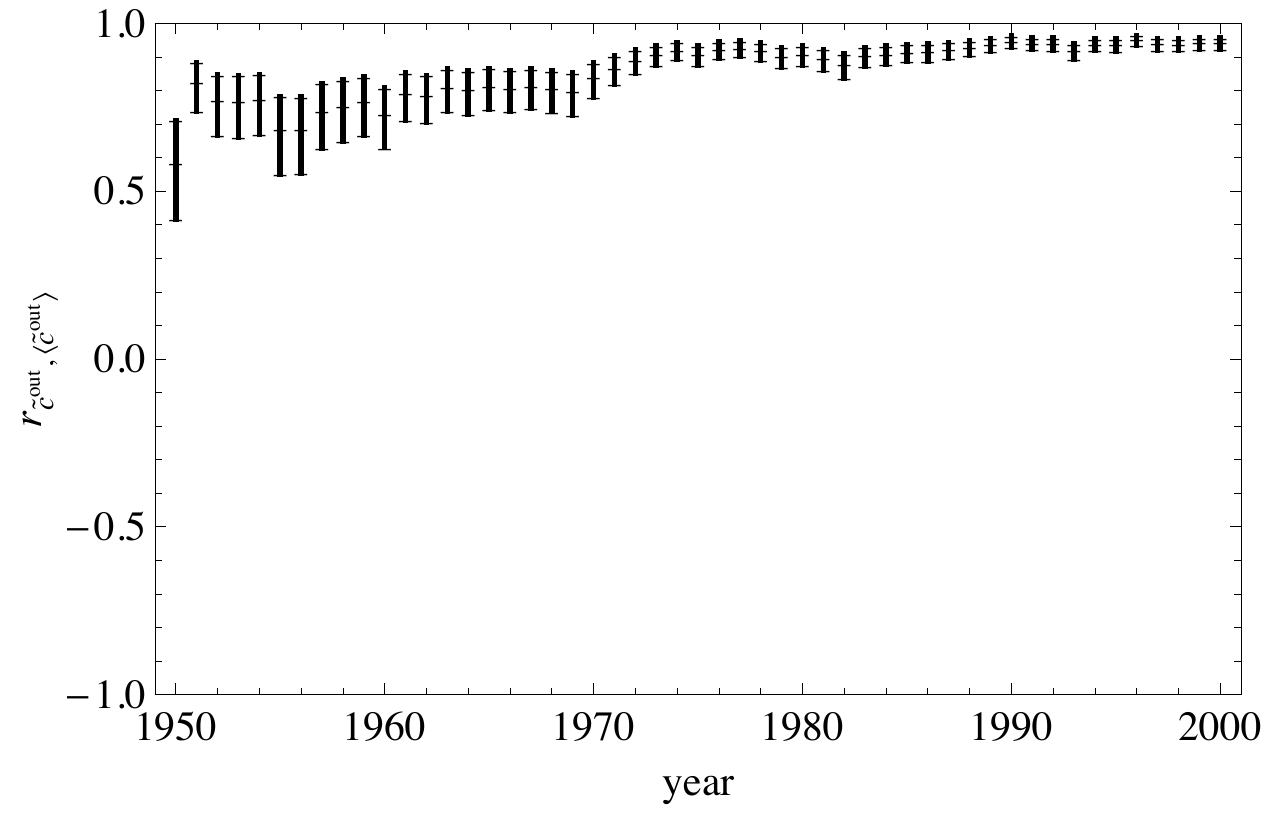}
	\end{minipage}
	\begin{minipage}[t]{6.5cm}
		\includegraphics[width=1\textwidth]{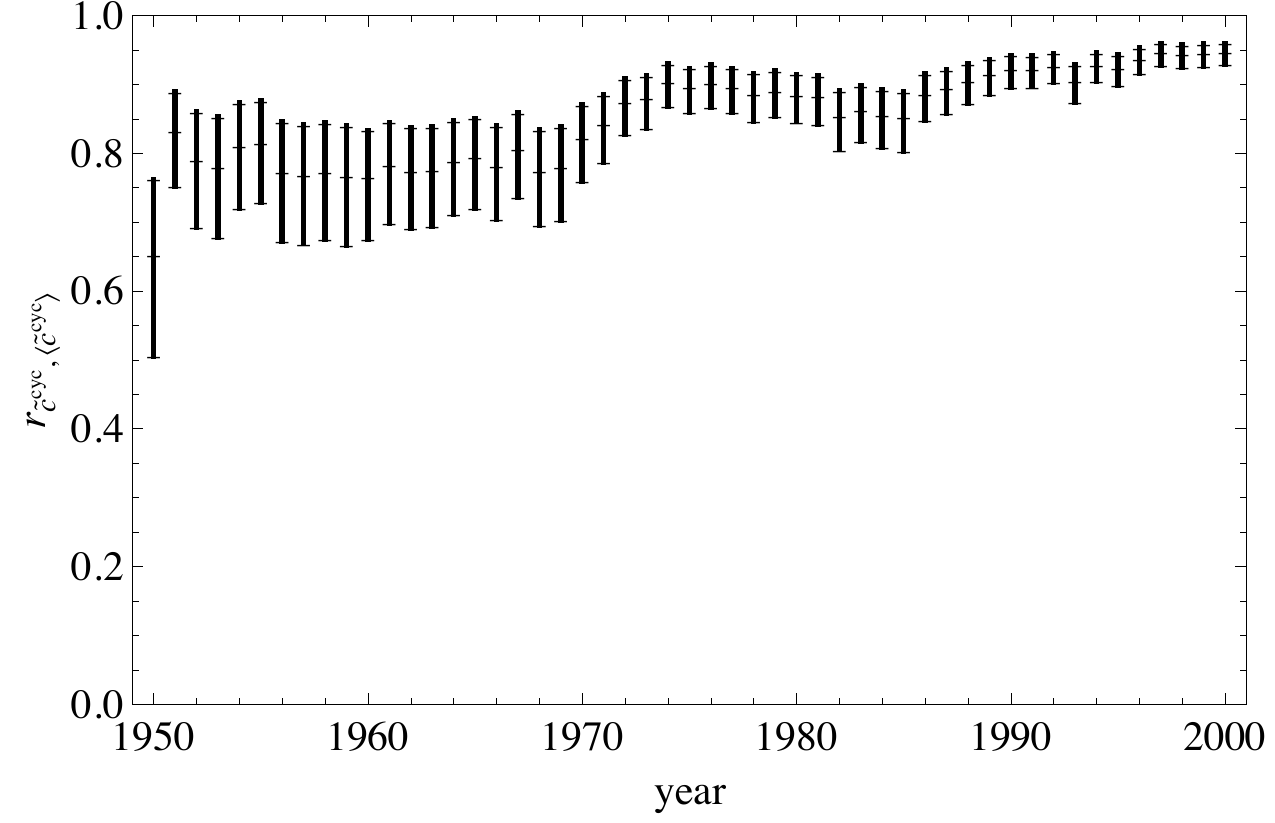}
	\end{minipage}
	\begin{minipage}[t]{6.5cm}
		\includegraphics[width=1\textwidth]{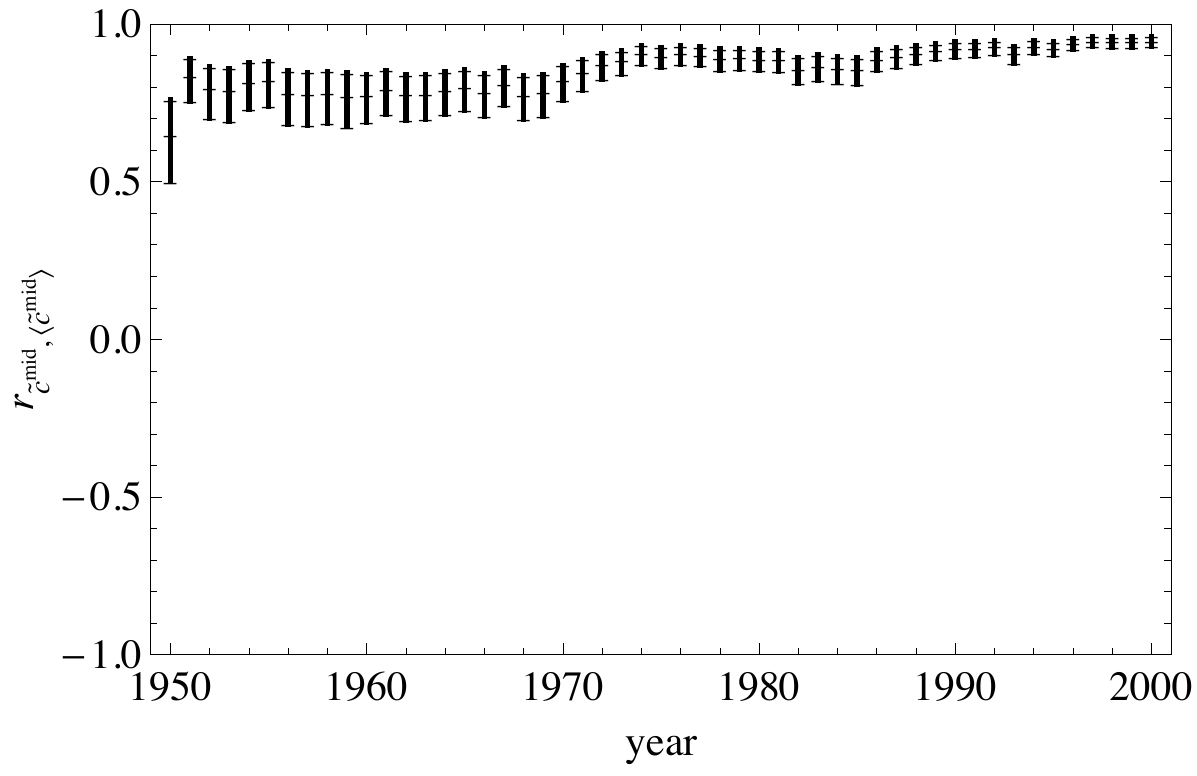}
	\end{minipage}
	\end{center}
	\caption{The weighted-directed WTW: Pearson correlation coefficient between observed and null-model node weighted-clustering coefficients and 95\% confidence bands. Top-left: WCC In. Top-right: WCC Out. Bottom-left: WCC Cycle. Bottom-right: WCC Middleman.} \label{Fig:corr_wcc}
\end{figure}

Weighted-clustering patterns convey a similar message. The null model persistently underestimates  average WCC values until we get to the very final part of the sample (Figure \ref{Fig:ave_wcc}). In particular, the disagreement is very strong in the 50's and 60's. Nevertheless, the null model is able to replicate, as it happened for ANNS, the decreasing trend in average clustering. Furthermore, as Figure \ref{Fig:corr_wcc} suggests, the agreement of the null model in replicating weighted-clustering patterns improves when we approach the last part of the sample. Not also that confidence bands tend to shrink over time, thus signaling a better fit to the data. Again, this is in accordance with results previously obtained for a shorter time window by \citet{Squartini_etal_2011a_pre,Squartini_etal_2011b_pre}.

\begin{figure}[htbp]
	\begin{center}
		\includegraphics[width=10cm]{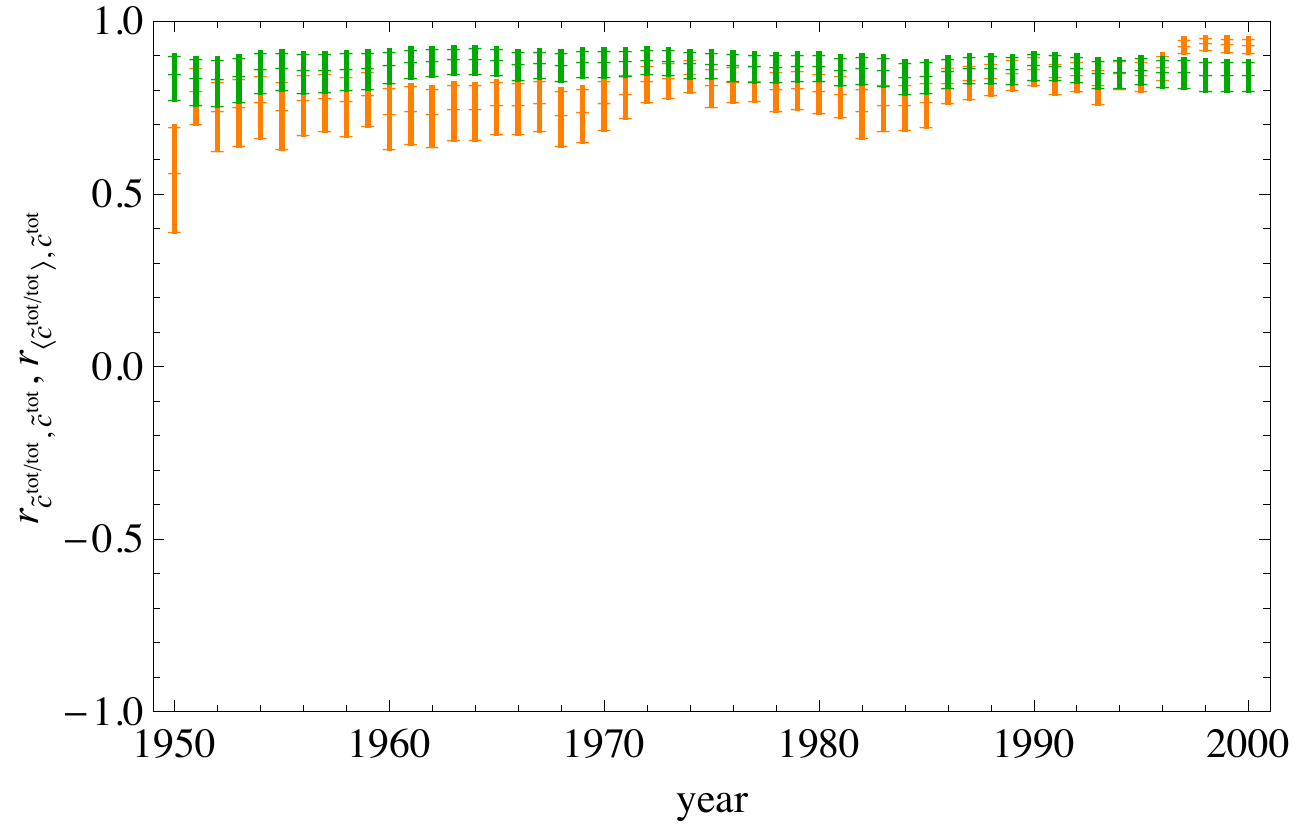}
	\end{center}
	\caption{Correlation between total weighted clustering coefficient and node strength in the weighted-directed WTW. Orange: Observed correlation between total WCC and total NS. Green: Correlation between expected total WCC and observed total NSD.} \label{Fig:weighted_cluststr}
\end{figure}

\begin{figure}[htbp]
	\begin{center}
	\begin{minipage}[t]{6.5cm}
		\includegraphics[width=1\textwidth]{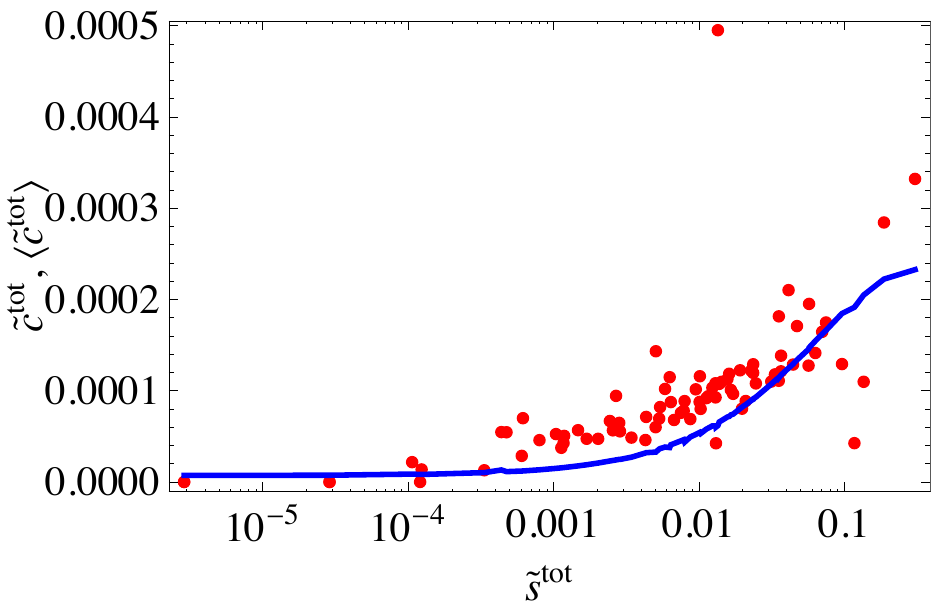}
	\end{minipage}
	\begin{minipage}[t]{6.5cm}
		\includegraphics[width=1\textwidth]{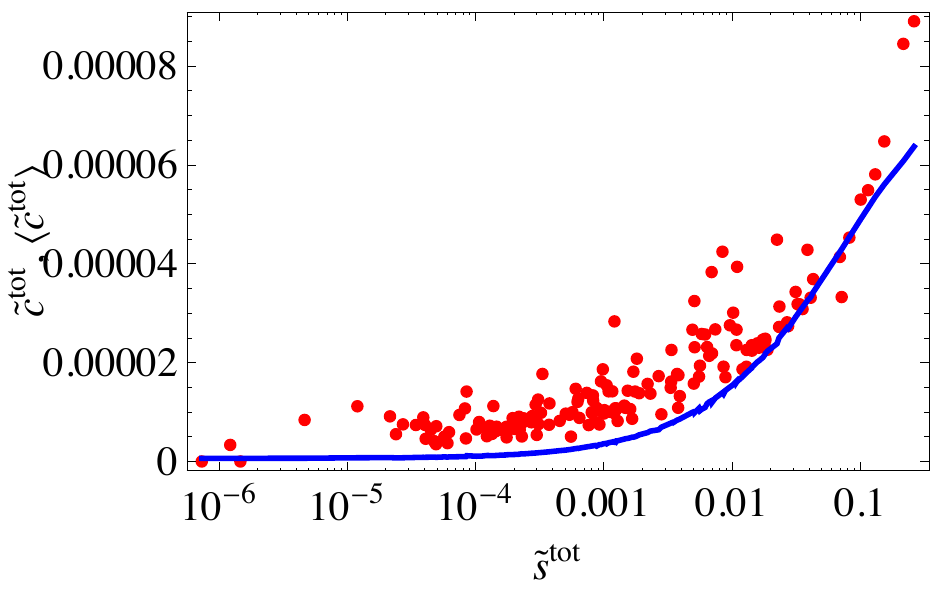}
	\end{minipage}
	\end{center}
	\caption{Weighted clustering coefficient vs. observed total node strength in the weighted WTW. Scatter plots of total WCC vs. NS in 1950 (left) and 2000 (right). Red: observed quantities. Blue: null-model fit.} \label{Fig:scatters_tot_wcc_s}
\end{figure}

Another well-known property that differentiate binary and weighted analysis of the WTW is the fact that, on the one hand, countries holding more partners are also more clustered, whereas countries better connected in terms of node strength typically trade with partners that are poorly connected between them (i.e., the correlation between WCC and NS is negative and high). This is because high-NS countries often entertain many weak trade relationships with countries that trade very poorly between them, therefore yielding low-weight triangles \citep{Fagiolo2008physa}. Figure \ref{Fig:weighted_cluststr} shows that the null model employed here persistently underestimates the high and positive correlation observed in the data between WCC and NS. The agreement improves after 1980, as expected values tend to increase over time and overestimate observed WCC-NS correlation in the very last years under analysis. Despite this improvement, however, estimated weighted clustering badly predicts observed WCC values in the entire node-strength range, as testified by Figure \ref{Fig:scatters_tot_wcc_s}.      

\section{Discussion}\label{Sec:Discussion}

The analysis presented so far aimed at exploring the ability of a family of random null-network models to replicate the observed topological properties of the WTW in the 1950-2000 period. 

Our results suggest that in the binary representation of the WTW, a null random model controlling only for observed in- and out-degree sequences does a good job in reproducing disassortativity and clustering patterns. This is true especially for the last part of the sample, thus confirming results already obtained in \citet{Squartini_etal_2011a_pre,Squartini_etal_2011b_pre}. 

However, the null model is not able to replicate the observed architecture before 1965, where however the binary WTW does not seem to be characterized by statistically-significant correlation relationships. This is a value added with respect to the study in \citet{Squartini_etal_2011a_pre,Squartini_etal_2011b_pre}. Indeed, in those papers, we have employed data covering a shorter time span (1992-2000) in order to allow for the largest variability in commodity-specific data as possible. That choice prevented us to fully analyze how the null model performed across a longer time span. Here, focusing on aggregate data, we are able to fully address this question. We shall go back to this point below.    
    
All this implies that, from a network perspective, disassortativity and clustering profiles observed in the binary WTW after 1965 arise as natural outcomes rather than genuine correlations, once the local topological properties are fixed to their observed values.

From an international-trade perspective, conversely, these results indicate that binary network descriptions of trade can be significantly simplified by considering the degree sequence(s) only. This implies that, in any binary representation of the WTW, knowing how many importing and exporting partners a given country holds, turns out to be maximally informative, since its knowledge conveys almost the entire information about the topology of the network. In other words, the patterns observed in the binary WTW do not require the presence of higher-order mechanisms as an additional explanation, beside knowledge of degree sequences. The fact that node degrees alone are enough to explain higher-order network properties means that the degree sequence is an important structural pattern in its own. This highlights the importance of explaining the observed degree sequence in international-trade models. 

Our weighted-network analysis, on the contrary, shows that the picture changes completely when explicitly considering heterogeneity in link weights. Indeed, most of observed topological properties cannot be reproduced by the corresponding null-random model where one controls for in- and out-strength sequences (i.e., total country imports and exports). This indicates that the WTW is an excellent example of a network whose higher-order weighted topological properties cannot be deduced from its local weighted properties. 

Taken together, these results have two important implications for international-trade models. First, the binary analysis, by indicating that degree sequences are maximally informative, suggests that trade models should be substantially revised in order to  explicitly include the degree sequence of the WTW among the key properties to reproduce. Note that standard international-trade models like the micro-founded gravity model \citep[which is the work-horse theoretical apparatus in international-trade theoretical analyses, cf.][]{GravityBook} do not aim at explaining or reproducing the observed degree sequence but focus more on the structure of bilateral weights. Our results suggest that one of the main focuses of international-trade theories should become explaining the determinants underlying the emergence and persistence of the very first trade relationship between any two countries previously not connected by trade links.\footnote{An earlier attempt to go in this direction has been recently made by \citet{HMR_2008,EKS_2012}, who develop international-trade models with heterogeneous firms that are consistent with a number of data facts, including the number of zeros (i.e., network density). However, they fall short of providing explanations of where exactly these zeros are, i.e. why some potential trade-link actually turns on.}

Second, the foregoing findings about weighted WTW statistics indicate that a weighted-network description of trade flows, by focusing on higher-order properties in addition to local ones, captures novel and fresh evidence. Indeed, local properties alone (e.g. knowledge of node in- and out-strengths) are not enough to reproduce observed patterns about weighted disassortativity and clustering. Therefore, traditional analyses of country trade profiles focusing only on local properties and country-specific statistics \citep[e.g., total trade, etc.][]{Feenstra2004} convey a partial description of the richness and detail of the WTW architecture. In turn, economic theories that, like the gravity model, only aim at explaining the local properties of the weighted WTW (i.e., the total values of imports and exports of world countries) are of a limited informative content, as such properties have no predictive power on the rest of the structure of the network.

The foregoing results extend the analysis in \citet{Squartini_etal_2011a_pre,Squartini_etal_2011b_pre} in three related ways. First, we employ a different source of data for bilateral-trade flows. Despite all existing trade-flow databases eventually derive from the COMTRADE dataset, they differ a lot in terms of year coverage, possibility to disaggregate the data according to product categories, and methods employing to clean the raw figures.\footnote{For example, existing databases differ in the way trade flows are reported according to the reporter (importer or exporter), whether zeroes are all considered as missing trade, etc..} The fact that our analysis finds a good match within the same time window employed in previous studies is itself a robustness test. Second, we employ a database which, despite being reliable only for aggregate trade figures, allows us to go back to 1950 as our starting year. This, as discussed above, entails a mismatch between the null model and observed measures for the first part of the sample in the binary case. Indeed, in that sample period the binary WTW does not exhibit any clearcut correlation patterns (e.g., in terms of disassortativity or clustering-degree). The reason why such a mismatch occurs may lie in the third way this paper extends previous analyses. In \citet{Squartini_etal_2011a_pre,Squartini_etal_2011b_pre} the size of the network (i.e., number of nodes) was kept constant, so as to have a balanced panel. This means that a relevant number of countries was systematically eliminated from the sample in more recent years. Conversely, here we focus on a non-balanced country panel. The fact that network size increases over time introduces some discrepancy between balanced and non-balanced topology in terms of binary links, therefore structurally modifying higher-order node statistics such as clustering.   

\section{Concluding Remarks}\label{Sec:Conclusions}

In this paper we have investigated the performance of a family of null random models for the WTW in the period 1950-2000. We have employed a method recently explored in \citet{Squartini_Garlaschelli_2011}, which allows to analytically obtain the expected value of a given network statistic across the ensemble of networks that preserve on average some local properties, and are otherwise fully random. 

We have studied both a binary and a weighted directed representation of the WTW, using as constraints, respectively, the observed node in/out-degree and in/out-strength sequences. This choice is motivated by two related considerations. First, we want to allow for sufficient randomness in the ensemble of null networks in order to provide a relatively loose benchmark model against which comparing observed statistics. Indeed, our null model should not embody too strict assumptions on the way links and weights are placed. At the same time, the null model should not generate with a positive probability variants of the WTW that are completely impossible from an economic point of view. Therefore, a good compromise is to control for either degree or strength sequences, i.e. fixing as constraints either the number of import/export trade partners of a country, or its total import and export values. Second, as already mentioned, by controlling for node degree and strength sequences, we are preserving the local structure of the WTW, and consequently information coming from standard international-trade statistics. Studying the performance of the null model as far as higher-order network statistics are concerned (e.g., assortativity and clustering) allows us to check whether a network approach to international trade can convey fresh insights.

The analysis presented in this work may be extended in many ways. First, one can explore the space of null models by considering alternative constraints. For example, one may study what happens in the binary case when only in- or out-degree sequences are kept fixed (and not the two together), to understand if import or export partnerships play a different role in explaining higher-order properties. In the weighted case, a null model where also in- and out-degree sequences are controlled for may be instead employed to investigate whether the joint knowledge of partnership number and trade value can better replicate assortativity and clustering also in the first part of the sample.\footnote{See \cite{Bhatta2007a} for an attempt in this direction.}


Second, one can study the extent to which our null models are able to replicate additional higher-order properties of the network, like geodesic distances, node centrality indicators, emergence of cliques, etc..

Third, one might think to explore in more details the properties of the dynamic process underlying network evolution. It must be indeed noticed that the nature of the analysis above was of a comparative-static nature. Every snapshot of the ITN is considered as isolated by the precedent and the following one. Our focus was not on explaining why the network has been changing over time, but much more on understanding the performance of the same enforced topological properties over subsequent snapshots. One interesting avenue of research would then require to think to dynamic null models, when one introduces some time-dependence into the set of constraints.    
 
Finally, the foregoing analysis intentionally focused on network statistics as the only candidate constraints. This may limit the scope of the study, as it is well-known from the gravity-equation literature \citep{GravityBook,Fagiolo2010jeic,Garla2004} that bilateral link weights and network properties are heavily influenced by country size and income (i.e. GDP and per-capita GDP), geographical distance, and a number of other country-related and bilateral interaction factors. Notice that by controlling for in- and out-strength one is already taking into account some size effect, as country total import and export is somewhat positively correlated with country size. Nevertheless, by directly considering country GDP and geographical distance in the analysis, an important and fruitful bridge between traditional international-trade analyses and complex-network approaches to trade may be hopefully established.

\singlespacing
\bigskip


\bibliographystyle{gfagsm_jae}
\bibliography{rewiring_jebo}

\newpage


\begin{table}
\begin{tabular}{p{2.5cm}|p{6cm}|p{8cm}}
\textbf{Topological Properties} &\textbf{Observed} & \textbf{Expected}\\
\hline
Binary Link & $a_{ij}\newline a_{ji}$ & $p_{ij}=\frac{x_{i}^{out}x_{j}^{in}}{1+x_{i}^{out}x_{j}^{in}}\newline p_{ji}=\frac{x_{j}^{out}x_{i}^{in}}{1+x_{j}^{out}x_{i}^{in}}$\\
\hline
Degrees & $k_{i}^{in}=\sum_{j}a_{ji}\newline k_{i}^{out}=\sum_{j}a_{ij}\newline k_{i}^{tot}=k_{i}^{in}+k_{i}^{out}$ & $\langle k_{i}^{in}\rangle=k_{i}^{in}\newline \langle k_{i}^{out}\rangle=k_{i}^{out}\newline \langle k_{i}^{tot}\rangle=\langle k_{i}^{in}\rangle+\langle k_{i}^{out}\rangle$\\
\hline
ANND & $k_{i}^{nn,\:inin}=\frac{\sum_{j}a_{ji}k_{j}^{in}}{k_{i}^{in}}$ \newline $k_{i}^{nn,\:inout}=\frac{\sum_{j}a_{ji}k_{j}^{out}}{k_{i}^{in}}$ \newline $k_{i}^{nn,\:outin}=\frac{\sum_{j}a_{ij}k_{j}^{in}}{k_{i}^{out}}$ \newline $k_{i}^{nn,\:outout}=\frac{\sum_{j}a_{ij}k_{j}^{out}}{k_{i}^{out}}$ \newline $k_{i}^{nn,\:tot}=\frac{\sum_{j}(a_{ij}+a_{ji})k_{j}^{tot}}{k_{i}^{tot}}$ & $\langle k_{i}^{nn,\:inin}\rangle=\frac{\sum_{j}p_{ji}k_{j}^{in}}{k_{i}^{in}}$ \newline $\langle k_{i}^{nn,\:inout}\rangle=\frac{\sum_{j}p_{ji}k_{j}^{out}}{k_{i}^{in}}$ \newline $\langle k_{i}^{nn,\:outin}\rangle=\frac{\sum_{j}p_{ij}k_{j}^{in}}{k_{i}^{out}}$\ \newline $\langle k_{i}^{nn,\:outout}\rangle=\frac{\sum_{j}p_{ij}k_{j}^{out}}{k_{i}^{out}}$ \newline $\langle k_{i}^{nn,\:tot}\rangle=\frac{\sum_{j}(p_{ij}+p_{ji})k_{j}^{tot}}{k_{i}^{tot}}$\\
\hline
Clustering & $c_{i}^{cyc}=\frac{\sum_{j}\sum_{k}a_{ij}a_{jk}a_{ki}}{k_{i}^{in}k_{i}^{out}-k_{i}^{\leftrightarrow}}$ \newline  $c_{i}^{mid}=\frac{\sum_{j}\sum_{k}a_{ik}a_{ji}a_{jk}}{k_{i}^{in}k_{i}^{out}-k_{i}^{\leftrightarrow}}$ \newline  $c_{i}^{in}=\frac{\sum_{j}\sum_{k}a_{ki}a_{ji}a_{jk}}{k_{i}^{in}(k_{i}^{in}-1)}$ \newline  $c_{i}^{out}=\frac{\sum_{j}\sum_{k}a_{ik}a_{jk}a_{ij}}{k_{i}^{out}(k_{i}^{out}-1)}$ \newline $c_{i}^{tot}=\frac{\sum_{j}\sum_{k}(a_{ij}+a_{ji})(a_{jk}+a_{kj})(a_{ki}+a_{ik})}{2\big[k_{i}^{tot}(k_{i}^{tot}-1)-2 k_{i}^{\leftrightarrow}\big]}$ & $\langle c_{i}^{cyc}\rangle=\frac{\sum_{j}\sum_{k}p_{ij}p_{jk}p_{ki}}{k_{i}^{in}k_{i}^{out}-\sum_{j}p_{ij}p_{ji}}$ \newline $\langle c_{i}^{mid}\rangle=\frac{\sum_{j}\sum_{k}p_{ik}p_{ji}p_{jk}}{k_{i}^{in}k_{i}^{out}-\sum_{j}p_{ij}p_{ji}}$ \newline $\langle c_{i}^{in}\rangle=\frac{\sum_{j}\sum_{k}p_{ki}p_{ji}p_{jk}}{2\sum_{j<k}p_{ji}p_{ki}}$ \newline $\langle c_{i}^{out}\rangle=\frac{\sum_{j}\sum_{k}p_{ik}p_{jk}p_{ij}}{2\sum_{j<k}p_{ij}p_{ik}}$ \newline $\langle c_{i}^{tot}\rangle=\frac{\sum_{j}\sum_{k}(p_{ij}+p_{ji})(p_{jk}+p_{kj})(p_{ki}+p_{ik})}{2\big[2\sum_{j<k}(p_{ji}p_{ki}+p_{ij}p_{ik})+2(k_{i}^{in}k_{i}^{out})-2\sum_{j}p_{ij}p_{ji}\big]}$\\
\hline
\end{tabular}
\caption{The Binary WTW: Observed and Expected Topological Properties. \textit{Note}: ANND stands for Average Nearest-Neighbor Degree.}\label{Tab:BinaryStats}
\end{table}

\newpage \clearpage

\begin{table}
\begin{tabular}{p{2.5cm}|p{6cm}|p{8cm}}
\textbf{Topological Properties} &\textbf{Observed} & \textbf{Expected}\\
\hline
Link Weights & $e_{ij} \newline e_{ji} \newline w_{ij}$ & $\langle e_{ij}\rangle=\frac{y_{i}^{out}y_{j}^{in}}{1-y_{i}^{out}y_{j}^{in}} \newline \langle e_{ji}\rangle=\frac{y_{j}^{out}y_{i}^{in}}{1-y_{j}^{out}y_{i}^{in}} \newline \langle w_{ij}\rangle=\frac{\langle e_{ij}\rangle}{\sum_{i\neq j}\langle e_{ij}\rangle}$\\
\hline
Strength & $s_{i}^{in}=\sum_{j}w_{ji} \newline s_{i}^{out}=\sum_{j}w_{ij} \newline s_{i}^{tot}=s_{i}^{in}+s_{i}^{out}$ & $\langle s_{i}^{in}\rangle=s_{i}^{in} \newline \langle s_{i}^{out}\rangle=s_{i}^{out} \newline \langle s_{i}^{tot}\rangle=s_{i}^{tot}$\\
\hline
ANNS & $s_{i}^{nn,\:inin}=\frac{\sum_{j}a_{ji}s_{j}^{in}}{k_{i}^{in}}$ \newline $s_{i}^{nn,\:inout}=\frac{\sum_{j}a_{ji}s_{j}^{out}}{k_{i}^{in}}$ \newline $s_{i}^{nn,\:outin}=\frac{\sum_{j}a_{ij}s_{j}^{in}}{k_{i}^{out}}$ \newline $s_{i}^{nn,\:outout}=\frac{\sum_{j}a_{ij}s_{j}^{out}}{k_{i}^{out}}$ \newline $s_{i}^{nn,\:tot}=\frac{\sum_{j}(a_{ij}+a_{ji})s_{j}^{tot}}{k_{i}^{tot}}$ & $\langle s_{i}^{nn,\:inin}\rangle=\frac{\sum_{j}p_{ji}s_{j}^{in}}{\langle k_{i}^{in}\rangle}$ \newline $\langle s_{i}^{nn,\:inout}\rangle=\frac{\sum_{j}p_{ji}s_{j}^{out}}{\langle k_{i}^{in}\rangle}$ \newline $\langle s_{i}^{nn,\:outin}\rangle=\frac{\sum_{j}p_{ij}s_{j}^{in}}{\langle k_{i}^{out}\rangle}$ \newline $\langle s_{i}^{nn,\:outout}\rangle=\frac{\sum_{j}p_{ij}s_{j}^{out}}{\langle k_{i}^{out}\rangle}$ \newline $\langle s_{i}^{nn,\:tot}\rangle=\frac{\sum_{j}(p_{ij}+p_{ji})s_{j}^{tot}}{\langle k_{i}^{tot}\rangle}$\\
\hline
Clustering & $c_{i}^{w,\:cyc}=\frac{\sum_{j}\sum_{k}w_{ij}^{1/3}w_{jk}^{1/3}w_{ki}^{1/3}}{k_{i}^{in}k_{i}^{out}-k_{i}^{\leftrightarrow}}$  \newline  $c_{i}^{w,\:mid}=\frac{\sum_{j}\sum_{k}w_{ik}^{1/3}w_{jk}^{1/3}w_{ji}^{1/3}}{k_{i}^{in}k_{i}^{out}-k_{i}^{\leftrightarrow}}$ \newline $c_{i}^{w,\:in}=\frac{\sum_{j}\sum_{k}w_{jk}^{1/3}w_{ji}^{1/3}w_{ki}^{1/3}}{k_{i}^{in}(k_{i}^{in}-1)}$  \newline  $c_{i}^{w,\:out}=\frac{\sum_{j}\sum_{k}w_{ik}^{1/3}w_{ij}^{1/3}w_{jk}^{1/3}}{k_{i}^{out}(k_{i}^{out}-1)}$  \newline $c_{i}^{w,\:tot}=\frac{\sum_{j}\sum_{k}(w_{ij}^{1/3}+w_{ji}^{1/3})(w_{jk}^{1/3}+w_{kj}^{1/3})(w_{ki}^{1/3}+w_{ik}^{1/3})}{2\big[k_{i}^{tot}(k_{i}^{tot}-1)-2 k_{i}^{\leftrightarrow}\big]}$  & $\langle c_{i}^{w,\:cyc}\rangle=\frac{\sum_{j}\sum_{k}\langle w_{ij}^{1/3}\rangle \langle w_{jk}^{1/3}\rangle \langle w_{ki}^{1/3}\rangle}{\langle k_{i}^{in}\rangle \langle k_{i}^{out}\rangle-\sum_{j}p_{ij}p_{ji}}$ \newline $\langle c_{i}^{w,\:mid}\rangle=\frac{\sum_{j}\sum_{k}\langle w_{ik}^{1/3}\rangle \langle w_{jk}^{1/3}\rangle \langle w_{ji}^{1/3}\rangle}{\langle k_{i}^{in}\rangle \langle k_{i}^{out}\rangle-\sum_{j}p_{ij}p_{ji}}$ \newline $\langle c_{i}^{w,\:in}\rangle=\frac{\sum_{j}\sum_{k}\langle w_{jk}^{1/3}\rangle \langle w_{ji}^{1/3}\rangle \langle w_{ki}^{1/3}\rangle}{2\sum_{j<k}p_{ji}p_{ki}}$ \newline  $\langle c_{i}^{w,\:out}\rangle=\frac{\sum_{j}\sum_{k}\langle w_{ik}^{1/3}\rangle \langle w_{ij}^{1/3}\rangle \langle w_{jk}^{1/3}\rangle}{2\sum_{j<k}p_{ij}p_{ik}}$ \newline $\langle c_{i}^{w,\:tot}\rangle=\frac{\sum_{j}\sum_{k}\langle w_{ij}^{1/3}+w_{ji}^{1/3}\rangle \langle w_{jk}^{1/3}+ w_{kj}^{1/3}\rangle \langle w_{ki}^{1/3}+w_{ik}^{1/3}\rangle}{2\big[2\sum_{j<k}(p_{ji}p_{ki}+p_{ij}p_{ik})+2(\langle k_{i}^{in}\rangle \langle k_{i}^{out}\rangle)-2\sum_{j}p_{ij}p_{ji}\big]}$\\

\hline
\end{tabular}
\caption{The Weighted WTW: Observed and Expected Topological Properties. \textit{Note}: ANNS stands for Average Nearest-Neighbor Strength.}\label{Tab:WeightedStats}
\end{table}


\newpage \clearpage


\bigskip \bigskip

\newpage \clearpage

\bigskip \bigskip

\newpage \clearpage

\bigskip \bigskip

\bigskip \bigskip

\newpage

\end{document}